\documentclass[10pt, reprint, twocolumn, amssymb, amsmath, nofootinbib, tightenlines, floatfix, longbibliography, citeautoscript, amsmath, groupedaddress, ajp]{revtex4-1}

\usepackage{fancyhdr}
\fancyheadoffset{0cm}
\usepackage{comment}
\usepackage{bm}
\usepackage{amsmath}
\usepackage{amsfonts}
\usepackage{amssymb}
\usepackage{float}
\usepackage{graphicx}
\usepackage{dcolumn}
\usepackage{mathtools}

\usepackage{hyperref}
\hypersetup{
    colorlinks=true,
    linktocpage=true,
    linkcolor=blue,
    filecolor=magenta,      
    urlcolor=blue,
    citecolor=blue,
    bookmarks=true,
}

\makeatletter
\renewcommand*\l@section{\@dottedtocline{1}{1.5em}{2.3em}}
\makeatother

\graphicspath{{Graphics/}}

\newcommand{\ep}{{\epsilon}}

\newcommand{\be}{\begin}
\newcommand{\en}{\end}
\newcommand{\eq}{equation}

\newcommand{\delep}{\delta\epsilon}
\newcommand{\etal}{\textit{et al.}}

\newcommand{\bdivd}{\beta / \delta}

\begin{document}

\lhead{\small{D. M. Riffe and R. B. Wilson}}
\rhead{\small{Ultrafast Relaxation dynamics of excited carriers...}}
\cfoot{--\thepage--}

\title{Ultrafast relaxation dynamics of excited carriers in metals:  Simplifying the intertwined dependencies upon scattering strengths, phonon temperature, photon energy, and excitation level}

\begin{abstract}
Using the Boltzmann transport equation (BTE), we study the evolution of nonequilibrium carrier distributions in simple ($sp$) metals, assumed to have been instantaneously excited by an ultrafast laser pulse with photon energy $h \nu$.  The mathematical structure of the BTE  scattering integrals reveals that $h \nu$ is a natural energy scale for describing the dynamics.  Normalizing all energy quantities by $h \nu$ leads to a set of three unitless parameters---$\bdivd$, $\gamma$, and $\alpha$---that control the relaxation dynamics:  $\bdivd$ is the normalized ratio of electron-phonon to electron-electron scattering strengths, $\gamma$ is the normalized phonon (lattice) temperature, and $\alpha$ is the normalized absorbed energy density.  Using this theory, we systematically investigate relaxation times for the high-energy part of the distribution ($\tau_H$), energy transfer to the phonon subsystem ($\tau_E$), and intracarrier thermalization ($\tau_{th}$).  In the linear region of response (valid when $\alpha$ is sufficiently small), we offer heuristic descriptions of each of these relaxation times as functions of $\bdivd$ and $\gamma$.  Our results as a function of excitation level $\alpha$ show that many ultrafast experimental investigations lie in a transition region between low excitation (where the relaxation times are independent of $\alpha$) and high excitation (where the two-temperature model of carrier dynamics is valid).  Approximate boundaries that separate these three regions are described by simple expressions involving the normalized parameters of our model.
\end{abstract}

\author{D. M. Riffe}
\email[Author to whom correspondence should be addressed; electronic mail: ]{mark.riffe@usu.edu}

\affiliation{Physics Department, Utah State University, Logan, UT 84322, USA}

\author{Richard B. Wilson}

\affiliation{Mechanical Engineering Department and Materials Science and Engineering Department, University of California, Riverside, CA 92521, USA}

\date{\today}

\maketitle

\section{Introduction}

The carrier relaxation dynamics of an ultrafast-laser-excited metal are important to a number of current scientific endeavors, including ultrafast magnetism \cite{Battiato2010,Schellekens2014,Choi2015,Wilson2017,Wilson2017b,Yang2017,Alekhin2017,Seifert2018}, hot-carrier induced photocatalysis \cite{Linic2015,Zhang2017,Aslam2018,Szczerbinski2018}, solar-energy conversion \cite{Clavero2014,Wu2015,Jang2016}, interpretation of ultrafast x-ray \cite{Chen2011b,Chen2011a,White2014,Lu2020} and electron \cite{Park2005,Nie2006,Ligges2009,Ligges2011,Chase2016,Sokolowski-Tinten2017,Mo2018} diffraction, and inelastic light-scattering and photoluminescence signals from plasmonic materials \cite{Szczerbinski2018,Carattino2018,Jollans2020,Hogan2020,Wu2021,Huang2014,Cai2019,Suemoto2019,Ono2020,Suemoto2020b,Suemoto2021,Sivan2023,Rodriguez2023}.  As we discuss in detail here, in any given experiment there are five key parameters that govern the relaxation:  the electron-phonon ($ep$) scattering strength $K_{ep}$, electron-electron ($ee$) scattering strength $K_{ee}$, phonon temperature $T_p$, laser photon energy $h \nu$, and absorbed energy density $u_a$.  Indeed, in any ultrafast-laser investigation of a metal these parameters (or some subset) are typically utilized to make quantitative sense of the results.

In addition to calculations aimed at describing the results of any particular experiment, a number of theoretical studies have made progress in discerning how these five parameters contribute to general trends associated with ultrafast carrier relaxation \cite{Groeneveld1992,Groeneveld1995,Gusev1998,Wilson2020,Kabanov2008,Baranov2014,Tas1994,Rethfeld2002,pietanza2007,Mueller2013}.  In general, this work utilizes the Boltzmann transport equation (BTE) to gain insight into the consequences of simultaneous electron-phonon and electron-electron scattering.  Taken separately, the basic effects of these two scattering mechanisms are simple:  electron-electron scattering evolves  a nonthermal carrier distribution into a thermal one, and electron-phonon scattering mediates energy transfer from the excited carriers to vibrational degrees of freedom.  Taken together, however, the two scattering mechanisms intertwine to produce nontrivial dynamics.  

Several studies explore these dynamics in the low-excitation limit.  The early BTE computations of Groeneveld \etal~reveal what is now a fundamental tenet of excited carrier dynamics:  strong $ee$ scattering enhances energy transfer to the phonons \cite{Groeneveld1992,Groeneveld1995}.  This effect is due to the increase in carrier number associated with intracarrier thermalization.  Gusev and Wright analytically study the BTE to extract details of this synergism and find the energy transfer time $\tau_E$ is proportional to $(\tau_0)^{1/3} (\tau_{S})^{2/3}$ (their notation), where $\tau_0$ is a characteristic $ee$ scattering time and $\tau_S$ is the time for a single carrier to lose some fixed amount of energy \cite{Gusev1998}.  Gusev and Wright also investigate intracarrier thermalization and deduce if $ep$ scattering is relatively weak, then the thermalization time $\tau_{th}$ is proportional to $T_p^{\,-2}$.  Wilson and Coh numerically integrate the BTE to investigate carrier dynamics as a function of the ratio of $ep$ to $ee$ scattering strengths in the low-excitation limit \cite{Wilson2020}.  They confirm energy relaxation has a nontrivial relationship to the two scattering mechanisms.  In a range of ratios that applies to many metals, they find $\tau_E \sim (\tau_0)^{1/4} (\tau_S)^{3/4}$, a result not unlike that of Gusev and Wright.  They also derive a more general hueristic relationship for $\tau_E$.  In addition, Wilson and Coh show relaxation of the highest energy carriers is typically controlled by $ee$ scattering.  Kabanov and coworkers explore the mathematical structure of the BTE scattering integrals in detail, deriving approximate solutions for the BTE in various limits \cite{Kabanov2008,Baranov2014}.  They show (for typical metals such as Au and Ru) that by the time excited carriers can be considered thermalized those carriers have already lost a large fraction of their energy to the phonons \cite{Baranov2014}.  Evidently, in at least some circumstances thermalization and energy relaxation take place on similar timescales.   

A few studies look at the dependence of carrier relaxation upon excitation level.  Via general consideration of an approximate $ee$ scattering integral, Tas and Maris infer $\tau_{th}$ is proportional to $u_a^{-1}$ \cite{Tas1994}.  Other calculations of intracarrier thermalization focus on specific materials.  The numerical studies of Al by Rethfeld \textit{et al.}~\cite{Rethfeld2002} and Al and Au by Mueller and Rethfeld \cite{Mueller2013} show a substantial decrease in $\tau_{th}$ as the absorbed energy density is increased, but the dependence on $u_a$ is not as simple as that suggested by Tas and Maris.  The study of Al by Rethfeld \textit{et al.}~\cite{Rethfeld2002} and a study of Ag by Pietanza \etal~\cite{pietanza2007} investigate the energy decay of several laser excited distributions.  Both studies show if $u_a$ is large enough, then energy decay of the laser excited distribution is indistinguishable from that of a thermal [i.e., Fermi-Dirac (FD)] distribution.  Additionally, the calculations of Rethfeld \textit{et al.}~show the timescale for carrier cooling becomes larger with increasing laser excitation \cite{Rethfeld2002}. 

To summarize, three important timescales are identified by these prior investigations.  The first is the timescale $\tau_H$ for relaxation of the high-energy part of the distribution \cite{Wilson2020}.  The second is the timescale $\tau_E$ for transfer of energy from the excited electrons to the phonons \cite{Groeneveld1992,Groeneveld1995,Gusev1998,Wilson2020,Kabanov2008,Baranov2014,Rethfeld2002,pietanza2007}.  The third is the the timescale $\tau_{th}$ for intracarrier thermalization of the laser excited distribution \cite{Gusev1998,Kabanov2008,Baranov2014,Tas1994,Rethfeld2002,pietanza2007,Mueller2013}.

Unfortunately, a systematic understanding of the dependence of these three timescales on the five key parameters---$K_{ep}$, $K_{ee}$, $T_p$, $u_a$, and $h \nu$---is still decidedly lacking.  Partly this is due to a lack of consistent nomenclature among different studies, which makes it challenging to compare results.  More importantly, though, is the fact that extant calculations do not encompass the full ranges of values for these five parameters that are relevant to experimental studies. 

The recent study by Wilson and Coh partially addresses this situation by investigating $\tau_H$ and $\tau_E$ as a functions of the $ep$ and $ee$ scattering strengths and laser photon energy \cite{Wilson2020} over pertinent values.  Overall, they show $\tau_H$ and $\tau_E$ can largely be characterized by the ratio of $K_{ep}$ to $K_{ee}$, with some dependence upon $h \nu$. However, they limit their study to low excitation (small $u_a$) and a room temperature (RT) lattice ($T_p = 300$ K).  Here we build upon this recent study---as well as the earlier studies---in the following ways.  

We begin by studying the structure of BTE scattering integrals (Sec.~\ref{SecII}).  We see that the photon energy $h\nu$ is a natural energy normalizing factor for these integrals, which leads to a reduction in the number of key parameters from five to just three:  a unitless ratio of $ep$ to $ee$ scattering $\bdivd$, a unitless phonon temperature $\gamma$, and a unitless level of laser excitation $\alpha$.   This energy normalization subsequently leads to natural normalization of time:  if any timescale is multiplied by either the normalized $ep$ scattering strength $\beta$ or the normalized $ee$ scattering strength $\delta$, then the resulting unitless timescale is a function of the scattering strengths only through the unitless scattering-strength ratio $\bdivd$.

Before getting to our main results regarding distribution relaxation, we briefly discuss single-electron scattering times (Sec.~\ref{SecIII}).  This discussion allows us to clearly distinguish these single-particle times from the relaxation times $\tau_H$, $\tau_E$, and $\tau_{th}$ that describe evolution of the distribution function $f(\ep,t)$.  We also briefly discuss the energy relaxation time $\tau_E^{th}$ for a hot thermal (FD) distribution, which is relevant to understanding $\tau_E$ for laser-pulse-excited distributions.

The core of the paper comprises our investigation of $f(\ep,t)$ and the associated distribution relaxations times $\tau_H$, $\tau_E$, and $\tau_{th}$ as functions of the three normalized parameters (Sec.~\ref{CoreSection}).  We study, first in the low-excitation limit (small $\alpha$), the evolution of $f(\ep,t)$ over the complete range (0 to $\infty$) of $\beta/\delta$.  We do this for three different values of the normalized phonon temperatures $\gamma$.  From our calculations of $f(\ep,t)$ we numerically derive values of $\tau_H$, $\tau_E$, and $\tau_{th}$.  The dependence of each (normalized) relaxation time on the ratio $\beta/\delta$ is clearly revealed, and for each we present a heuristic expression that accurately describes the $\beta / \delta$ dependence.  The dependence of the three relaxation times is then investigated over ranges of $\bdivd$ and $\alpha$ appropriate to room-temperature ($T_p = 300$ K) experimental studies of ultrafast carrier dynamics in simple metals.  The dependence of the relaxation times on $\alpha$ reveal three distinct dynamical regions:  a low-excitation ($\alpha$ independent) region, an intermediate (transition) region, and a high-excitation region where the two-temperature (2T) model \cite{anisimov1974} of carrier dynamics is approximately valid.  We find that most experimental studies reside in the intermediate excitation region.

A number of approximate relations for $\tau_E$ and $\tau_{th}$ appear in the literature \cite{Tas1994,Gusev1998,Kabanov2008,Baranov2014,Wilson2020}.  As part of our discussion we re-express these relations in terms of our normalized parameters, which allows us to readily assess their accuracy via comparison with our numerical results.

\section{BTE Model of Carrier Excitation and Relaxation}
\label{SecII}

The derivation of our theoretical model, which we briefly review here, has been previously described in detail \cite{Riffe2023}.  First, we assume excitation is uniform in space so that carrier transport can be neglected.  Second, the scattering integrals in the BTE are averaged over $k$ space, which produces a formulation of the BTE that only depends upon the carrier energy $\ep$, 
\be{\eq}
\label{1}
\frac{\partial f(\ep,t)}{\partial t} = \Gamma_{\! ep}[f] + \Gamma_{\! ee}[f] + \Gamma_{\! e\gamma}[f],
\en{\eq}
where the three terms on the right side of this equation are the electron-phonon ($ep$), electron-electron ($ee$), and electron-photon ($e\gamma$) scattering integrals.  The $e\gamma$ integral describes excitation by a laser source, while the $ep$ and $ee$ integrals describe subsequent relaxation of the carriers.  Our general formulations of the scattering integrals depend upon averaged scattering strengths and the electronic density of states $g(\ep)$.  Because for the $sp$ metals of interest here $g(\ep)$ is largely featureless for several eV about the Fermi level, we  simply assume $g(\ep) = g_0$ is constant.  With these approximations the three scattering integrals can be expressed as
\begin{widetext}
\be{\eq}
\label{2}
\Gamma_{\! ep}[f] = - K_{ep} \, \Big\{ \big[ 2f(\ep) - 1 \big] f'(\ep)  -   k_B T_p \, f''(\ep)   \Big\},
\en{\eq}

\begin{align}
\label{3}
\Gamma_{\! ee}[f] &= - K_{ee} \, \int d\ep_3 \int d\ep_4 \Big(  \big\{ f(\ep) f(\ep_3 + \ep_4 - \ep) [1 - f(\ep_3)] [1- f(\ep_4)]  \nonumber \\
	&  - [1 - f(\ep)] [1-f(\ep_3 + \ep_4 - \ep)] f(\ep_3) f(\ep_4)  \big\} \Big),
\end{align}
and
\be{\eq}
\label{4}
\Gamma_{\! e\gamma}[f] = K_{e\gamma}(t)  \, \big\{f(\ep - h \nu) [ 1 - f(\ep) ]  - f(\ep) [ 1 - f(\ep + h \nu) ]    \big\}.
\en{\eq}
Equation~(\ref{2}) was derived under the additional assumption $k_B T_p \gtrsim \hbar \Omega$, where $\Omega$ is any phonon frequency.  In our analysis we assume $T_p$ does not vary in time.  Additionally, the electron-photon integral assumes photon absorption is indirect \cite{Riffe2023}, which is primarily applicable to intraband transitions.
\end{widetext}

A key quantity in our analysis is the rate of energy transfer from the excited carriers to the phonons.  This is obtained from the $ep$ scattering integral [Eq.~(\ref{2})] and can be written as \cite{Riffe2023}
\be{\eq}
\label{4b}
\frac{d \langle \ep \rangle }{dt}  = - K_{ep} \, g_0 \int d\ep  \big\{ f'(\ep) k_B T_p + f(\ep) [ 1 - f(\ep)  ]  \big\},
\en{\eq} 
If the excited carriers are internally thermalized---so that they are described by a FD distribution with temperature $T_e$---then Eq.~(\ref{4b}) simplifies to
\begin{equation}
\label{4c}
\frac{d \langle\ep \rangle_{\scriptscriptstyle \! F\!D}}{dt} = - K_{ep} g_0 \, k_B \, (T_e - T_p).
\end{equation}
This last relation is a key result of the 2T model of electron-phonon dynamics \cite{anisimov1974}, where it is assumed that the electronic and vibrational subsystems are characterized by temperatures $T_e$ and $T_p$, respectively.

The interaction strengths $K_{ep}$, $K_{ee}$, and $K_{e\gamma}$ are all related to physical quantities that have been characterized for many metals.  The $ep$ strength can be written as \cite{Riffe2023}
\be{\eq}
\label{5}
K_{ep} = \pi \hbar \lambda \langle \Omega^2 \rangle,
\en{\eq}
where
\be{\eq}
\label{6}
\lambda \langle \Omega^n \rangle = 2 \int d\Omega \, \alpha^2F(\Omega) \, \Omega^{n-1}.
\en{\eq}
Here $\alpha^2F(\Omega)$ is the $ep$ spectral density function and $\lambda = \lambda \langle \Omega^0 \rangle$ is the is the mass enhancement factor from superconductivity theory.  The $ee$ scattering strength is related to the single-carrier $ee$ scattering rate $1/\tau_{ee}$ via \cite{Kabanov2008}
\begin{equation}
\label{7}
\frac{1}{\tau_{ee}(\delep)} = \frac{K_{ee}}{2} (\delep)^2 ,
\end{equation}
where $\delep = \ep - \ep_F$ is the energy of a carrier with respect to the Fermi energy $\ep_F$.  (This scattering rate is discussed in more detail in Sec.~\ref{SecIIIb}.)  The excitation scattering strength $K_{e\gamma}(t)$ is a function of $h \nu$, the volume normalized density of states $g_v = g_0 / V$ [J$^{-1}$m$^{-3}$], and the power density $p_d(t)$ [J$\,$s$^{-1}\,$m$^{-3}$] absorbed by the metal,\footnote{The time dependent $K_{e\gamma}(t)$ defined here differs from the constant $K_{e \gamma}$  given in Ref.~\cite{Riffe2023}.  For a constant density of states $g_v$ the two quantities are related via $K_{e \gamma}(t) = K_{e\gamma} n_{h \nu}(t) / g_v$, where $n_{h \nu}(t)$ is a function proportional to the laser intensity.  See Ref.~\cite{Riffe2023} for further details.}
\be{\eq}
\label{8}
K_{e\gamma}(t) = \frac{p_d(t)}{(h \nu)^2 g_v}.
\en{\eq}

Because our primary goal is to characterize relaxation of the carriers, we treat laser excitation as simply as possible:  we assume (i) the laser pulse is infinitely short and (ii) any given electron can only absorb one photon.  This approximation decouples the excitation and relaxation processes, although---as we see below---relaxation is dependent upon both photon energy and laser excitation strength.  We assume the pre-excitation carriers are characterized by a FD distribution $f_{\scriptscriptstyle \! F\!D}(\ep,T_p)$. It then follows that laser-pulse excitation produces the nascent distribution
\begin{align}
\label{9}
f_{n}(\ep) 	&=  \alpha \, \big\{  1 - f_{\scriptscriptstyle \! F\!D}(\ep - h \nu,T_p) [ 1 - f_{\scriptscriptstyle \! F\!D}(\ep,T_p) ]  \nonumber \\
		&- f_{\scriptscriptstyle \! F\!D}(\ep,T_p) [ f_{\scriptscriptstyle \! F\!D}(\ep + h \nu,T_p) ]   \big\},
\end{align}
which hence is the initial distribution for the BTE driven solely by $ep$ and $ee$ scattering.  Here
\be{\eq}
\label{10}
\alpha = \int \!\! K_{e\gamma}(t) \, dt = \frac{u_d}{(h \nu)^2 g_v} = \frac{n_e}{h\nu \, g_v},
\en{\eq}
where $n_e$ [m$^{-3}$] is the density of excited carriers.  Because $\alpha$ is unitless, it is a natural descriptor for the carrier excitation level.  As we see below, $\alpha$ ($- \alpha$) is (essentially) the laser induced change in $f$ above (below) $\ep_F$ at energies $|\delep|  < h \nu$.  

Because the photon energy $h \nu$ sets the range of energies over which the distribution is initially excited and because (typically) $h \nu \gg k_B T_p$, it is natural (and quite convenient) to normalize any energy quantities by $h \nu$.  Doing so yields the normalized interaction strengths   
\be{\eq}
\label{11}
\beta = \frac{K_{ep}}{h \nu}
\en{\eq}
and
\be{\eq}
\label{12}
\delta = K_{ee} (h \nu)^2,
\en{\eq}
and the normalized phonon and electron temperatures
\be{\eq}
\label{13}
\gamma = \frac{k_B T_p}{h \nu}, \quad \gamma_e = \frac{k_e T_e}{h \nu}.
\en{\eq}
Notice the normalized $ep$ and $ee$ strengths $\beta$ and $\delta$ have the same units [s$^{-1}$].  These normalized quantities thus provide a natural (unitless) descriptor for the ratio of $ep$ to $ee$ interaction strengths,
\be{\eq}
\label{14}
\frac{\beta}{\delta} = \frac{K_{ep}}{K_{ee} (h \nu)^3}.
\en{\eq}  
We use this ratio as the primary parameter in organizing our results below.  Notice $\partial f(\ep,t) / \partial t$ is additive in $K_{ep}$ and $K_{ee}$, and thus is additive in $\beta$ and $\delta$.  Consequently, \textit{any} timescale $\tau$ associated with $f(\ep,t)$ has the property that either (unitless combination) $\tau \beta$ or $\tau \delta$ is a function of $\beta$ and $\delta$ \textit{only} via the unitless interaction strength ratio $\beta / \delta$.

\begin{figure*}[t]
\centerline{\includegraphics[scale=0.64]{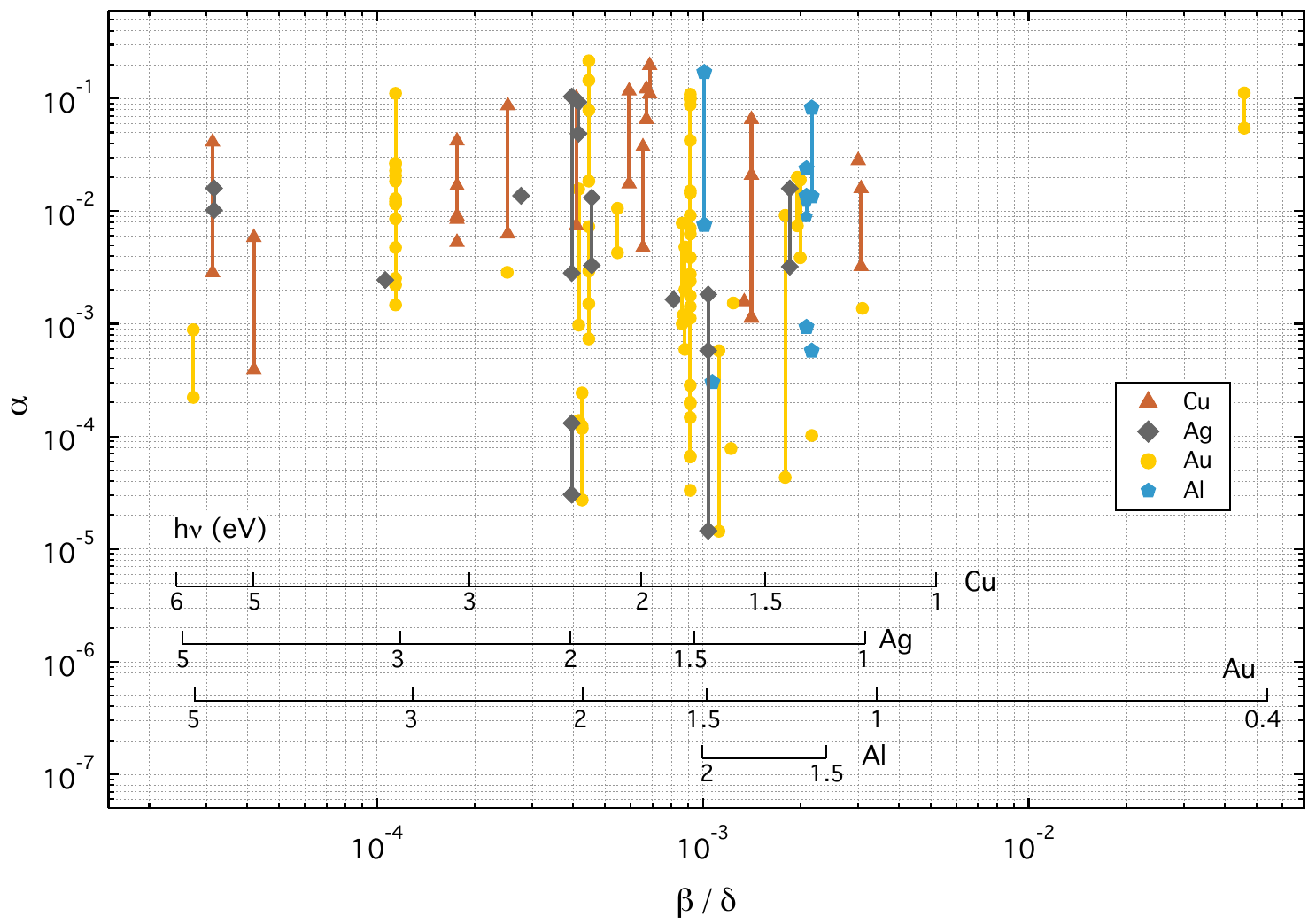}}
\caption{Concurrent values of normalized excitation level $\alpha$ and normalized ratio of $ep$ to $ee$ coupling strengths $\beta / \delta$ = $K_{ep}/ [K_{ee} (h \nu)^3]$ in sub-damage-threshold RT experimental measurements (from the literature) on Cu, Ag, Au, and Al.  Each set of two markers connected by a vertical line indicate the range of $\alpha$ for a particular experiment.  Experiments that involve only one excitation level are indicated by a single marker.  The individual $h \nu$ axes for the four metals indicate the laser-pulse photon energy used in each experiment.}
\label{Fig1}
\end{figure*}

So what values of the controlling parameters $\bdivd$ and $\alpha$ are germane to experimental conditions utilized in ultrafast investigations of simple metals?  The answer is provided in Fig.~\ref{Fig1}, which plots concurrent values of $\alpha$ and $\beta / \delta$ that apply to an extensive collection of experiments\footnote{Details of the experiments represented in Fig.~\ref{Fig1} can be found in the Supplemental Material \cite{SuppMat}.} on Cu, Ag, Au, and Al \cite{Elsayed1987,Hohlfeld1995,Hohlfeld1996,Papadogiannis1997,Bartoli1997,Farztdinov1999,Bonn2000,Kruglyak2005,Ligges2009,Ligges2011,Shen2015,Nakamura2016,Mo2018,Obergfell2020,Suemoto2021,Riffe1993,Groeneveld1995,Fatti1998,Fatti2000,Owens2010,Er2013,Schoenlein1987,Brorson1987,Elsayed-Ali1991,Fann1992,Juhasz1992,Juhasz1993,Girardeau-Montaut1993,Sun1993,Sun1994,wang1994,Hohlfeld1997,Maznev1997,Moore1999,Hibara1999,Hostetler1999,Smith2000,Smith2001,Guo2001,Ibrahim2004,Hopkins2007,Ma2010,Hopkins2011,Chen2011a,DellaValle2012,Kolomenskii2013,Hopkins2013,Guo2014,White2014,Chase2016,Sokolowski-Tinten2017,Heilpern2018,Yao2018,Sielcken2020,Lu2020,Tomko2021,Grauby2022,Tas1994,Stevens2005,Park2005,Nie2006,Ma2013,Waldecker2016}.   We note all of the represented experiments were performed on RT samples at excitation levels below the damage threshold.  Because $\beta / \delta \! \sim \! (h \nu)^{-3}$, this ratio strongly depends upon the photon energy $h \nu$.  Strikingly, ultrafast experiments on Au have been performed for photon energies from 0.42 to 5.0 eV, which corresponds to a range of $\beta  /  \delta$ from $\sim$3$\times 10^{-5}$ to $\sim$5$\times 10^{-2}$.  As can also be deduced from Fig.~\ref{Fig1}, the excitation strength $\alpha$ experimentally ranges from $\sim$$10^{-5}$ to a few times $10^{-1}$.  As our calculations reveal below, this range crosses over from the region where relaxation rates are independent of $\alpha$ (at lowest excitation levels) to a limit where the 2T model \cite{anisimov1974} is a reasonable approximation to the dynamics (at highest excitation levels).

\section{Carrier Dynamics Timescales}
\label{SecIII}

The timescales important to carrier dynamics can be divided into two types:  (i) those associated with scattering of individual carriers and (ii) those associated with relaxation of an excited distribution.  In anticipation of analyzing the results for the relaxation of the nascent distribution $f_{n}(\ep)$ in Sec.~\ref{CoreSection}, here we discuss several of these timescales.  A point of emphasis is the relationship between these timescales and the two material constants---$K_{ep}$ and $K_{ee}$---that govern the strength of each type of scattering.  We also offer a physical interpretation of the normalized scattering-strength ratio $\bdivd$.

\subsection{Single-carrier timescales}
\label{SecIIIb}

Here we consider both $ep$ and $ee$ scattering of a singly excited carrier.  We assume the initial energy $\ep = \ep_F + \delep$ of the carrier is such that $\delep \gg k_BT_p$ and $\delep \gg \hbar \Omega$.  Under these conditions any possible final state resulting from $ep$ scattering (with energy $\ep \pm \hbar \Omega$) will initially be empty.  Regarding $ee$ scattering of this single electron, the condition $\delep \gg k_BT_p$ enables the approximation $T_p = 0$, in which case all possible final states are at energies between $\ep_F$ and $\ep_F + \delep$, and these states are also initially empty.

The first relevant single-carrier time is the spontaneous phonon-emission time \cite{Gusev1998}
\begin{equation}
\label{15} 
\tau^0_{ep} = \frac{1}{\pi \lambda \langle \Omega \rangle},
\end{equation}
which is related to the $ep$ interaction strength $K_{ep}$ [see Eq.~(\ref{5})] via
\begin{equation}
\label{16}
\tau^0_{ep}  = \frac{\hbar \bar{\Omega}}{K_{ep}},
\end{equation}
where $\bar{\Omega} = \langle \Omega^2 \rangle / \langle \Omega \rangle$ is the scattering-strength-weighted average phonon frequency [see Eq.~(\ref{6})].  The scattering strength $K_{ep}$ thus has a key interpretation:  because one phonon is spontaneously emitted at the rate $1/\tau^0_{ep}$, the average energy-loss rate $\dot{q}_0$ for a singly excited electron is simply
\be{\eq}
\label{17}
\dot{q}_0 = K_{ep}.
\end{\eq}

In Table \ref{table1} we list values of $K_{ep}$ and $\tau^0_{ep}$ for a selection of simple metals.  These calculations require three parameters---$\lambda$, $\langle \Omega \rangle$, and $\langle \Omega^2 \rangle$---for any given metal.  For $\lambda$ we use the recommended values tabulated by Grimvall \cite{Grimvall1981}.  We estimate the values of $\langle \Omega \rangle$ and $\langle \Omega^2 \rangle$ using $\langle \Omega \rangle = \frac{2}{3} (k_B \hbar)  \, \Theta_{-1}$ and $\langle \Omega^2 \rangle = \frac{1}{2} ( k_B/\hbar)^{2}  \, \Theta_{-1} \Theta_{1}$; here $\Theta_m$ is the $m^{th}$-moment Debye temperature \cite{Riffe2023}.  For all but Ag, we obtain values of $\Theta_{m}$ from Wilson \cite{Wilson2011}.  The values for Ag are derived from an analysis of the phonon spectra of Kamitakahara and Brockhouse \cite{Kamitakahara1969}  by Antonov \textit{et al.} \cite{Antonov1990}  As the results in Table \ref{table1} show, $K_{ep}$ ranges from $\sim$$5 \times 10^{-3}$ to $\sim$$1$ meV/fs, while $\tau^0_{ep}$ lies between $\sim$20 and $\sim$600 fs.

The second relevant single-carrier time is the total $ep$ scattering time \cite{Grimvall1981,allen1987}
\begin{equation}
\label{18}
\tau_{ep}(T_p) = \frac{\hbar}{2 \pi \lambda \, k_B T_p},
\end{equation}
which in terms of $K_{ep}$ can be written as
\begin{equation}
\label{19}
\tau_{ep}(T_p) =\frac{\hbar^2 \langle \Omega^2 \rangle}{2 K_{ep} \, k_B T_p}.
\end{equation}

In Table \ref{table1} we also include values for this scattering time for $T_p = 300$ K for the indicated metals.  As shown there, $\tau_{ep}(300 \, {\rm K})$ lies between $\sim$10 fs and $30$ fs.

We point out a key feature common to both $\tau^0_{ep}$ and $\tau_{ep}(T_p)$.  In contrast to $K_{ep}$, which is proportional to $\lambda \langle \Omega^2 \rangle$, the scattering rates $1/\tau^0_{ep}$ and $1/\tau_{ep}(T_p)$ are proportional to $\lambda \langle \Omega \rangle$ and  $ \lambda \langle \Omega^0 \rangle$ (= $\lambda$), respectively [see Eqs.~(\ref{15}) and (\ref{18})].   However, Eq.~(\ref{8}) for $(\partial f / \partial t)_{ep}$ simply involves $K_{ep} \sim \lambda \langle \Omega^2 \rangle$, sans any other moments of $\Omega$.  Therefore, neither of these two single-particle scattering times is fundamentally related to relaxation of the distribution function $f(\ep)$.


We now turn our attention to $ee$ scattering.  For a single electron initially in a state with energy $\ep = \ep_F + \delep$ above a zero-temperature Fermi sea the $ee$ scattering rate of this carrier is given by Eq.~(\ref{7}).  Owing to our interest in the early-time decay of a laser excited solid---with excitation energies on the order of an eV---in Table \ref{table1} we compile values of $\tau_{ee}(1 \, \rm{eV})$ for each of the metals, and from these values we use Eq.~(\ref{7}) to calculate the likewise listed $K_{ee}$ values.

We determine these values of $\tau_{ee}(1 \, \rm{eV})$ in two different ways, depending upon the metal in question.  For Cu, Ag, Au, and Al, $\tau_{ee}(\delep)$ has been experimentally assessed via time-resolved two-photon photoemission (TR2PPE) measurements \cite{Bauer2015}; values of $\tau_{ee}(1 \, \rm{eV})$ are extracted from these measurements for these metals.  No such data have been reported for the alkali metals.  For estimating $\tau_{ee}(1 \, \rm{eV})$ for these metals we turn to the Fermi-liquid-theory (FLT) expression of Pines and Nozieres for the $ee$ scattering rate \cite{Pines1966},
\begin{equation}
\label{20}
\frac{1}{\tau_{ee}(\delep)} = \frac{\sqrt{3} \pi^2 \omega_p}{128} \bigg( \! \frac{\delep}{\ep_F} \! \bigg)^{\! \! 2}.
\end{equation}
Here $\omega_p$ is the plasma frequency.  Using theoretical band-structure derived values of $\omega_p$ and $\ep_F$ for the alkali metals \cite{papaconstantopoulos1986}, we calculate the values of $\tau_{ee}(1 \, \rm{eV})$ listed in Table \ref{table1}.  

Interestingly, Ag and Au have the weakest $ee$ scattering, with the smallest values of $K_{ee} \approx$ 0.03 eV$^{-2}$s$^{-1}$ and concomitantly $\tau_{ee}(1 \, \rm{eV}) \approx$ 70 fs.  This weak scattering has been attributed to strong screening of the $n \, sp$ ($n = $ 5 or 6) valence electrons by the rather delocalized $(n-1)\,d$ electrons \cite{Bauer2015}.  Conversely, the two heaviest alkali metals (Rb and Cs) have the strongest $ee$ scattering with $K_{ee} \approx$ 0.4 eV$^{-2}$s$^{-1}$, resulting in $\tau_{ee}(1 \, \rm{eV}) =$ 5 fs.  From the point of view of FLT, this strong scattering can be traced to the relatively small values of $\ep_F$ for these two metals.

Insight into the normalized scattering strength ratio $\beta / \delta$ is provided by considering the effects of each scattering process on the average energy-loss rate of a singly excited electron.  Let us assume such an electron has excess energy $\delep$ equal to a photon energy $h \nu$.  If it were subject to only $ee$ scattering, then it can be shown its initial average energy-loss rate is given by $\dot{q}_{ee} = K_{ee} (h\nu)^3 / 3$.  Conversely, if this carrier were subject to only $ep$ scattering, then its average energy-loss rate would be  $\dot{q}_{ep} = K_{ep}$ [see Eq.~(\ref{17})].  The ratio of these two rates is simply $\dot{q}_{ep} / \dot{q}_{ee} = 3 \beta / \delta$.  The ratio $\bdivd$ thus gives some measure of the relative effectiveness of the two scattering processes in decreasing the energy of any single carrier within a laser excited distribution.

In most cases the ratio $\beta / \delta$ is quite small.  As shown in Table \ref{table1}, for $h \nu =$ 1.5 eV the values of $\bdivd$ are all less than  $3\times 10^{-3}$.  Similarly, Fig.~\ref{Fig1} shows for Cu, Ag, Au, and Al that in nearly all experiments $\beta / \delta \lesssim 3 \times 10^{-3}$ (the exception being the experiment on Au with $h \nu$ = 0.42 eV, for which $\beta / \delta$ is significantly larger but still $< 0.1$).  Such small $\beta / \delta$ values indicate that $ee$ scattering is the dominant contributor in relaxing the high energy end of a (typical) laser excited distribution.  Indeed, this is quantitatively  born out in our calculations of the hot-carrier relaxation time $\tau_H$, reported in Sec.~\ref{SubSecVC2}.


\begin{table*}[t]
\caption{Material parameters and resulting $ep$ and $ee$ interaction parameters for several simple metals.  The mass enhancement factor $\lambda$ and phonon-frequency moments $\langle \Omega \rangle$ and $\langle \Omega^2 \rangle$ are used as input parameters for calculating $\tau^0_{ep}$ and $\tau_{ep}$ (the spontaneous-emission and total $ep$ single-carrier scattering times, respectively), the $ep$ coupling strength $K_{ep}$, the thermalized-distribution energy relaxation time $\tau_E^{th}$, and the number of phonons $N_{ph}$ emitted during the time $\tau_E^{th}$.  The $ee$ scattering time $\tau_{ee}$ (at an excitation energy $\delta \ep =$ 1 eV) is used as input to calculate the $ee$ scattering strength $K_{ee}$.  The ratio $\beta / \delta = (K_{ep} / K_{ee}) (h \nu)^{-3}$ is the normalized ratio of $ep$ to $ee$ scattering strengths (here tabulated for photon energy $h \nu =$ 1.5 eV). \label{table1}}
\vspace{0.2cm}
\begin{tabular}{l@{\hspace{0.3cm}}r@{\hspace{0.3cm}}r@{\hspace{0.3cm}}r@{\hspace{0.3cm}}r@{\hspace{0.3cm}}r@{\hspace{0.4cm}}r@{\hspace{0.4cm}}r@{\hspace{0.3cm}}r@{\hspace{0.3cm}}r@{\hspace{0.3cm}}r@{\hspace{0.3cm}}r@{\hspace{0.3cm}}r@{\hspace{0.3cm}}r}
\hline
\hline      

Metal		& $\lambda$\footnote{Reference [\onlinecite{Grimvall1981}].}  &  $\hbar \langle \Omega \rangle $\footnote{See text for sources of $\langle \Omega \rangle$ and $\langle \Omega^2 \rangle$.}	& $\hbar \langle \Omega^2 \rangle^{\! 1/2} $  & $K_{ep}$ & $\tau_{ep}^{0}\,$ & $\tau_{ep}$\footnote{at $T_p =$ 300 K.} & $\tau_E^{th}\,$\footnote{in the low excitation limit at $T_p =$ 300 K.} & $N_{ph}$ & $K_{ee}$  	  & $\tau_{ee}$\footnote{at $\delep$ = 1 eV.} & $\beta / \delta$ \footnote{for $h \nu = 1.5$ eV.}  \\
	 	&		& (meV) 	& (meV) 	 & (meV$\!$/fs) & (fs) 	& (fs)	& (fs)	& 	& (eV$^{-2}  \rm{fs}^{-1}$)	& (fs) & ($10^{-6}$) \\
\hline
Li 		& 0.40	& 19.9	& 22.2	& 0.940	& 26 		& 10		& 91		& 2.5		& 0.167  	& 12		& 1660 \\	

Na  		& 0.16  	& 8.8		& 9.7		& 0.072	& 148	& 25		& 1190	& 5.8		& 0.292 	& 7		& 73 \\

K  		& 0.13  	& 5.3		& 5.8		& 0.021	& 305 	& 31		& 4050	& 9.6		& 0.359	& 6		& 17 \\

Rb 		& 0.16  	& 3.3		& 3.6		& 0.010	& 400	& 25		& 8490	& 15.3	& 0.426 	& 5		& 7 \\

Cs 		& 0.15  	& 2.3		& 2.6		& 0.005	& 608	& 27		& 18200	& 21.6	& 0.423	& 5 		& 3 \\


 \rule{0pt}{4ex}Cu 		& 0.15  	& 18.0	& 19.1	& 0.262	& 78		& 27		& 325	& 3.0		& 0.050 	& 40 		& 1550 \\

Ag 		& 0.13  	& 11.6	& 12.5	& 0.098	& 139	& 31		& 872	& 4.5		&  0.031 	& 65 		& 940 \\

Au 		& 0.17  	& 9.7		& 10.7	& 0.092	& 127	& 24		& 923	& 5.2		& 0.027 	& 75 		& 1020 \\


 \rule{0pt}{4ex}Al 		& 0.43  	& 21.0	& 22.8	& 1.071	& 23		& 9		& 79		& 2.5		& 0.133 	& 15 		& 2380 \\

\hline
\hline

\end{tabular}
\smallskip
\end{table*}

\subsection{Thermalized-distribution relaxation}

Our discussion in Sec.~\ref{CoreSection} focuses on relaxation times associated with the nascent (just after laser excitation) distribution given by Eq.~(\ref{9}).  To appreciate those results, however, it behooves us to first consider the dynamics of a thermal (FD) distribution of carriers, which are characterized by an electron temperature $T_e$.  In particular, here we look at the energy relaxation time $\tau_E^{th}$ of a FD distribution described by the electron temperature $T_e$.  This relaxation time is defined via the relation $1 / \tau_E^{th} = - (d \langle \ep \rangle / dt ) / \langle \ep \rangle$, where $\langle \ep \rangle$ is the excess energy in the carriers.  In terms of $K_{ep}$ this time can be written as \cite{allen1987}
\be{\eq}
\label{21}
\tau_E^{th}(T_e,T_p) = \frac{\pi^2}{6} \frac{k_B (T_e + T_p)}{K_{ep}}.
\en{\eq}
For high excitation levels where $T_e \gg T_p$ this relaxation time is not constant, as $T_e$ varies in time.  Therefore, energy decay is generally nonexponential.  However, in the low-excitation limit, where $T_e \approx T_p$, Eq.~(\ref{21}) simplifies to
\be{\eq}
\label{22}
\tau_E^{th}(T_p) = \frac{\pi^2}{3} \frac{k_B T_p}{K_{ep}},
\en{\eq}
in which case the decay is well approximated by a simple exponential (under the constant-$T_p$ approximation assumed here).  As expected---and as opposed to the single-particle times $\tau^0_{ep}$ and $\tau_{ep}(T_p)$---this distribution-associated timescale depends upon the phonon spectrum solely through $K_{ep} \sim  \lambda \langle \Omega^2 \rangle$ (and no other moments of $\Omega$).  In Table \ref{table1} we list values of $\tau_E^{th}(300 \, \rm{K})$ as given by Eq.~(\ref{22}) for all of the considered metals.  Although $\lambda$ has some influence, the dramatic range of $\tau_E^{th}$ values is largely a consequence of $K_{ep}$ being proportional to $\langle \Omega^2 \rangle$.

\begin{figure}[t]
\centerline{\includegraphics[scale=0.45]{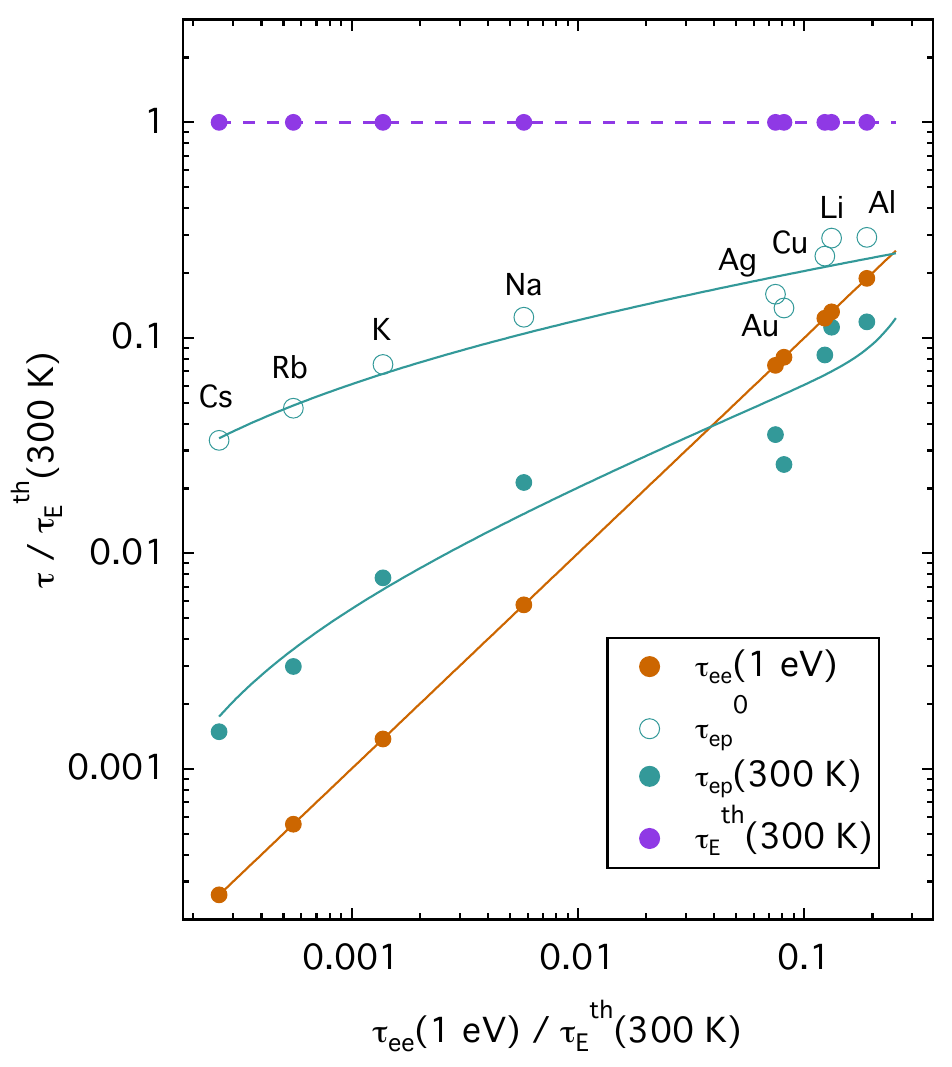}}
\caption{Relaxation times relevant to excited carrier dynamics for the metals in listed in Table \ref{table1}.  The time $\tau_{ee}$(1$\,$eV) is the single-carrier $ee$ scattering time for a carrier 1 eV above the Fermi energy, $\tau^0_{ep}$ is the spontaneous phonon-emission scattering time, and $\tau_{ep}$(300 K) is the total $ep$ scattering time at $T_p  = $ 300 K.  All times are normalized to the thermalized-distribution energy relaxation time $\tau_{E}^{th}(300 {\rm K})$.  Solid lines are guides to the eye.}
\label{Fig1A}
\end{figure}

In order to coherently organize the set of scattering times presented in Table \ref{table1}, we plot  in Fig.~\ref{Fig1A} $\tau^0_{ep}$, $\tau_{ep}(300 \, {\rm K})$, and $\tau_{E}^{th}(300 \, {\rm K})$ vs $\tau_{ee}(1 \, {\rm eV})$.  For reference, $\tau_{ee}(1 \, {\rm eV})$ is also plotted against itself.  Because $\tau_{E}^{th}(300 \, {\rm K})$ is the largest of all of these timescales, we have normalized all of the scattering times by this relaxation time.  As can be seen in the figure (or deduced from Table \ref{table1}), the three $ep$ driven timescales are related by the condition
\be{\eq}
\label{23}
\tau_{ep}(300 \, {\rm{K}}) < \tau^0_{ep} < \tau_E^{th}(300 \, \rm{K})
\en{\eq}
for all of the metals listed.  Furthermore, we see that the ratio of $ee$ to $ep$ rates is quite different for the four heaviest alkali metals in comparison with the Li, the noble metals, and Al.

Recognizing that the relationship between $\tau_{ep}(300 \, \rm{K})$ and $\tau_E^{th}(300 \, \rm{K})$ is typically a strong inequality---i.e., $\tau_{ep}(300 \, \rm{K}) \ll \tau_E^{th}(300 \, \rm{K})$---Allen remarked ``Electron energy loss at high $T$ by phonon emission is a multistep process.'' \cite{allen1987}  However, because $\tau_{ep}(300 \, \rm{K})$ involves scattering events that include both phonon emission and absorption, this relationship is perhaps not as interesting as that between $\tau^0_{ep}$ and $\tau_E^{th}(300 \, \rm{K})$, owing to $\tau^0_{ep}$ being directly related to the energy-loss rate $\dot{q}_0 = K_{ep}$ of a singly excited carrier [see Eq.~(\ref{16})].

However, an even better connection between relaxation at the single-carrier and the thermalized-distribution levels comes about by considering the average energy-loss rate $\dot{q}_{th}$ for a single carrier that is itself part of the thermal distribution.  As we now show,  $\dot{q}_{th} \ne \dot{q}_0$.  Because $\dot{q}_{th}$ is equal to $ - (d \langle \ep \rangle / dt ) / N_{ex}$, we need the number of excess excitations (both electrons and holes)
\begin{align}
\label{24}
N_{ex}&= 2 \, g_0 \int_{\ep_F}^\infty [f_{\scriptscriptstyle \! F\!D}(\ep,T_e) - f_{\scriptscriptstyle \! F\!D}(\ep,T_p)] \, d\ep \nonumber \\
		&= 2 \ln(2) \, g_0 \, k_B (T_e - T_p),
\end{align}
and the thermalized-distribution energy-loss rate given by Eq.~(\ref{4c}).  We hence obtain
\be{\eq}
\label{30F}
\dot{q}_{th} = \frac{K_{ep}}{2 \ln(2)} = \frac{\dot{q}_0}{2 \ln(2)}.
\en{\eq}
Thus, for a carrier in a thermalized distribution the effective spontaneous phonon-emission time is
\be{\eq}
\label{30G}
2 \ln(2) \, \tau^0_{ep} \approx 1.39 \, \tau^0_{ep}.
\en{\eq}
That this time constant is longer than that for a singly excited carrier is not surprising, as not all possible final states are initially empty for an electron within a FD distribution.  Succinctly put, Pauli blocking slows down the energy decay rate compared to that for a singly excited electron.  It appears that this difference between $\dot{q}_0$ and $\dot{q}_{th}$ has not always been recognized \cite{Tas1994,Gusev1998}.

The ratio of Eq.~(\ref{22}) to Eq.~(\ref{30G}) provides the answer to an interesting question:  how many phonons $N_{ph}$ are emitted (on average) by each carrier within an energy-decay time of the distribution?  Indeed, in the low-excitation limit this number is identically this ratio,
\be{\eq}
\label{30H}
N_{ph} = \frac{\tau_E^{th}(T_p)}{2 \ln(2) \, \tau^0_{ep}} = \frac{\pi^2}{6 \ln(2)} \frac{k_BT_p}{\hbar \bar{\Omega}}.
\en{\eq}
In Table \ref{table1} we tabulate this quantity for $T_p = 300 \, \rm{K}$.  As Eq.~(\ref{30H}) shows, the range of $N_{ph}$ values simply reflects the variation of characteristic phonon frequencies across this range of metals.

\begin{figure*}[t]
\centerline{\includegraphics[scale=0.59]{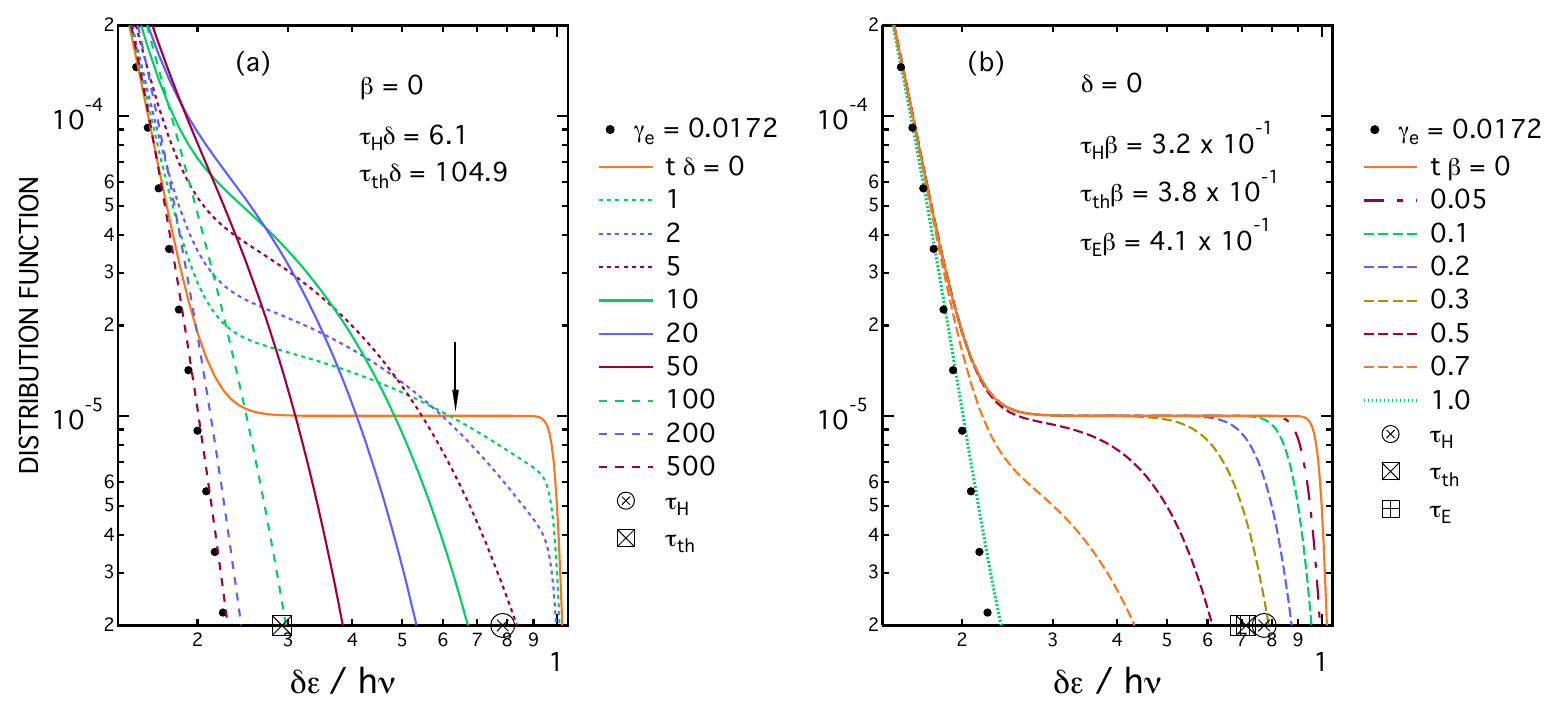}}
\caption{Time series of (a) distribution function $f(\delep,t)$ during relaxation for $\beta = 0$ ($ee$ scattering only) and (b)  during relaxation for $\delta = 0$ ($ep$ scattering only).  The vertical arrow in (a) is located at $\delep / (h \nu) = (3 - \sqrt{3}) / 2$.  Here the normalized excitation level and phonon temperature are $\alpha = 1 \times 10^{-5}$ and $\gamma = 0.0172$, respectively.}
\label{Fig3}
\end{figure*}

\begin{figure*}[t]
\centerline{\includegraphics[scale=0.59]{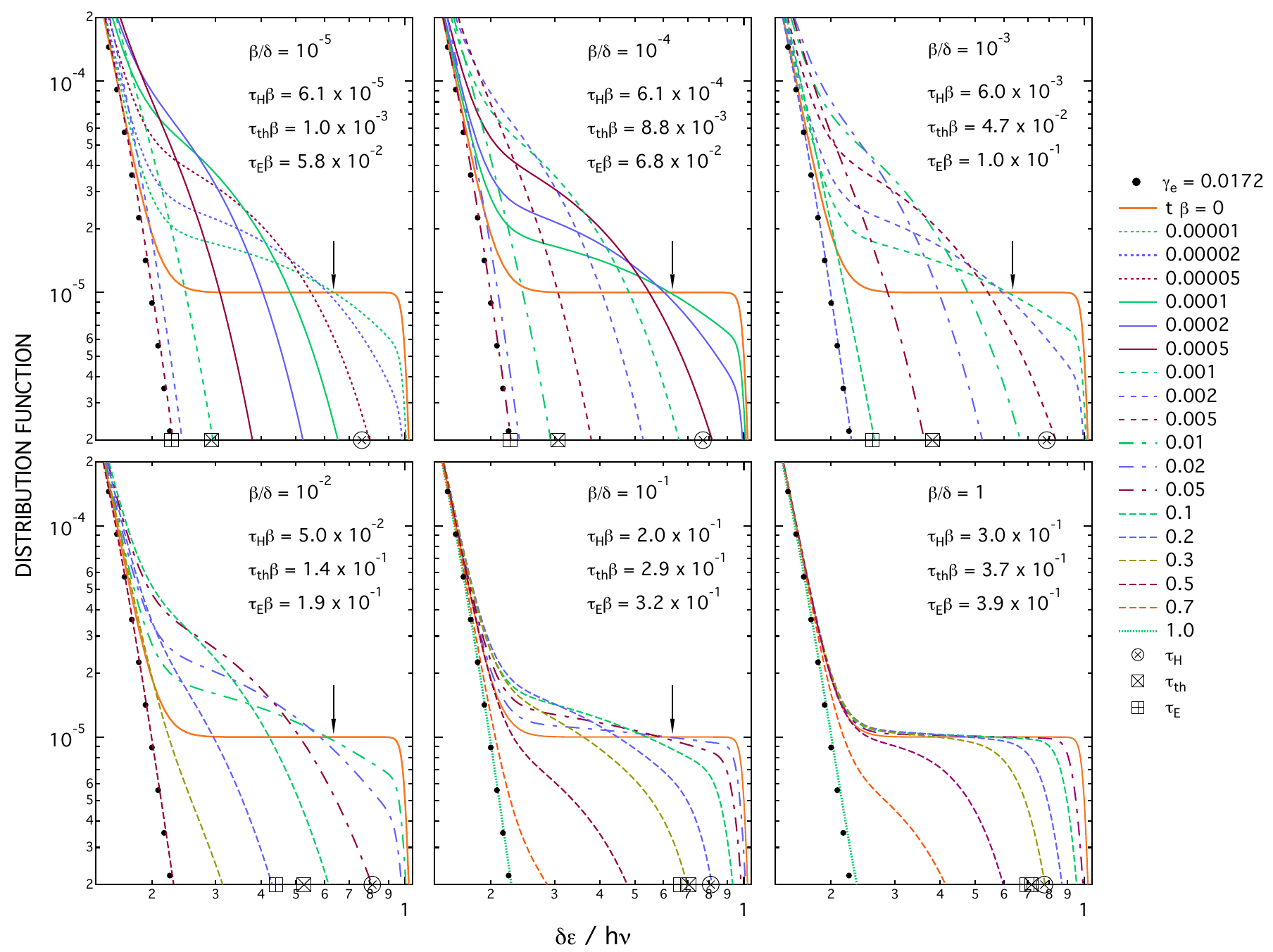}}
\caption{Time series of laser-excited distributions $f(\delep,t)$ during relaxation for six different values of the ratio $\beta / \delta$ of $ep$ to $ee$ interaction strengths.  The vertical arrows are located at $\delep / (h \nu) = (3 - \sqrt{3}) / 2$. Here the normalized excitation level and phonon temperature are $\alpha = 1 \times 10^{-5}$ and $\gamma = 0.0172$, respectively.}
\label{Fig4}
\end{figure*}

\section{Distribution Relaxation}
\label{CoreSection}

We now report results pertaining to relaxation of a laser excited carrier distribution, calculated using the model described in Sec.~\ref{SecII}.  As discussed there, our calculations are aimed at quantitatively discerning how the relative strength $\beta / \delta$ of $ep$ to $ee$ scattering, relative phonon temperature $\gamma$, and excitation level $\alpha$ govern the dynamics.  As we shall see, for typical experimental conditions the metals in Table \ref{table1} are characterized by moderate to strong $ee$ scattering or---equivalently---by weak to moderate $ep$ scattering.  Furthermore, we derive phenomenological expressions for timescales related to (i) hot-carrier relaxation ($\tau_H$), (ii) energy transfer from the excited carriers to the phonon subsystem ($\tau_E$), and (iii) thermalization of the nascent distribution ($\tau_{th}$) that provide insight into the dynamics.

\subsection{Low-excitation limit}

In this section we study carrier relaxation in the low-excitation limit.  Specifically, our results here are for $\alpha = 10^{-5}$ and values of $\gamma$ equal to 0.0086, 0.0172, and 0.0344.  These $\gamma$ values were chosen because for $h \nu = 1.5$ eV (a typical Ti:Al$_2$O$_3$ laser photon energy) and $T_p = 300$ K, we have $\gamma = k_B T_p / h \nu = 0.0172$.  The other two values of $\gamma$ thus correspond to the same photon energy with $T_p$ equal to either 150 or 600 K.  (Note however, these results also correspond any combination of $T_p$ and $h \nu$ with the same ratios.)  We note for $\alpha = 10^{-5}$ and $\gamma = 0.0172$, if the initial distribution were thermalized, then its temperature $T_e$ would only be 1.0\% greater than $T_p$.

\subsubsection{Relaxation of $f(\ep, t)$}
\label{SecVB1}

We being our discussion of $f(\ep, t)$ relaxation by first examining the two extremes:  $ee$ scattering only ($\beta = 0$) and $ep$ scattering only ($\delta = 0$).  Results for these two cases are illustrated in Fig.~\ref{Fig3}.  Note that $\alpha = 10^{-5}$ is readily ascertained from the curves in the two panels via the large plateau in $f(\delta \ep,0)$ at this value.  

As has been noted in the literature \cite{Tas1994,Groeneveld1995,Gusev1998,Wilson2020}, a primary consequence of $ee$ scattering is the creation of more excitations as the distribution initially relaxes.  This behavior is manifest in part (a) of Fig.~\ref{Fig3}. Although $f(\delta \ep,t)$ initially decreases at the highest energies, it increases to a much greater extent at lower energies, owing to (further) excitation of carriers that are initially below $\ep_F$.   At the earliest times ($t \delta \lesssim 1$) the division between $f(\delep,t)$ increasing and decreasing occurs at the energy $\delep / (h \nu) = (3 - \sqrt{3})/2 \approx 0.634$ (see \cite{SuppMat}); the vertical arrow in the figure marks this energy.

As is apparent in part (b) of Fig.~\ref{Fig3}, relaxation of the distribution under the sole influence of $ep$ scattering is quite different from that due to $ee$ scattering.  The behavior of this distribution time series can be understood by recalling that a singly excited carrier loses energy at the rate $\dot{q} = K_{ep}$.  Because $f \ll 1$ for $\delta \ep/ h \nu \gtrsim 0.25$, each carrier in this part of the distribution also loses (on average) energy at this same rate.  Therefore, to good approximation the excited part of the distribution uniformly moves to lower energy at exactly this rate.  In terms of normalized quantities we have $\dot{q}/ h \nu = K_{ep}/ h \nu = \beta$.  Therefore, at time $t \beta$ the top end of the distribution will roughly reside at $1 - t \beta$.  This behavior, first noted by Gusev and Wright \cite{Gusev1998}, is evident in the figure.  Further details concerning $ep$ driven relaxation are discussed in Appendix \ref{Appendix D}.

The transition from dominant $ee$ scattering to dominant $ep$ scattering is illustrated in Fig.~\ref{Fig4}, which comprises a set of $f(\delep,t)$ curves for $\beta/\delta$ ranging between 10$^{-5}$ to 1.  First, we note as a function of $t \delta$ (rather than $t \beta$), the sets of curves for $\beta/\delta$ = 10$^{-5}$ and 10$^{-4}$ are are essentially identical to the curves in part (a) of Fig~\ref{Fig3}.  Thus, for  values of $\beta/\delta \lesssim 10^{-4}$ $ee$ scattering is the main driver of changes in $f$.  Differences in behavior become apparent beginning with the $\delta/\beta = 10^{-3}$ curves:  as $\delta/\beta$ increases from 10$^{-3}$ to 10$^{-1}$ the dominance of $ee$ over $ep$ scattering clearly diminishes, and by $\delta/\beta = 10^{-1}$ the effects of $ep$ scattering are apparent as early as $t \delta = 1$ (or equivalently $t \beta = 0.1$).  For $\beta/\delta = 1$ $ep$ scattering is clearly dominant:  these curves are nearly identical to those in part (b) of Fig.~\ref{Fig3} where $\delta = 0$.  

\begin{table}[b]
\caption{Relaxation times associated with hot electrons, energy, and thermalization in the strong $ee$ and strong $ep$ scattering limits at low excitation ($\alpha = 10^{-5}$).  Analytic approximations in the strong $ep$ scattering limit are indicated in the footnotes.  For $h \nu = 1.5$ eV, the values of $\gamma$ correspond to $T_p =$ 150, 300, and 600 K.    Values of the Gusev and Wright thermalization time $\tau_{th}^{G}  \delta$ are calculated using Eq.~(\ref{35}). \label{table2a}}
\vspace{0.2cm}
\begin{tabular}{c@{\hspace{0.4cm}}d@{\hspace{0.4cm}}d@{\hspace{0.4cm}}d@{\hspace{0.5cm}}r@{\hspace{0.5cm}}r@{\hspace{0.5cm}}r@{\hspace{0.5cm}}r@{\hspace{0.5cm}}r@{\hspace{0.5cm}}r}
\hline
\hline  
    
Relaxation &\multicolumn{3}{c} {$\gamma \, (=k_B T_p / h \nu)$}   \\

times &0.0086  & 0.0172 & 0.0344   \\

\hline

\multicolumn{4}{c} {Hot Electrons}		 \\	

$\tau_H^{ee}  \delta$  	& 6.098	& 6.146	& 6.356	\\

$\,\,\,\, \tau_H^{ep}  \beta$ \footnote{Analytic approximation: $\tau_H^{ep} \beta = 0.3161$.} 	& 0.3162	& 0.3182	& 0.3275		\\

\\

\multicolumn{4}{c} {Energy}		  					 \\

$\tau_E^{th}  \beta$	  	& 0.0283	& 0.0566	& 0.1131		\\

$\,\,\,\, \tau_E^{ep}  \beta$ \footnote{Analytic approximation: $\tau_E^{ep} \beta = 0.3935$.} 		& 0.3966	& 0.4060	& 0.4233	\\

\rule{0pt}{2.5ex}  $\tau_E^{ep} / \tau_E^{th}$ 	& 14.0	& 7.2		& 3.7	\\

\\

\multicolumn{4}{c} {Thermalization} 		\\

$\tau_{th}^{ee}  \delta$	 &365.2	& 104.9	& 31.83	\\

$\,\,\,\,\,\, \tau_{th}^{ep}  \beta$ \footnote{Analytic approximations: $\tau_{th}^{ep} \beta =$ 0.3819, 0.3694, and 0.3415 for $\gamma =$ 0.0086, 0.0172, and 0.03444, respectively.}	& 0.3944	& 0.3848	& 0.3746	\\

 \rule{0pt}{3ex} $\tau_{th}^{\scriptscriptstyle G}  \delta$	 &679	& 170	& 43	\\

\hline
\hline

\end{tabular}
\smallskip
\end{table}

\subsubsection{Relaxation times}
\label{SubSecVC2}

\begin{figure}[t]
\centerline{\includegraphics[scale=0.55]{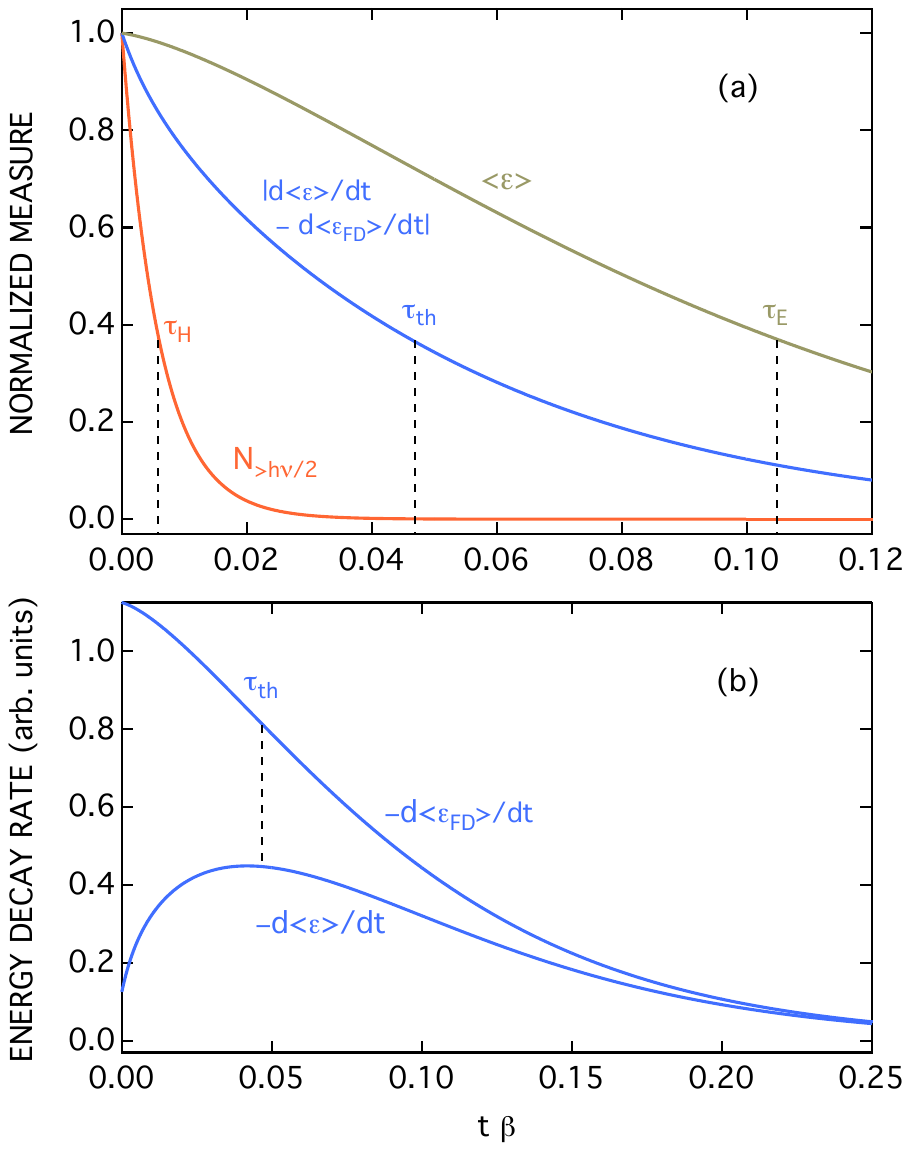}}
\caption{Illustration of the measures for $\tau_H$, $\tau_E$, and $\tau_{th}$.  Each relaxation time is defined to be the time that its corresponding measure decays to $1/e$ of its initial value.  For $\tau_H$, $\tau_E$, and $\tau_{th}$ these measures are $N_{\scriptscriptstyle >h \nu \!/ 2}$ (the number of carriers with energy $\delta \epsilon \ge h \nu /2$), $\langle \epsilon \rangle$ (the excess energy in the carriers), and $|d \langle \epsilon \rangle / dt - d \langle \epsilon_{\scriptscriptstyle F\!D} \rangle / dt|$ (the difference in energy loss rates between the laser excited distribution and an equal-energy FD distribution), respectively.  Panel (a) shows the time dependence of each of these measures, with the corresponding $\tau$ indicated.  Panel (b) separately shows the two energy decay rates $-d \langle \epsilon \rangle / dt$ and $-d \langle \epsilon_{\scriptscriptstyle F\!D} \rangle / dt$ used to determine $\tau_{th}$.  Here $\alpha = 10^{-5}$, $\bdivd = 10^{-3}$, and $\gamma = 0.0172$, which corresponds to the set of $f(\delta\epsilon,t)$ curves in the top right panel of Fig.~\ref{Fig4}. }
\label{Fig5}
\end{figure}

We now investigate the three relaxation times $\tau_H$, $\tau_E$, and $\tau_{th}$.  Each time is determined via a suitable measure of our numerically calculated distribution functions $f(\ep,t)$ (such as those shown in Figs.~\ref{Fig3} and \ref{Fig4}).  In each case the relaxation time is defined to be the time it takes for its measure to decay to $1/e$ of its initial value (subsequent to laser-pulse excitation).

\begin{figure*}[t]
\centerline{\includegraphics[scale=0.60]{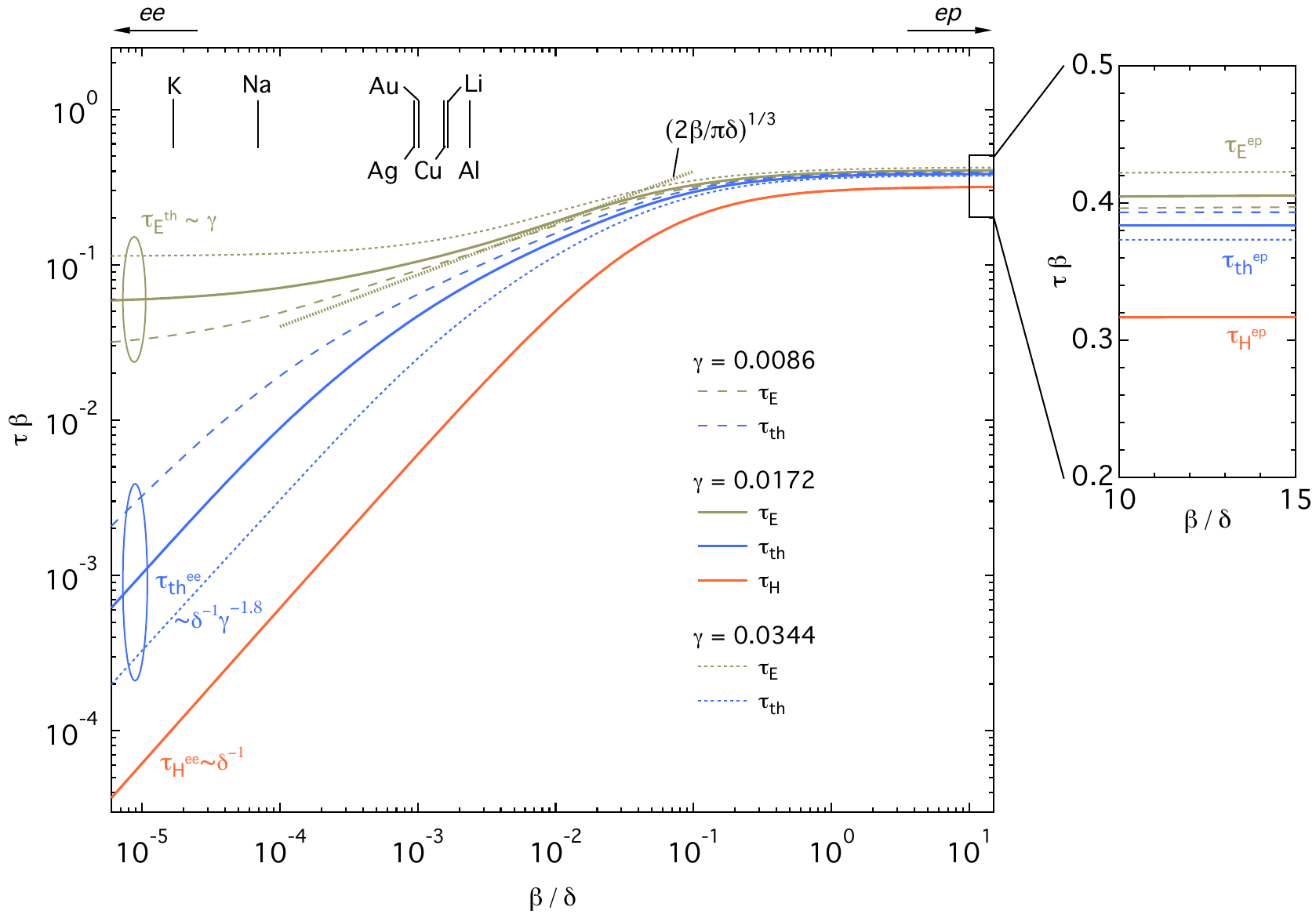}}
\caption{(a) Normalized relaxation times $\tau_H \beta$, $\tau_{th} \beta$, and $\tau_E \beta$ vs ratio $\beta / \delta$ of $ep$ to $ee$ interaction strengths.  Strong $ee$ ($ep$) coupling is on the left (right) side of the main graph.  The small graph to the right is a detailed look at the $10 < \bdivd < 15$ region.  These results are at low excitation level ($\alpha = 1 \times 10^{-5}$), and the values of $\bdivd$ indicated for the seven metals are for photon energy $h \nu$ = 1.5 eV. The dotted line labeled $(2 \beta/\pi \delta)^{1/3}$ is the result $\tau^G_{E} \beta$ derived by Gusev and Wright \cite{Gusev1998}.}
\label{Fig7}
\end{figure*}

So what measures do we employ?  For $\tau_H$ and $\tau_E$ they are quite simple (and identical to those used by Wilson and Coh \cite{Wilson2020}):  the measure for $\tau_H$ is  $N_{\scriptscriptstyle >h \nu \!/ 2}$, the number of excess carriers with energy $\delta \ep > h \nu /2$, and the measure for $\tau_E$ is $\langle \ep \rangle$, the excess energy in the carriers.  Conversely, any measure that might be used to define $\tau_{th}$ is not so simple, owing to the need for some comparison between $f(\ep,t)$ and a thermal (FD) distribution $f_{\scriptscriptstyle F\!D}(\ep,T_e)$.  A number of choices for this measure can be found in the literature, including  (i) the number of excited carriers at small values of $\delta \ep$ ($\delta \ep \ll h \nu$) \cite{Sun1994,Fatti2000,Kabanov2008,Baranov2014}, (ii) the energy per excited carrier \cite{Gusev1998}, (iii) carrier entropy \cite{Mueller2013}, and (iv) the rate of energy transfer from the carriers to the phonon subsystem \cite{Rethfeld2002}.  Given that the degree of intracarrier thermalization profoundly affects the transfer of energy from the electrons to the phonons, we define $\tau_{th}$ (in a manner similar to Rethfeld \textit{et al}.~\cite{Rethfeld2002}) via the rate of energy exchange between the carrier and phonon subsystems.  Specifically, our measure is $|d \langle \epsilon \rangle / dt - d \langle \epsilon_{\scriptscriptstyle FD} \rangle / dt|$, the (magnitude of the) difference in the energy-exchange rates exhibited by the laser-excited nonequilibrium distribution and an \textit{equal-energy} thermal distribution.

The time dependence of our three relaxation-time measures are illustrated in Fig.~\ref{Fig5}(a).  Not unexpectedly, each measure decays in time toward zero.  The dashed vertical line connected to each curve indicates when that measure has decayed to $1/e$ of its initial value, thence giving the associated relaxation time.  Several properties exhibited by these curves are worth noting.  (i) The ordering $\tau_H < \tau_{th} < \tau_E$ is exhibited for all cases studied at low excitation level.  At high excitation levels we shall see that $\tau_{th}$ becomes smaller than $\tau_H$.  (ii) The curve for $\langle \ep \rangle$ vs $t$ is clearly not exponential in nature.  Although not obvious in the figure, this is also true of the thermalization measure.  Decay of $N_{\scriptscriptstyle >h \nu \!/ 2}$ is typically quite close to exponential at low excitation, but becomes rather more complicated at high excitation.  (iii) It is worth considering the two components associated with the degree of thermalization; these components are displayed in panel (b) of Fig.~\ref{Fig5}.  Unsurprisingly, $-d \langle \epsilon_{\scriptscriptstyle F\!D} \rangle / dt$ decreases monotonically, consistent with the the fact that this quantity monotonically increases as a function of $\langle \ep_{\scriptscriptstyle F\!D} \rangle$ \cite{anisimov1974}.  That $-d \langle \epsilon \rangle / dt$ initially increases is due early-time dominance of $ee$ scattering, which initially drives an increase in the number of carriers that are available to exchange energy with the phonons \cite{Groeneveld1992,Groeneveld1995,Gusev1998,Wilson2020}.  This comparison of $-d \langle \epsilon \rangle / dt$ and  $-d \langle \epsilon_{\scriptscriptstyle F\!D} \rangle / dt$ also readily shows that several time constants must pass before the distribution is effectively thermalized.

An overview of our numerical results for the relaxation times is shown in Fig.~\ref{Fig7}, which plots values (normalized by $\beta$) for all three times as a function of $\bdivd$.  The left (right) side of the graph corresponds to relatively strong $ee$ ($ep$) scattering.  The vertical lines labeled with the set of $sp$ metals indicate the corresponding $\bdivd$ values for $h \nu = 1.5$ eV.  Only one curve for $\tau_H$ is shown, owing to its extremely weak dependence upon $\gamma$ (a consequence of $\gamma \ll 1/2$ for all three cases).   As mentioned with regard to Fig.~\ref{Fig5}, the relation $\tau_{H} < \tau_{th} < \tau_{E}$ is exhibited by all of these data.  In what follows we discuss each of these relaxation times in detail.

\begin{table}[h]
\caption{Distribution relaxation times $\tau_H$, $\tau_{th}$, and $\tau_{E}$ for the metals listed in Table~\ref{table1}.  Calculated times are for $h\nu = 1.5$ eV and $ T_p = 300$ K ($\gamma = 0.0172$) at low excitation ($\alpha = 10^{-5}$).  These times correspond to the solid curves in Fig.~\ref{Fig7}.  Also tabulated are the ratios $\tau_H / \tau_E$, $\tau_E / \tau_E^{th}$, and $\tau_{th} / \tau_E$.   \label{table3}}

\vspace{0.2cm}
\begin{tabular}{l@{\hspace{0.3cm}}r@{\hspace{0.4cm}}r@{\hspace{0.4cm}}r@{\hspace{0.3cm}}d@{\hspace{0.3cm}}d@{\hspace{0.3cm}}d@{\hspace{0cm}}r@{\hspace{0.3cm}}r@{\hspace{0.3cm}}r@{\hspace{0.3cm}}r@{\hspace{0.3cm}}r@{\hspace{0.3cm}}r@{\hspace{0.3cm}}r}
\hline
\hline      

\rule{0pt}{2.2ex}Metal		& $\tau_H$  &  $\tau_{th}$ & $\tau_E$  & \tau_H /  . \!\! \tau_E & \tau_E /  . \!\! \tau_E^{th}  & \tau_{th} / . \!\! \tau_E \\

	 	& (fs) & (fs) 	& (fs) 	 &   \\
\hline
Li 		& 16		& 99		& 187		& 0.08	& 2.08		& 0.53 		 \\	

Na  		& 9  		& 140	& 1430		& 0.006	& 1.2	1	& 0.098	 \\

K  		& 8  		& 125	& 4370		& 0.002	& 1.08	& 0.028 	 \\

Rb 		& 6  		& 107	& 8860		& 0.0007	& 1.04	& 0.012	 \\

Cs 		& 6  		& 109	& 18600		& 0.0003	& 1.03	& 0.006	 \\


 \rule{0pt}{4ex}Cu 	& 53  	& 343	& 662	& 0.08	& 2.04	& 0.52		 \\

Ag 		& 87  	& 691	& 1590	& 0.05	& 1.83	& 0.43	 \\

Au 		& 100  	& 770	& 1720	& 0.06	& 1.86	& 0.45	 \\


 \rule{0pt}{4ex}Al 	& 19  	& 105	& 179	& 0.1	1	& 2.26	& 0.58		 \\

\hline
\hline

\end{tabular}
\smallskip
\end{table}

\subsubsection*{a. Hot-carrier relaxation}

As can be inferred from Fig.~\ref{3}, both $ee$ and $ep$ scattering contribute to the decrease in $N_{\scriptscriptstyle >h \nu \!/ 2}$, the number of carriers with energy $\delta \ep > h \nu/2$.  As we now show, the contributions of the two scattering mechanisms are nearly independent of each other.  Complete independence would be confirmed if the rate $1/\tau_H$ were to exhibit Matthiessen's rule, $1/\tau_H = 1/\tau_H^{ee} + 1/\tau_H^{ep}$,
where here (and in subsequent relaxation-time expressions) the superscripts $ee$ and $ep$ indicate the limit of purely $ee$ scattering or $ep$ scattering, respectively.  This relationship almost describes our data.  However, we find that a slightly better description is given by an extended Matthiessen's rule,
\be{\eq}
\label{29a}
\bigg( \frac{1}{\tau_H} \bigg)^{\!\! a} = \bigg (\frac{1}{\tau_H^{ee}} \bigg)^{\!\! a} + \bigg( \frac{1}{\tau_H^{ep}} \bigg)^{\!\! a},
\en{\eq}  
where from fitting we find an exponent $a = 0.95$.\footnote{Analysis of the numerical $\tau_H$ data, as well as the $\tau_E$ and $\tau_{th}$ data, is described in detail in \cite{SuppMat}.}  Values of $\tau_H^{ee} \delta$ and $\tau_H^{ep} \beta$ are given in Table \ref{table2a} for the three values of $\gamma$ investigated.  As is evident there, both $\tau_H^{ee} \delta$ and $\tau_H^{ep} \beta$ are largely independent of $\gamma$.  Indeed, all of the $\tau_H$ data are well described by the relation
\be{\eq}
\label{29b}
\bigg( \frac{1}{\tau_H} \bigg)^{\!\! 0.95} = \bigg (\frac{\delta}{6.2} \bigg)^{\!\! 0.95} + \bigg( \frac{\beta}{0.32} \bigg)^{\!\! 0.95}.
\en{\eq}  

The result expressed by Eq.~(\ref{29b}) is quite illuminating.  First, from this equation we infer the contributions of the two scattering mechanism are equal at $\beta / \delta \approx 5 \times 10^{-2}$.  Second, as shown by the $\beta / \delta$ locations of the seven metals in Fig.~\ref{Fig7}, we see for these (and the heavier alkali metals, see Table \ref{table1}) that $ee$ scattering is the dominant relaxation mechanism associated with $\tau_H$.

The strong $ee$ scattering limit is thus interesting in its own right.  First, we note $\tau_H^{ee} \, \delta \approx 6.2$ is in agreement with the results of Wilson and Coh \cite{Wilson2020}.  Second, using Eqs.~(\ref{7}) and (\ref{12}) it is trivial to show
\be{\eq}
\label{30}
\tau_H^{ee} = 3.1 \, \tau_{ee}(h \nu).
\en{\eq}
That is, in this limit the hot-electron relaxation time is only three times longer than the single-particle scattering time for the highest energy electrons in the distribution.

At the other extreme---dominant $ep$ scattering---we have $\tau_H^{ep} \beta \approx 0.32$.  As discussed in Appendix \ref{Appendix D}, this result can be simply understood by solely considering the consequences of spontaneous phonon emission.  Indeed, via an analytic calculation (see the appendix) we find $\tau_H^{ep} = (1 - 1/e)/2 = 0.3161$, which is remarkably close to the numerical results displayed in Table \ref{table2a}.

Because $\beta / \delta \! \propto \! (h \nu)^{-3}$ [see Eq.~(\ref{14})], the $\bdivd$ values associated with the seven metals indicated in Fig.~\ref{Fig7} can be shifted parallel to the abscissa by changing $h \nu$.  For example, a change in $h \nu$ from 1.5 eV (the value used in the figure) to 0.5 eV shifts the value of $\beta / \delta$ for Al from $2.4 \times 10^{-3}$ to $6.5 \times 10^{-2}$.  This change puts Al into the region where $ep$ scattering very slightly dominates $ee$ scattering in determining $\tau_H$, but  for the other metals in Fig.~\ref{Fig5}, $\tau_H$ is still primarily controlled by the $ee$ interaction.

To visually connect $\tau_H$ with the electron distributions shown in Fig.~\ref{Fig3} and \ref{Fig4}, we have indicated $\tau_H$ on these figures:  on the abscissa of each graph we mark (using the symbol $\otimes$) the location where the distribution function at time $t = \tau_H$ crosses the abscissa.  Because in all of these graphs $f = 2 \times 10^{-6}$ along the abscissa, this point is determined by the condition $f(\delep,\tau_H) = 2 \times 10^{-6}$.  As is evident, the energy of this point is universally quite close to $\delta \ep / h \nu$ = 0.8.

Using these normalized results, we have calculated $\tau_H$ values for each of the metals listed in Table~\ref{table1}.  The results for $\tau_H$ for input parameters $h\nu = 1.5$ eV and $ T_p = 300$ K ($\gamma = 0.0172$) are reported in Table~\ref{table3}.  As can be seen there, $\tau_H$ ranges from a few fs (for the four heaviest alkali metals) to 100 fs (for Au).  

\subsubsection*{b. Energy relaxation}

\begin{figure*}[t]
\centerline{\includegraphics[scale=0.48]{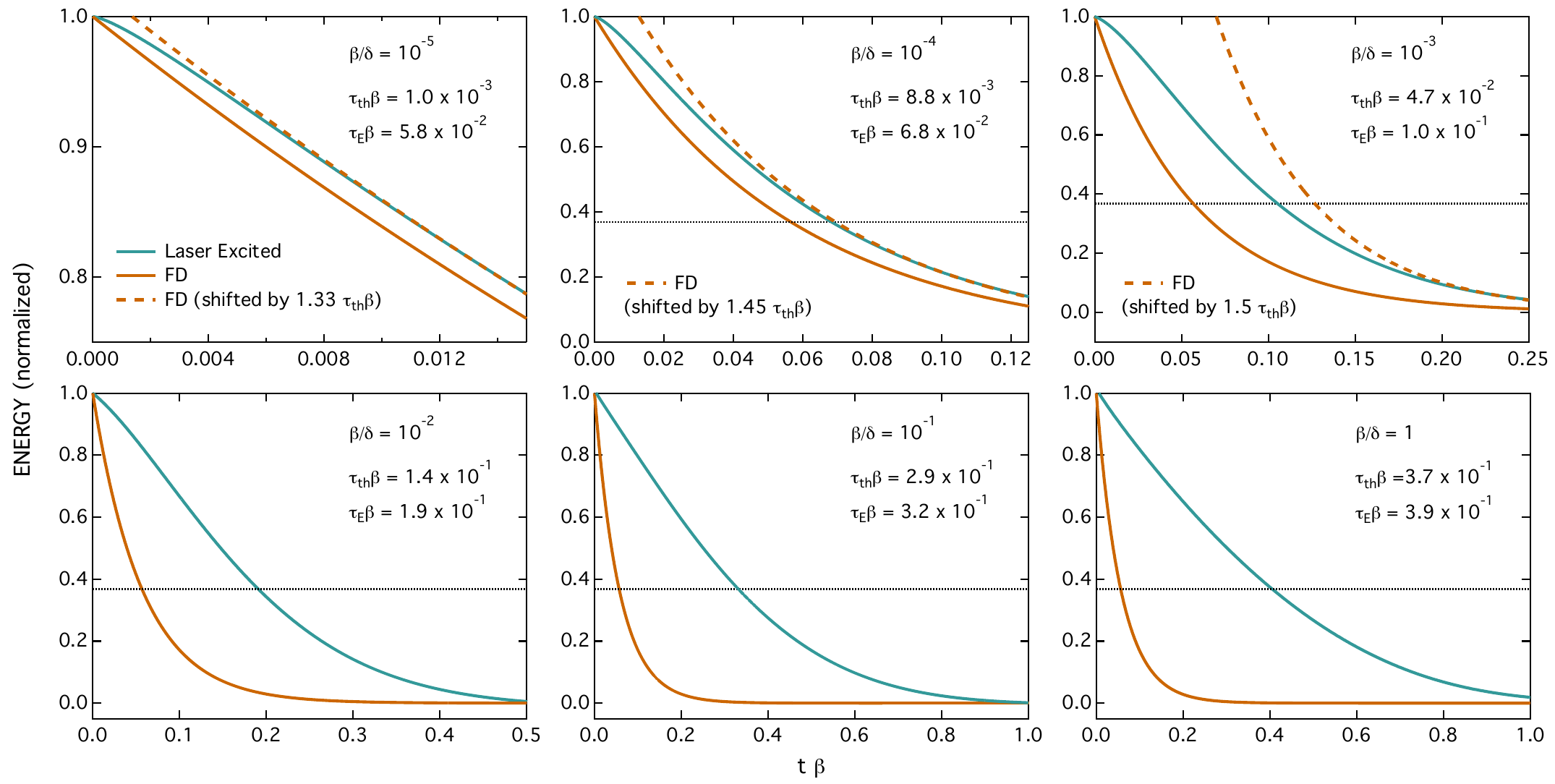}}
\caption{Carrier energy vs time for six different values of the ratio $\beta / \delta$ of $ep$ to $ee$ interaction strengths.  Each panel (as labeled by $\beta / \delta$) corresponds to the analogous panel of Fig.~\ref{Fig4}.  The energy decay for each laser-excited distribution is compared to the decay of a thermalized (FD) distribution.  The dashed-line graph in the three top panels is the thermalized-distribution decay shifted by a small factor times $ \, \tau_{th} \beta$.  Energy is normalized to the initial energy in each distribution, and the horizontal dotted line marks $1/e$ on the energy axis.  Here the normalized excitation level and phonon temperature are $\alpha = 1 \times 10^{-5}$ and $\gamma = 0.0172$, respectively.}  
\label{Fig6}
\end{figure*}

We now consider the energy relaxation time $\tau_E$.  A point of comparison is the relationship of $\tau_E$ to the energy-relaxation time $\tau_E^{th}$ of a thermalized (FD) distribution.  Because the present calculations are for very low excitation, the energy decay of a thermalized distribution is purely exponential, and so $\tau_E^{th}$ as given by Eq.~(\ref{22}) is the same as the $1/e$ decay time.

A comparative overview of $\tau_E$ and $\tau_E^{th}$ is provided by Fig.~\ref{Fig6}, which plots in each panel two energy-decay curves, one for a laser-excited distributions and one for a FD distribution with the same initial energy.  The laser-excited decay curves correspond to the set of distributions in Fig.~\ref{Fig4}.  Noting the horizontal dotted line in a panel in Fig.~\ref{Fig6} marks the $1/e$ point for the energy remaining in the distribution, we readily observe (i) $\tau_E / \tau_E^{th} > 1$ and (ii) $\tau_E / \tau_E^{th}$ increases as $\beta / \delta$ increases. 

Our numerical results for $\tau_E$ vs $\bdivd$ are shown in Fig.~\ref{Fig7}.  As the figure illustrates, each $\tau_E \beta$ curve has two clear asymptotic limits, with a halfway point close to $\bdivd = 10^{-2}$.  As $\beta / \delta$ becomes increasingly smaller $\tau_E$ approaches $\tau_E^{th}$ [see Eq.~(\ref{22})], which in normalized parameters can be expressed as 
\be{\eq}
\label{31}
\tau_E^{th} \, \beta= \frac{\pi^2}{3} \, \gamma.
\en{\eq}
In this very small $\beta/\delta$ regime $ee$ scattering is so strong that the timescale $\tau_{th}$ for the distribution to become thermalized (which we discuss in detail shortly) is much shorter than the energy relaxation timescale $\tau_E$.  Therefore, energy transfer to the phonons essentially occurs from a thermalized carrier distribution.  Consequently, in this regime the standard 2T model \cite{anisimov1974} is an excellent description of the energy dynamics.  At increasingly larger values of $\beta / \delta$---where $ep$ scattering dominates---the second asymptote is approached.  From our numerical calculations we find this limit is given by
\be{\eq}
\label{32}
\tau_E^{ep} \, \beta  \approx 0.40. \frac{}{}
\en{\eq}
More precise values of $\tau_E^{ep} \beta$ for each value of $\gamma$ are given in Table \ref{table2a}.  We note those numerical results are quite close to the analytic approximation $\tau_E^{ep} \, \beta= 1 - \sqrt{1/e} = 0.3935$, which is derived in Appendix \ref{Appendix D}.  The ratio $\tau_E^{ep} / \tau_E^{th} \approx 0.12 / \gamma$ (values are tabulated in Table \ref{table2a}) indicates the degree to which strong intraelectron thermalization enhances energy decay.  This ratio is very nearly exhibited by the energy-decay curves in the $\beta / \delta = 1$ panel of Fig.~\ref{Fig6}.

We note the $ee$ interaction does not appear (via $\delta$) in either of the asymptotic-limit expressions for $\tau_E$.  At the smallest values of $\beta / \delta$ this occurs because the effects of $ee$ scattering are over well before significant energy relaxation occurs.  Conversely, in the large $\beta / \delta$ limit, any potential contribution from $ee$ scattering is much too slow to impact relaxation of the energy. 

We have developed a heuristic equation for $\tau_E$ that improves upon several equations offered by Wilson and Coh \cite{Wilson2020}.  In our notation their most accurate expression (their Eq.~S17) can be written as 
\begin{widetext}
\begin{align}
\label{32B}
\tau_{E}	&= \tau_E^{th} \bigg[ 1 + (A_0 + A_1 h \nu) \Big\{ 1 - \tanh \ln \big(  A_2  [(h \nu)^{2.86} K_{ee} / K_{ep}]^{0.42} \big) \Big\} \nonumber \\
		& \quad + \big[A_3 + A_4 (h \nu)^{1.1} \big] \,  {\rm sech} \ln \big( A_5 \,  [(h \nu)^{2.38} K_{ee} / K_{ep}]^{0.42}  \big)    \bigg],
\end{align}
\end{widetext}
where the $A_i$'s are fitted constants.  There are two issues with this equation.  First, given the structure of the BTE, the parameters $h \nu$, $K_{ep}$, and $K_{ee}$ should appear together as the (unitless) combination $(h \nu)^3 K_{ee} / K_{ep}$ (= $\delta / \beta$).  This combination almost appears correctly in the argument of the $\tanh \ln$ function, but in the $\rm{sech} \ln$ function the combination is rather far off.  Second, the other occurrences of $h \nu$ should appear divided by phonon temperature $k_B T_p$ (recall $h \nu / k_B T_p = \gamma^{-1}$).  That $h \nu$ appears by itself in Eq.~(\ref{32B}) can be attributed to the fact that Wilson and Coh only study materials at one temperature (RT).  Our updated version of this equation---which we have used to fit our numerical data---is 
\begin{align}
\label{32C}
\tau_{E} \beta	\!&=\! \frac{\pi^2}{3} \gamma \bigg[ 1 \! + \! \big(B_0 \! + \! B_1 \gamma^{-B_2} \big) \! \Big\{ 1 \! - \! \tanh \ln \big(  B_3  [\bdivd]^{-B_4}  \big) \Big\} \nonumber \\
		& \quad + \big(B_5 + B_6 \gamma^{-B_7} \big) \,  {\rm sech} \ln \big( B_8 \, [\bdivd]^{-B_9}   \big)    \bigg].
\end{align}
The best-fit values of the $B_i$'s are listed in Table \ref{table4}.  The typical deviation between the numerical data and Eq.~(\ref{32C}) is $\sim$1\%.

Aside from the analysis by Wilson and Coh, there has been one other expression previously offered for the energy relaxation time.  From their analytic approximations Gusev and Wright  \cite{Gusev1998} derive an expression for $\tau_E$, which in our notation is given by
\be{\eq}
\label{34}
\tau_E^{\scriptscriptstyle G}  \beta = \bigg( \frac{2}{\pi} \frac{\beta}{\delta} \bigg)^{\!\! 1/3}. 
\en{\eq}
This relation is plotted in Fig.~\ref{Fig7} as the dotted line labeled $(2 \beta / \pi \delta)^{1/3}$.  In the range of $\bdivd$ values germane to all but the four heaviest alkali metals, this expression is remarkably close to our numerical results for the smallest value of $\gamma = 0.0086$ (lowest temperature) we have studied.  Furthermore, it reproduces in reasonable fashion the trend in $\tau_E \beta$ over the range $5 \times 10^{-4} < \bdivd < 5 \times 10^{-2}$.  However, for the smallest values of $\bdivd$ (relevant to the heaviest alkali metals) this expression significantly underestimates $\tau_E$.  It also becomes less accurate as $\gamma$ increases, in that the dependence of $\tau_E \beta$ upon $\bdivd$ increasingly deviates from a power law with exponent $1/3$.  Indeed, for $\gamma = 0.0172$, the simple expression $\tau_E \beta\approx 0.66 \, (\bdivd)^{1/4}$ offered by Wilson and Coh is much more accurate for $5 \times 10^{-4} < \bdivd < 10^{-1}$ \cite{Wilson2020}.

In Table~\ref{table3} we also list values of $\tau_E$ for the simple metals.  Like the $\tau_H$ values, these $\tau_E$ values are calculated using input parameters $T_p = 300$ K and $h \nu = 1.5$ eV.  Most interesting is a comparison of $\tau_E$ and $\tau_E^{th}$ values, the latter quantity being the energy-decay time for an already thermalized distribution.  The ratio of these two quantities is also reported in Table~\ref{table3}.  Excepting the four heaviest alkali metals, this ratio is close to 2, a consequence of the relatively slow intracarrier thermalization (discussed in detail below) for these metals.

A comparison of $\tau_H$ and $\tau_E$ is also informative. (Here we also consider the specific case $\gamma = 0.0172$.)  When $\beta / \delta \lesssim 10^{-4}$ we have $\tau_H / \tau_E \approx 100 \, \beta / \delta \lesssim 10^{-2}$.  Therefore, for the four heaviest alkali metals hot-carrier relaxation is much faster than energy relaxation.  For $\beta / \delta$ values up to $2.5 \times 10^{-3}$ we have $\tau_H / \tau_E \lesssim 0.1$ .  Thus, for all of the other metals in Table \ref{table1} hot-carrier relaxation is at least somewhat faster than energy relaxation (see Table~\ref{table3} for specific ratio values).  For $\beta / \delta \gtrsim 1$ (where $ep$ scattering controls both relaxation times) the ratio is constant, with $\tau_H / \tau_E \approx 0.8$.  That the ratio is close to unity is not surprising, owing to $\sim$3/4 of the excess energy being initially contained in the hot electrons [those with $\delta \ep / (h \nu) > 0.5$].  The overall behavior of $\tau_H / \tau_E$ vs $\beta / \delta$ is also evident in the distribution-function plots in Fig.~\ref{Fig4}, where $\tau_E$ is indicated (via the symbol $\boxplus$) in the same manner as $\tau_H$.

\begin{table}[t]
\caption{Best-fit values of the parameters $B_i$ in Eq.~(\ref{32C}).  \label{table4}}
\vspace{0.2cm}

\begin{tabular}{l@{\hspace{1cm}}d@{\hspace{0.4cm}}d@{\hspace{0.4cm}}d@{\hspace{0.5cm}}r@{\hspace{0.5cm}}r@{\hspace{0.5cm}}r@{\hspace{0.5cm}}r@{\hspace{0.5cm}}r@{\hspace{0.5cm}}r}
\hline
\hline

$i$ & \,  B_i   \\

\hline

0  & -0.346314	\\

1  & 0.059041	\\

2  & 1.000961	\\

3  & 0.232987	\\

4  & 0.390694	\\

5  & -0.272834	\\

6  & 0.004287	\\

7  & 1.241227	\\

8  & 0.116529	\\

9  & 0.341017	\\

\hline
\hline

\end{tabular}
\smallskip
\end{table}

\subsubsection*{c. Carrier thermalization}

Along with the relaxation times $\tau_H$ and $\tau_E$, our numerical results for $\tau_{th}$ are shown in Fig.~\ref{Fig7}.  Not unexpectedly, for the smallest values of $\beta/\delta$ we have $\tau_{th} \ll \tau_E$.  Importantly, as the normalized temperature parameter $\gamma$ decreases these two relaxation approach each other.  At the smallest values of $\bdivd$ this approach as a function of $\gamma$ is dramatic, owing to $\tau_{th}^{ee} / \tau_E^{th} \sim \gamma^{2.8}$.  (The dependence of $\tau_{th}^{ee}$ on $\gamma$ is discussed in more detail below.)
  
At the extremes of $\beta / \delta$ the relaxation time $\tau_{th}$ has the same simple behavior as the hot-carrier relaxation time $\tau_H$.  For $\beta / \delta \lesssim 5 \times 10^{-5}$, $\tau_{th}$ is controlled by $ee$ scattering and is given by 
\be{\eq}
\label{37}
\tau_{th} \, \delta \approx \tau_{th}^{ee} \, \delta,
\en{\eq}
where the value of $\tau_{th}^{ee} \, \delta$ depends upon $\gamma$.\footnote{In the $ee$-only scattering limit, there is of course no energy transfer to the lattice, and so strictly speaking our thermalization measure $|d \langle \epsilon \rangle / dt - d \langle \epsilon_{\scriptscriptstyle F\!D} \rangle / dt|$ is undefined.  However, at each time step of the calculation we can compare the requisite two relaxation rates by assuming an infinitesimal $ep$ coupling.}  Values of $\tau_{th}^{ee} \, \delta$ for the three values of $\gamma$ studied here are given in Table \ref{table2a}.  Conversely, for $\beta / \delta \gtrsim 5 \times 10^{-1}$, $\tau_{th}$ is dominated by $ep$ scattering.  In this regime we have
\be{\eq}
\label{38}
\tau_{th} \, \beta\approx \tau_{th}^{ep} \, \beta \approx 0.38.
\en{\eq}
More accurate values of $\tau_{th}^{ep} \, \beta$ are given in Table \ref{table2a}, but as is evident there, $\tau_{th}^{ep} \, \beta$ is only weakly dependent upon $\gamma$.  As we have done with $\tau_H$ and $\tau_E$, we indicate $\tau_{th}$ (via the symbol $\boxtimes$) on the graphs in Figs.~\ref{Fig3} and \ref{Fig4}.  

Given that $\tau_{th}$ is controlled by $ee$ ($ep$) scattering at very small (large) values of $\bdivd$, one might wonder if their contributions at intermediate values of $\bdivd$ are approximately independent, as is the case for $\tau_H$.  We have thus also fit this data with our power-law extension of Matthiessen's rule,
\begin{align}
\label{88}
\bigg( \frac{1}{\tau_{th}}\bigg)^{\! \! b} &= \bigg( \frac{1}{\tau_{th}^{ee}} \bigg)^{\! \! b} + \bigg( \frac{1}{\tau_{th}^{ep}} \bigg)^{\! \! b} \nonumber \\
							&= \bigg( \frac{\delta}{\tau_{th}^{ee}  \delta} \bigg)^{\! \! b} + \bigg( \frac{\beta}{\tau_{th}^{ep}  \beta} \bigg)^{\! \! b},
\end{align}
where the constant values of $\tau_{th}^{ee}  \delta$ and $\tau_{th}^{ep}  \beta$ (for each value of $\gamma$) can be found in Table \ref{table2a}.  We find that Eq.~(\ref{88}) describe the data quite well for values of $\bdivd < 10^{-1}$, with best-fit value of the exponent $b$ equal to 0.42, 0.52, and 0.66 for $\gamma$ equal to 0.0086, 0.0172, and 0.034, respectively.  Because $b$ is not close to 1 for any of the cases, the two scattering mechanisms cannot be considered to be independent as far as thermalization is concerned.  As is the case with energy relaxation, $ee$ and $ep$ scattering intertwine in a nontrivial manner to effect carrier thermalization.  Unlike hot-carrier relaxation, however, the magnitude of the contribution from $ep$ scattering to thermalization is quite important:  for all but the four heaviest alkali metals, the thermalization rate is at least twice as fast as it would be if only $ee$ scattering were operating.  For Li and Al, for example, the thermalization rate is 2.8 and 3.3 times faster, respectively.  

\begin{figure}[t]
\centerline{\includegraphics[scale=0.65]{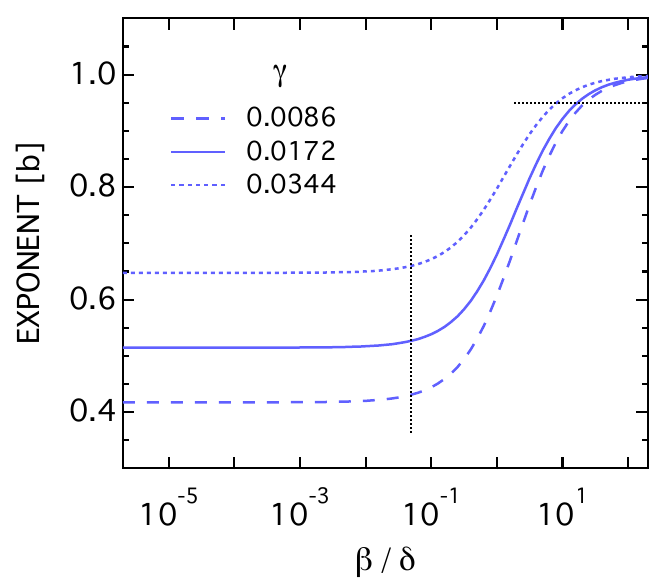}}
\caption{Dependence of the thermalization time exponent $b$ [see Eqs.~(\ref{88}) and (\ref{88B})] on the ratio $\beta / \delta$ of $ep$ to $ee$ interaction strengths.  }
\label{Fig8B}
\end{figure}

As $\bdivd$ increases beyond $10^{-1}$ we find that Eq.~(\ref{88}) with a constant value of $b$ becomes increasingly poor at describing the dynamics.  However,  by allowing the exponent $b$ to vary with $\bdivd$ as
\begin{equation}
\label{88B}
b = \frac{b_0 + b_1 \, \beta/\delta}{1 + b_1 \, \beta / \delta}
\end{equation}
a quite accurate description of the data is obtained for all values of $\bdivd$.  In Fig.~\ref{Fig8B} we plot fitted values of $b$ as a function of $\bdivd$.\footnote{Fitted values of $b_0$ and $b_1$ are reported \cite{SuppMat}.}  For $\bdivd \lesssim 5 \times 10^{-2}$, $b$ is approximately constant (and nearly equal to the values obtained from the constant-$b$ analysis above).  However, a transition ensues in the range $5 \times 10^{-2} \lesssim \bdivd \lesssim 10$, above which the exponent is not far from 1.  Hence, in the $ep$ dominant region ($\bdivd \gtrsim 10$) the contributions of the two scattering mechanisms can be considered to be approximately independent. 

Because rapid thermalization leads to efficient energy transfer, it is relevant to ask under what circumstances does thermalization precede significant energy relaxation?  To investigate this question we have again plotted in the top three panels of Fig.~\ref{Fig6} the FD decay curve, but shifted in time by a small factor times $\tau_{th} \beta$.  The small factor (1.33, 1.45, and 1.5 for $\beta / \delta$ = $10^{-5}$, $10^{-4}$, and $10^{-3}$, respectively) in each case is chosen so that the laser-excited and FD distribution curves overlap at later times.   This comparison allow us to see that it is only for $\beta / \delta = 10^{-5}$ that this overlap occurs before substantial energy decay has occurred.  Therefore, only for the $\beta / \delta \lesssim 10^{-5}$ data can we say that thermalization precedes energy relaxation.  A comparison of the ratio $\tau_{th} / \tau_E$ at $\bdivd = 10^{-5}$ and $10^{-4}$ is illuminating.  The two ratios are 0.017 and 0.12, respectively.  Thus, a ratio well under 0.1 is necessary for thermalization to precede energy decay.  

Table~\ref{table3} also reports values of $\tau_{th}$, calculated using the same input parameters as $\tau_H$ and $\tau_{E}$.   Relevant to the present discussion we also tabulate the ratio $\tau_{th} / \tau_E$.  For Rb and Cs this ratio 0.012 and 0.006, respectively.  Thus the 2T model is applicable to these two metals.  For K $\tau_{th} / \tau_E = 0.028$; the 2T model should thus be used with caution in this case.  For all of the other metals the ratio is $\gtrsim 0.1$, which invalidates use of the 2T model to describe the dynamics.  Our assessment agrees with the linearized BTE analysis of Baranov and Kabanov, who show that thermalization and energy decay happen on similar timescales in Au \cite{Baranov2014}.  Pietanza \textit{et al.} also note similar behavior at lower excitation levels in their BTE study of Ag \cite{pietanza2007}.

Two sets of researchers offer simple expressions for $\tau_{th}$ when $ee$ scattering dominates  \cite{Gusev1998,Kabanov2008,Baranov2014}.  Gusev and Wright define a thermalization time to be the time when the average energy per excitation becomes equal to that of a FD distribution \cite{Gusev1998}.\footnote{In reality, the average energy per excitation only asymptotically approaches that of a FD distribution.  However, in the approximate theory that Gusev and Wright develop there is a point in time when such equality does occur.}  Written in our notation, their equation for $\tau_{th}$ is 
\be{\eq}
\label{35}
\tau_{th}^{\scriptscriptstyle G}  \delta = \bigg( \! \frac{2}{\pi} \! \bigg)^{\!\! 5} \bigg[  \frac{\ln(2)}{\gamma}  \bigg]^{\! 2}  \approx \frac{1}{20 \, \gamma^2}.
\en{\eq}
Kabanov and coworkers divide the excited carrier distribution into thermal and nonthermal parts \cite{Kabanov2008,Baranov2014}.  They consider relaxation of the nonthermal component and identify an asymptotic thermalization time $\tau_{th}^{\scriptscriptstyle K}$.  Specifically, they show at long times [$t/\tau_{th}^{\scriptscriptstyle K} \gg (1,\gamma^2)$] the nonthermal component of the carrier density has the time dependence $e^{-t/\tau_{th}^{\scriptscriptstyle K}}/t$, where
\be{\eq}
\label{36}
\tau_{th}^{\scriptscriptstyle K} \, \delta = \frac{2}{\pi^2 \gamma^2}  \approx \frac{1}{5 \, \gamma^2}.
\en{\eq}
We note this relation has the same $1 / \gamma^2$ dependence as Eq.~(\ref{35}), but is about four times longer.  This difference can be attributed (at least in part) to the fact that at long times most of the electrons have relaxed to low-energy states ($\delta \ep \ll h \nu$), which have much reduced $ee$ scattering rates.

Numerically we find the $\gamma^{-2}$ behavior exhibited by Eqs.~(\ref{35}) and (\ref{36}) is not quite obeyed.  Our calculations instead reveal in the very strong $ee$ scattering regime that the thermalization time is accurately described by the expression $\tau_{th}^{ee} \delta = 3.33 + 0.059 \, \gamma^{-1.83}$.  Interestingly, values calculated using the approximate Gusev and Wright relation Eq.~(\ref{35}) are within at least a factor of 2 of our numerical results, as can be seen by comparing the values of $\tau_{th}^{ee} \delta$ and $\tau_{th}^{\scriptscriptstyle G} \delta$ listed in Table \ref{table2a}.\footnote{See also Fig.~S2 in \cite{SuppMat}.}

\subsection{Higher excitation levels}
\label{Sec4D}

\begin{figure}[t]
\centerline{\includegraphics[scale=0.50]{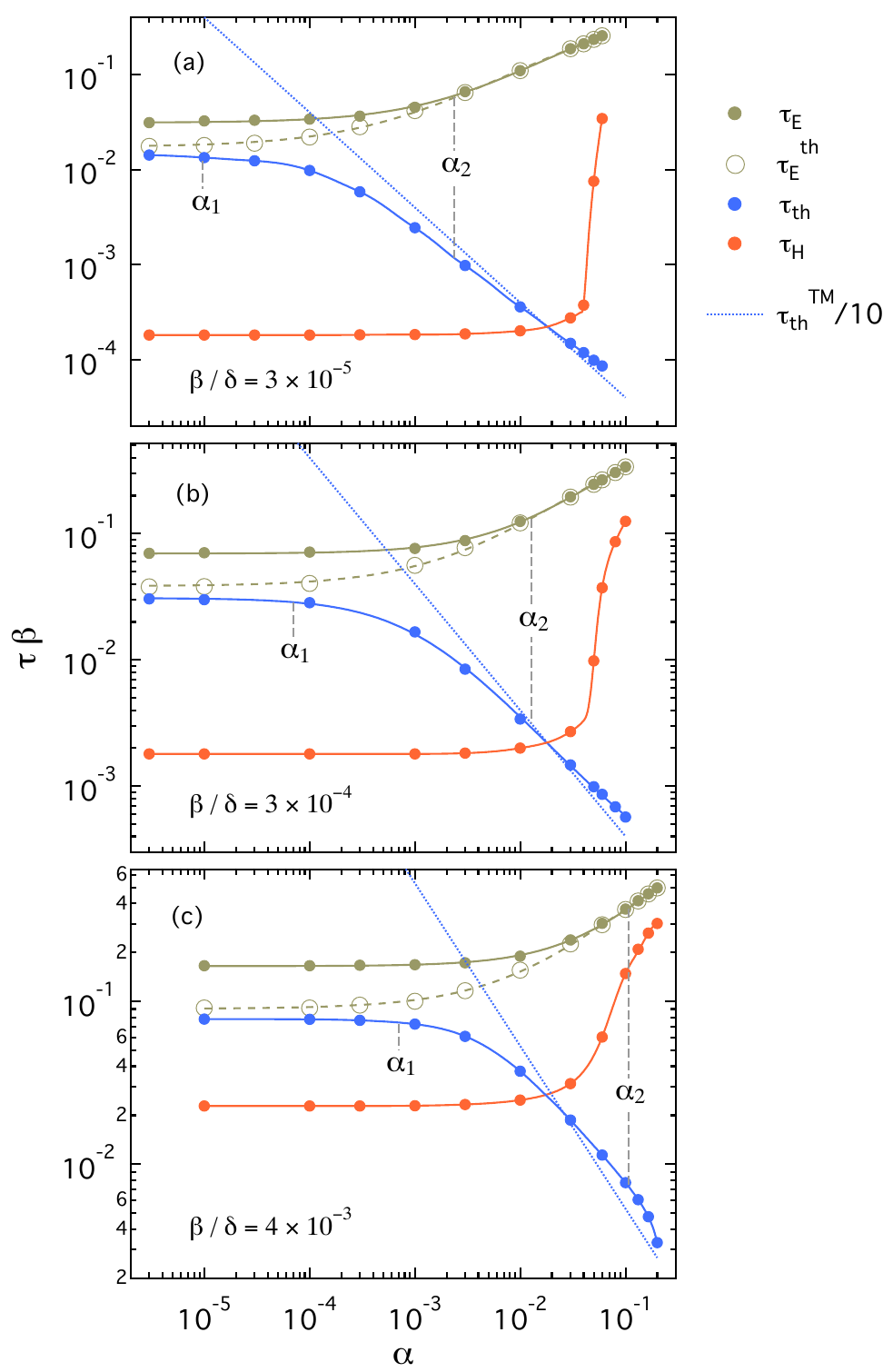}}
\caption{Relaxation times $\tau_H$, $\tau_{th}$, $\tau_E^{th}$, and $\tau_E$ in RT idealized Ag/Au vs excitation level $\alpha$.  Circles are calculated values, and curves are guides to the eye.  Calculations for $\bdivd = 3 \times 10^{-5}$, $3 \times 10^{-4}$, and $4 \times 10^{-3}$ (which correspond to $h \nu = 4.83$, 2.24, and 0.94 eV, respectively) are shown.  The dotted blue lines are 1/10 of the Tas and Maris \cite{Tas1994} thermalization time $\tau_{th}^{\scriptscriptstyle T \! M}$ [see Eq.~(\ref{45})].  The boundaries $\alpha_1$ and $\alpha_2$ are defined by Eqs.~(\ref{37b}) and (\ref{38b}), respectively.}
\label{Fig9}
\end{figure}

\begin{figure}[t]
\centerline{\includegraphics[scale=0.48]{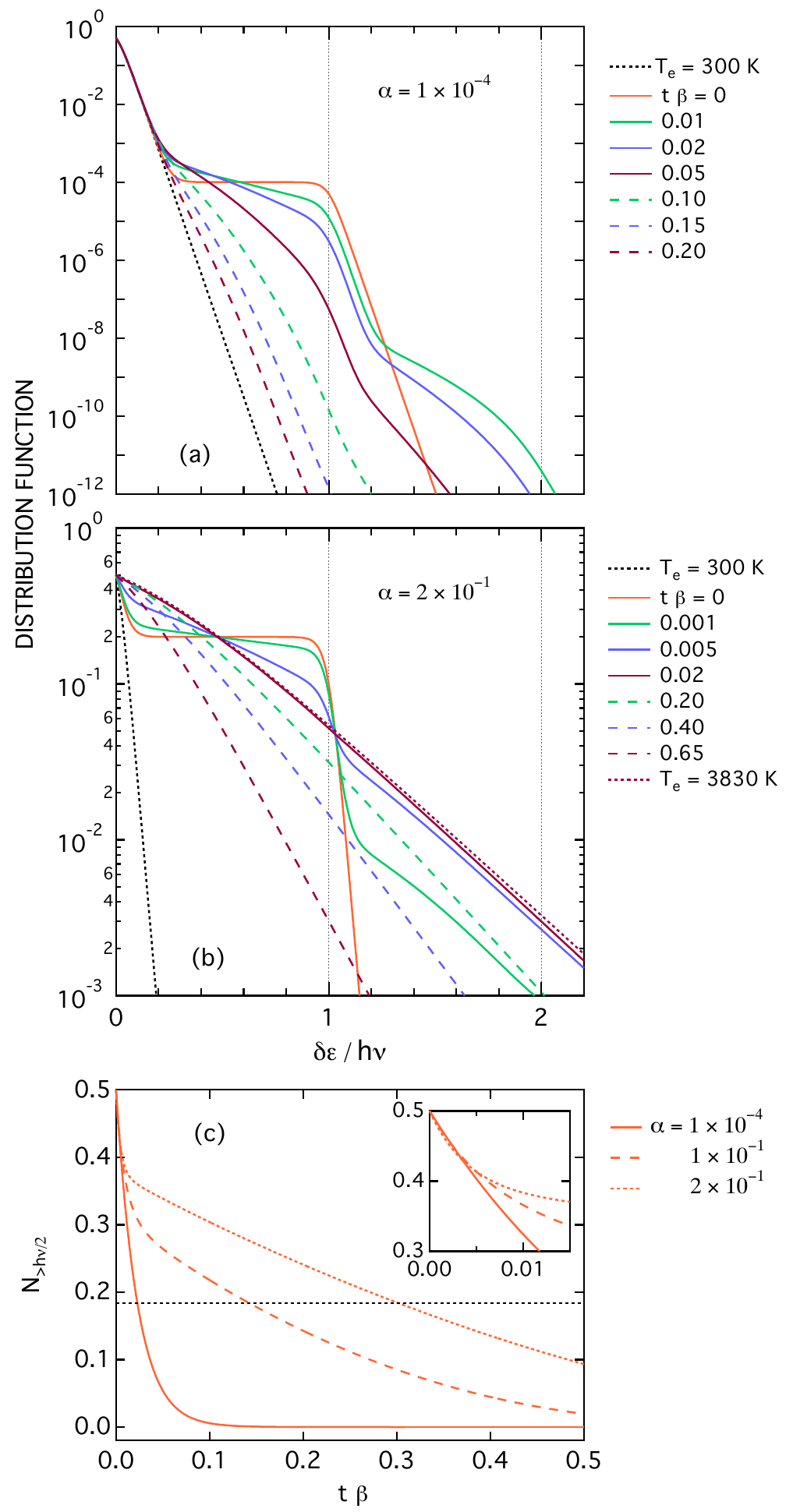}}
\caption{Dependence of hot-carrier relaxation on normalized excitation level $\alpha$ in idealized Ag/Au.  Panels (a) and (b) plot sequences of the distribution function $f(\delep,t)$ at low and high excitation levels, respectively.  In panel (b) the curve labeled $T_e = 3830$ K is a FD distribution at the initial (maximum) temperature in the limit of the 2T model.  Panel (c) plots the time dependence of $N_{\scriptscriptstyle >h\nu \!/2}$ for three values of $\alpha$.  (Here $N_{\scriptscriptstyle >h\nu \!/2}$ is normalized to an initial value of $1/2$, which is the fraction of initially excited carriers with energy $\delep > h \nu / 2$.)  Calculations are for $\bdivd = 4 \times 10^{-3}$, which corresponds to the data in Fig.~\ref{Fig9}(c).  }
\label{Fig9B}
\end{figure}

\begin{figure*}[t]
\centerline{\includegraphics[scale=0.54]{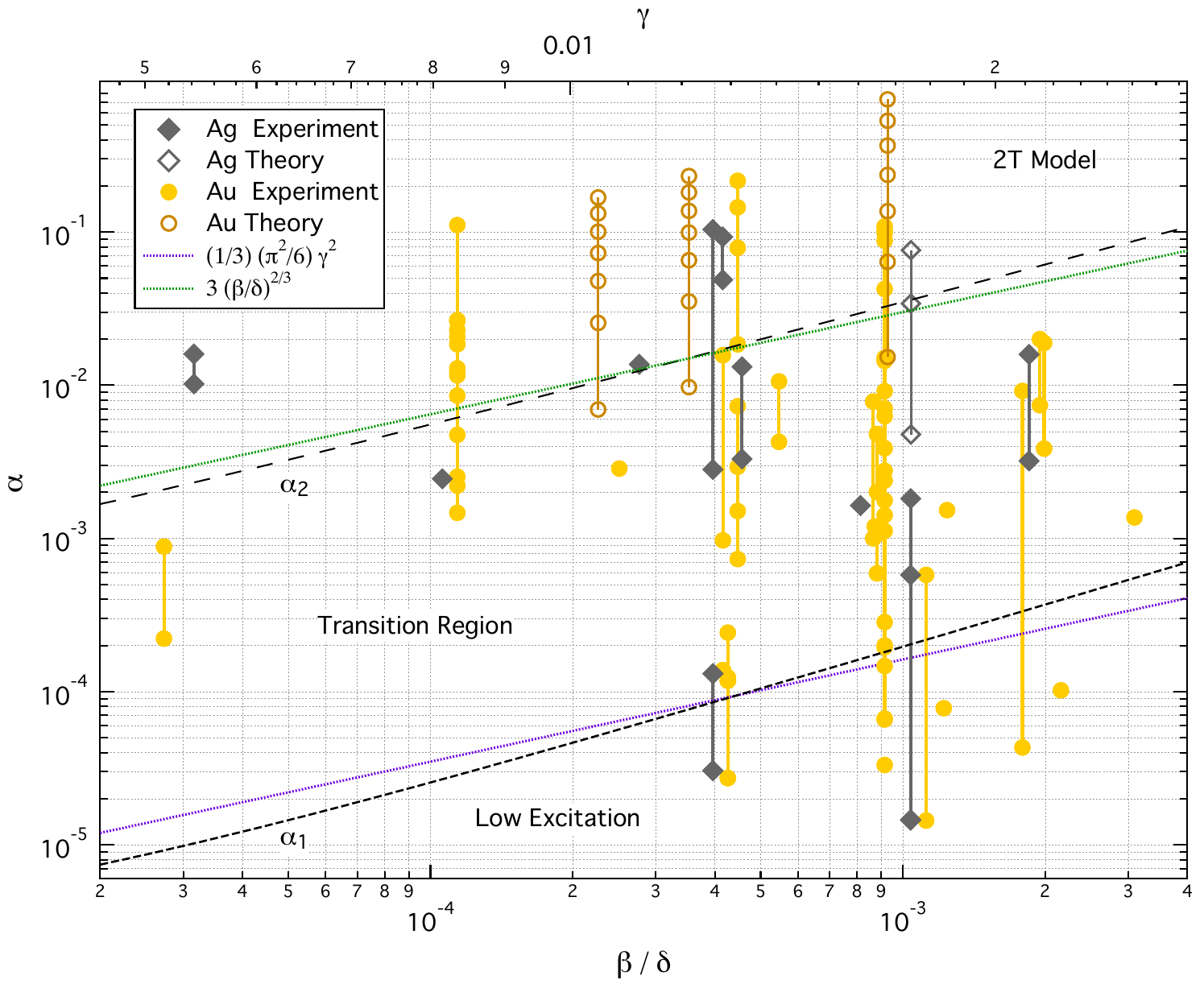}}
\caption{Three regions of carrier relaxation behavior (shown here for Ag and Au).  In the low excitation region (below the short-dashed line labeled $\alpha_1$) the relaxation dynamics are independent of the excitation level $\alpha$.  At the highest levels of excitation (above the long-dashed line labeled $\alpha_2$) the 2T model of relaxation is appropriate.  In the intermediate transition region the dynamics depend upon the extent of laser excitation.  (See the text for precise definitions of the $\alpha_1$ and $\alpha_2$ boundaries.)  The values of $\gamma$ indicated on the top axis correspond to RT (which applies to all of the experimental data).  The experimental data are the same as those shown in Fig.~\ref{Fig1} for Ag and Au (modulo the Au data  near $\bdivd = 5 \times 10^{-2}$). The theoretical data are from Refs.~\cite{pietanza2007} (Ag) and \cite{Mueller2013,Mueller2014} (Au).  The two dotted lines are approximate $\alpha_1$ and $\alpha_2$ boundaries defined by Eqs.~(\ref{48}) and (\ref{52}), respectively.}
\label{Fig12}
\end{figure*}

We now investigate carrier relaxation as a function of excitation level $\alpha$.  As we discuss in detail below, a transition region exists between the small $\alpha$ behavior discussed above and large $\alpha$ behavior (where the 2T model is a reasonable approximation).  As far as we have been able to ascertain, our analysis is the first to identify and characterize this transition region.  

For the following reasons, our $\alpha$ dependent calculations focus on Ag and Au.  First, of all the simple metals, Au has been most extensively studied in the ultrafast time domain (this fact is reflected in the data plotted in Fig.~\ref{Fig1}).  Second, the (unnormalized) ratio $K_{ep} / K_{ee}$ of $ep$ to $ee$ scattering strengths is nearly the same for the two metals:  $K_{ep} / K_{ee}$ = $3.17 \times 10^6$ meV$^3$ and  $3.46 \times 10^6$ meV$^3$ for Ag and Au, respectively.  We can thus simultaneously model both metals in reasonable fashion.  To do this we employ $K_{ep} / K_{ee}$ = $3.38 \times 10^6$ meV$^3$ (chosen so that $h \nu = 1.5$ eV conveniently corresponds to $\bdivd = 10^{-3}$).  To match the other relevant parameters associated with the Au and Ag data in Fig.~\ref{Fig1}, we carry out calculations for $T_p = 300$ K over a range of photon energies 0.9 eV $\lesssim h \nu \lesssim$ 5.0 eV.\footnote{These calculations exclude the one outlying experiment in Fig.~\ref{Fig1}, carried out at $h \nu$ = 0.42 eV.}  We note that because $T_p$ is fixed and $h \nu $ varies, the normalized temperature $\gamma = T_p / h \nu$ varies with the normalized ratio $\bdivd = K_{ep} / [K_{ee} (h \nu)^3]$.  As above, we discuss our results in terms of these normalized quantities.

In Fig.~\ref{Fig9} we display results for $\tau_E$, $\tau_E^{th}$, $\tau_{th}$ and $\tau_H$ (all normalized by $\beta$) as functions of $\alpha$ for three different values of $\bdivd$.\footnote{Analogous to the other relaxation times, the thermalized distribution energy relaxation time $\tau_E^{th}$ is defined as the $1/e$ decay time for energy relaxation by a FD distribution.}  As the figure shows, the variation of each individual relaxation time with $\alpha$ is similar at all three values of $\bdivd$.  The following discussion shows  that these variations are all directly related to the temperature $T_{e}$ of a thermalized distribution with energy close to that of the nascent distribution.

The decrease in $\tau_{th}$ vs $\alpha$ is due to the increase in $T_e$ of the target FD distribution.  Associated with larger $T_e$ values is an increase in the FD distribution $f_{\scriptscriptstyle \! F\!D}(\delta \ep, T_e)$ across the initial region of laser excitation.  Therefore, fewer scattering events are required before the relaxing distribution mimics a thermal one.  Additionally, these thermalizing scattering events occur (on average) more rapidly, owing to the electron-electron scattering rate $\tau_{ee}$ being proportional to $(\delta \ep)^{-2}$ [see Eq.~(\ref{20})].  The following discussion shows this decrease of $\tau_{th}$ vs $\alpha$ is consequential to the behavior of both $\tau_E$ and $\tau_H$ vs $\alpha$.

With regard to energy decay, recall the discussion of Fig.~\ref{Fig6} in Sec.~\ref{SubSecVC2}.  There we observe if the ratio of $\tau_{th}$ to $\tau_E$ is small enough, then intracarrier thermalization effectively precedes energy decay and the 2T model is approximately valid.  At the largest values of $\alpha$ in Fig.~\ref{Fig9} this ratio indeed becomes small enough, as reflected in the fact that at the largest values of $\alpha$ the times constants $\tau_E$ and $\tau_E^{th}$ coalesce. 

To gain insight into the dramatic upswing in $\tau_H$ at higher values of $\alpha$, we plot in Fig.~\ref{Fig9B} sequences of the distribution function $f(\delta\epsilon,t)$ at both low [panel (a)] and high [panel (b)] levels of excitation.  As discussed in Sec.~\ref{SecVB1}, the main consequence of $ee$ scattering is to scatter the highest energy electrons to lower energy states.  However, both panels (a) and (b) also show the $ee$ interaction scatters some carriers to states with energy $\delep / h \nu \! > \! 1$.  At low excitation the number of these higher-energy carries is negligible, as reflected in the positive change in $f(\delep,t)$ in this region being $\lesssim 10^{-8}$.  Conversely, in the high-excitation case the number of higher energy carriers is non-negligible.  Furthermore, because the distribution thermalizes very quickly (notice by the time $t \beta = 0.02$ the distribution is nearly equal to a FD distribution with $T_e = 3830$ K, the maximum electron temperature in the 2T-model limit), the subsequent decay of $N_{\scriptscriptstyle >h\nu/2}$ (the measure for $\tau_H$) becomes governed by energy transfer to the phonons, which is characterized by the much longer time $\tau_E$.  This behavior is readily observed in panel (c) of Fig.~(\ref{Fig9B}), where we plot $N_{\scriptscriptstyle >h\nu/2}$ for three (one low and two high) levels of excitation. The initial ($t \beta \lesssim 0.005$) decay of $N_{\scriptscriptstyle >h\nu/2}$ is essentially identical in all three cases (see the inset in the figure), but at the two higher levels of excitation the rapid intracarrier thermalization at large $T_e$ results in dramatically slower decay of $N_{\scriptscriptstyle >h\nu/2}$ at longer times.

The interrelated behaviors of $\tau_{th}$ and $\tau_E$ in Fig.~\ref{Fig9} suggest three distinct regions of carrier dynamics: (i) a low excitation region independent of the level of excitation, (ii) a transition region, and (iii) a high excitation region where the 2T model is approximately valid.  To quantitatively delineate these three regions we define two boundaries.  The boundary between the low excitation and transition regions we designate $\alpha_1$, which we define via 
\be{\eq}
\label{37b}
\tau_{th}(\alpha_1) = 0.95 \, \tau_{th}(\alpha \rightarrow 0).
\en{\eq} 
That is, this boundary is defined as the value of $\alpha$ where the thermalization time $\tau_{th}$ has decreases to 95\% of its value at vanishingly small $\alpha$.  The second boundary is designated by $\alpha_2$, which we define by the condition
\be{\eq}
\label{38b}
\tau_{th}(\alpha_2) = 0.02 \, \tau_{E}(\alpha_2).
\en{\eq} 
That is, $\alpha_2$ is the value of $\alpha$ where $\tau_{th}$ equals 2\% of $\tau_E$.  This choice of 2\% is guided by our prior analysis of Fig.~\ref{Fig6}, where we observed that a ratio of 1.7\% is sufficiently small to ensure that thermalization occurs before significant energy decay (insofar that only $\sim$10\% of the excess energy in the carriers is lost to the phonons by the time the energy decay matches that of a thermal distribution).  In each panel of Fig.~\ref{Fig9} both $\alpha_1$ and $\alpha_2$ are indicated.

The identification of these three regions is significant with regard to ultrafast experimental studies of simple metals.  In Fig.~\ref{Fig12} we reproduce the Ag and Au data from Fig.~\ref{Fig1}.  Also indicated are the two boundaries $\alpha_1$ and $\alpha_2$ (and thence the three regions of dynamical behavior).  As is evident, these regions have significant dependence on the normalized ratio of $ep$ to $ee$ scattering $\bdivd$.  Clearly, a majority of the Ag and Au studies take place in the transition region between low and high excitation.

In passing, we note our numerical identification of an $\alpha$ independent region at low excitation levels justifies the previous BTE scattering-integral linearizations of Gusev and Wright \cite{Gusev1998}, Kabanov \textit{et al.} \cite{Kabanov2008,Baranov2014}, and Wilson and Coh \cite{Wilson2020}.

There have been several other BTE based investigations of carrier relaxation as a function of excitation level.  Tas and Maris \cite{Tas1994} consider $ee$ scattering only and thence propose a simple relation for $\tau_{th}$.  Expressed in our notation, their relation is 
\be{\eq}
\label{45}
\tau_{th}^{\scriptscriptstyle T\! M} \delta = \frac{4}{3} \frac{1}{\alpha}.
\en{\eq}
While this relation correctly predicts that $\tau_{th}$ decreases vs $\alpha$, it does not capture the transitional behavior of $\tau_{th}$ displayed by the curves in Fig.~\ref{Fig9}, which consists of a near independence from $\alpha$ at low excitation changing over to a strong decrease with $\alpha$ at high excitation.   At higher values of excitation, $\tau_{th}$ does decrease with behavior that is not too different from $1/ \alpha$.  However, in this region of excitation the relaxation time predicted by Eq.~(\ref{45}) is approximately an order of magnitude too large.  This can be seen in Fig.~\ref{Fig9} by comparing our calculated values of $\tau_{th}$ with the dotted curves, which are equal to  one tenth of $\tau_{th}^{\scriptscriptstyle T \! M}$.

Other BTE based studies of excitation dependent relaxation focus on numerical results.  Rethfeld \textit{et al}.~study Al \cite{Rethfeld2002}, and Mueller and Rethfeld investigate Al and Au \cite{Mueller2013}.  Both studies show the rate of thermalization has a strong dependence upon the absorbed energy density $u_a$.  However, their calculated rates for Au are not directly comparable to our calculations in Fig.~\ref{Fig9}, owing to (i) entropy being the basis for their thermalization measure and (ii) the neglect of $ep$ scattering in those calculations \cite{Mueller2013}.  

Furthermore, the parameter space investigated by these two studies is rather limited.  In terms of our notation, the parameters of the Rethfeld \textit{et al}.~\cite{Rethfeld2002} study of Al encompass a single value of $\bdivd$ ($= 1.05 \times 10^{-3}$) and  an order-of-magnitude range of $\alpha$ ($4.5 \times 10^{-3}  \le  \alpha  \le  5.1 \times 10^{-2}$).  In the Mueller and Rethfeld study \cite{Mueller2013} these parameters are $\bdivd = 2.16 \times 10^{-3}$ and $2.4 \times 10^{-2} \le \alpha \le 0.98$ for Al and  $\bdivd = 9.3 \times 10^{-4}$ and $1.5 \times 10^{-2} \le \alpha \le 0.73$ for Au. In a slightly later paper Mueller and Rethfeld  expand their investigation of Au to two other values of $\bdivd$ ($2.3 \times 10^{-4}$ and $3.5 \times 10^{-2}$) \cite{Mueller2014}.  That study, however, does not report carrier relaxation times.  The locations in $\bdivd$ -- $\alpha$ parameter space of the Mueller and Rethfeld calculations for Au \cite{Mueller2013,Mueller2014} are shown in Fig.~\ref{Fig12} (as the open circles).\footnote{Although Mueller and Rethfeld ignore $ep$ scattering in their thermalization calculations (which is equivalent to $\bdivd = 0$) \cite{Mueller2013}, they do include it in calculations of excitation-dependent $ep $ coupling \cite{Mueller2013,Mueller2014}.}  As those data show, most of their analysis resides in the high excitation region where the 2T model is applicable.  

A study of Ag by Pietanza \textit{et al}.~\cite{pietanza2007} investigates energy decay, but also at limited parameter values:  $\bdivd = 1.03 \times 10^{-3}$ and $\alpha = 4.8 \times 10^{-3}$, $3.46 \times 10^{-2}$, and $7.6 \times 10^{-2}$.  In Fig.~\ref{Fig12} we also indicate (via open diamonds) these values.  Their three sets of energy-decay curves (see Fig.~9 of \cite{pietanza2007}) align with our analysis: (i) at the lowest level of excitation (which is within the transition region of Fig.~\ref{Fig12}), energy decay of their laser-excited distribution is significantly slower than that of the corresponding FD distribution;  (ii) at the intermediate level of excitation (which nearly resides on the $\alpha_2$ border), energy decay of their laser excited distribution is only slightly slower than the corresponding FD distribution; and (iii) at the highest level of excitation (which is well within the 2T-model region), their two energy decay curves are indistinguishable.  Pietanza \textit{et al}., however,  do not report energy relaxation times associated with their calculations.

We finish our discussion by developing approximate expressions for the boundaries $\alpha_1$ and $\alpha_2$.  For the lower boundary $\alpha_1$ we begin by considering the 2T-model expression that relates the absorbed energy density $u_a$ to the peak electron temperature $T_e^{\scriptscriptstyle(p)}$ and phonon temperature $T_p$ \cite{Ashcroft1976},
\be{\eq}
\label{46}
\frac{u_a}{g_v} = \frac{\pi^2}{6} \big[  (k_B T_e^{\scriptscriptstyle(p)})^2 - (k_B T_p)^2 \big].
\en{\eq}
Using Eqs.~(\ref{10}) and (\ref{13}) this equation is readily rewritten in terms of normalized variables as
\be{\eq}
\label{47}
\frac{\alpha}{\gamma^2} = \frac{\pi^2}{6} \big[  (\gamma_e^{\scriptscriptstyle(p)} / \gamma)^2  -1 \big].
\en{\eq}
Even though the 2T model is not generally valid at small excitation levels, this equation suggests the low excitation region can be roughly defined by $\alpha_1 / \gamma^2 = c_1 \pi^2 / 6$, where $c_1$ is some small factor.  By comparing this equation with our numerical results we find
\be{\eq}
\label{48}
\alpha_1 \approx \frac{1}{3} \frac{\pi^2}{6} \gamma^2
\en{\eq}
gives a decent description of our numerical boundary. This approximate boundary is illustrated in Fig.~\ref{Fig12} as the dotted violet line.  We note $c_1 = 1/3$ is equivalent to $\gamma_e^{\scriptscriptstyle (p)} / \gamma = 1.15$.  Therefore, the low-excitation region roughly corresponds to the excited carriers having an effective temperature no greater than 15\% of their pre-excitation temperature. 

The upper boundary $\alpha_2$ is defined Eq.~(\ref{38b}).  To approximate this relation we derive simple expressions for $\tau_{th}$ and $\tau_E$ at high levels of excitation.  For $\tau_{th}$ we simply employ $\tau_{th}^{\scriptscriptstyle T\!M} / 10$, which yields values not too far from our numerical results (see Fig.~\ref{Fig9}).  We thus have [see Eq.~(\ref{45})]
\be{\eq}
\label{49}
\tau_{th} \approx \frac{2}{15} \frac{1}{\alpha \, \delta}.
\en{\eq}
For $\tau_E$ we first note for excitations near the $\alpha_2$ boundary that $\tau_E \approx \tau_E^{th}$.  Hence, Eq.~(\ref{21}) is reasonably accurate in this excitation regime. Furthermore, because $T_e \gg T_p$ for these excitation levels we can write
\be{\eq}
\label{50}
\tau_E \approx \frac{\pi^2}{6} \frac{k_B T_e^{\scriptscriptstyle(p)}}{K_{ep}} = \frac{\pi^2}{6} \frac{\gamma_e^{\scriptscriptstyle(p)}}{\beta}.
\en{\eq}
From Eq.~(\ref{47}) we have $\alpha = (\pi^2 / 6) (\gamma_e^{\scriptscriptstyle(p)})^2$ at high excitation levels.  We can thus express $\tau_E$ in terms of $\alpha$ and $\beta$ as
\be{\eq}
\label{51}
\tau_E \approx \frac{\pi}{\beta} \sqrt{\frac{\alpha}{6}}.
\en{\eq}
Our approximate condition for $\alpha_2$ is obtained by setting the ratio between Eqs.~(\ref{49}) and (\ref{51}) to 0.02 [see Eq.~(\ref{38b})], which yields
\be{\eq}
\label{52}
\alpha_2 \approx \bigg(  \frac{20 \sqrt{6}}{3 \pi} \frac{\beta}{\delta}  \bigg)^{\! \! 2/3} \approx 3 \bigg( \frac{\beta}{\delta} \bigg)^{\! \! 2/3}
\en{\eq}
This approximate boundary is illustrated in Fig.~\ref{Fig12} as the green dotted line, which shows that for most relevant values of $\bdivd$ this simple expression is quite accurate.

\section{Summary}

In this paper we accomplish a number of key objectives for understanding ultrafast carrier dynamics in laser excited metals. Here we review these accomplishments, which systematically extend previous general considerations of these dynamics using the Boltzmann transport equation \cite{Groeneveld1992,Groeneveld1995,Gusev1998,Kabanov2008,Baranov2014,Tas1994,Wilson2020,Rethfeld2002,pietanza2007,Mueller2013}.  

(i)  We employ an energy normalization of the BTE that streamlines investigation of relaxation times associated with laser-pulse-excited electron distributions.  Arising from this normalization are three unitless parameters:  the $ep$ to $ee$ scattering-strength ratio $\bdivd$ [see Eqs.~(\ref{14})], excitation level $\alpha$ [see Eq.~(\ref{10})], and phonon temperature $\gamma$ [see Eq.~(\ref{13})].  We assume laser excitation is instantaneous, after which the nascent distribution relaxes to an equilibrium FD distribution.  From this approximation of decoupled excitation and relaxation it follows that any timescale multiplied by either $\beta$ or $\gamma$ is necessarily a function of only $\bdivd$, $\alpha$, and $\gamma$.  This normalization allows our results to be readily applied to any simple metal.

(ii)  It is important to distinguish single-carrier scattering times from relaxation times associated with an excited carrier distribution.  To this end we discuss expressions for the single-carrier spontaneous phonon-emission time $\tau^0_{ep}$ [see Eqs.~(\ref{15}) and (\ref{16})], total $ep$ scattering time $\tau_{ep}$ [Eqs.~(\ref{17}) and (\ref{18})], and $ee$ scattering time $\tau_{ee}(\delta \ep)$ [Eq.~(\ref{7})].  The energy relaxation time $\tau_{E}^{th}$ [Eqs.~(\ref{18}) and (\ref{19})] for a hot thermalized distribution is also discussed.  Values for all four of these times are presented in Table \ref{table1} and Fig.~\ref{Fig1A} for a representative selection of simple metals.  

(iii)  Utilizing our BTE scattering integrals [see Eqs.~(\ref{2}), (\ref{3}), and (\ref{4})] we thoroughly investigate the behavior of three essential distribution relaxation times---hot carrier ($\tau_H$), energy ($\tau_E$), and thermalization ($\tau_{th}$)---as functions of $\beta / \delta$ at low excitation level [$\alpha = 10^{-5}$] and three different normalized phonon temperatures $\gamma$.  In Fig.~\ref{Fig4} we report six sets of time dependent distribution functions.  From these (and other) distribution functions we extract values for these three relaxation times.  We are able to describe all three relaxation times using heuristic functions of $\beta/\delta$ [see Eqs. (\ref{29b}), (\ref{32C}), and (\ref{88})/(\ref{88B})].  Because $\tau_H$ is nearly described by Matthiessen's rule [see Eq.~(\ref{29b})], these two interactions act nearly  independently as far as hot-carrier relaxation is concerned.  A nontrivial intertwining of the two scattering mechanisms is reflected in the more complicated expressions for $\tau_E$ and $\tau_{th}$ [Eqs. (\ref{29b}) and (\ref{88})/(\ref{88B})].  With regard to the phonon temperature, we find (i) $\tau_H$ is nearly insensitive to $\gamma$, (ii) changes in $\tau_E$ are loosely allied with $\tau_E^{th} \propto \gamma$ [see Eq.~(\ref{31})], and (iii) $\tau_{th} \propto 1/\gamma^{1.8}$ when $ee$ scattering is dominant.

(iv)  These general low-excitation results are relevant to the metals listed in Table \ref{table1}.  The two heaviest alkali metals (Rb and Cs), which are characterized by the smallest values of $\beta / \delta$ (typically $10^{-6} \lesssim \beta / \delta \lesssim 10^{-5}$), exhibit the simplest dynamical behavior.  For these metals $ee$ scattering is so strong (relative to $ep$ scattering) that it controls both $\tau_H$ and $\tau_{th}$ [see Figs.~\ref{Fig5} and \ref{Fig7}].  Because $\tau_{th} \lesssim 0.02 \,\, \tau_{E}$ for these metals, thermalization occurs before significant energy transfer to the phonons.  Consequently, the two-temperature (2T) model of $ep$ dynamics  \cite{anisimov1974}  is a reasonable approximation.  The other metals listed in Table \ref{table1} (excluding K and Na) are characterized by significantly larger values of $\beta / \delta$ (typically $10^{-4} \lesssim \beta / \delta \lesssim 5 \times 10^{-3}$).  In this regime $\tau_H$ is still dominated by $ee$ scattering, almost to the same extent as at smaller values of $\beta / \delta$ [see Fig.~\ref{Fig5}].  By contrast, $\tau_{th}$ has significant contribution from $ep$ scattering.  In addition, because $ee$ scattering is relative weaker, secondary-electron generation is relatively slower; energy relaxation is consequently slower [see Fig.~\ref{Fig7}].  Lastly, in this $\beta / \delta$ regime $\tau_{th}$ and $\tau_E$ are similar enough that the processes of thermalization and energy relaxation happen on comparable timescales, with the end result that by the time the distribution has thermalized most of the excitation energy has transferred to the phonons (see Fig.~\ref{Fig6}).  Therefore, for these metals the 2T model is a very poor approximation at low levels of excitation.

(v) We investigate the three distribution relaxation times vs excitation level $\alpha$ in the range $3 \times 10^{-6} \le \alpha \le 2 \times 10^{-1}$ (see Figs.~\ref{Fig9}, \ref{Fig9B}, and \ref{Fig12}). The calculations are appropriate for Ag and Au at RT ($T_p = 300$ K) and photon energies $h\nu$ between 0.9 and 5 eV. The hot-carrier relaxation time $\tau_H$ is nearly invariant up to $\alpha \approx 10^{-2}$.  However, in the region $10^{-2} < \alpha < 10^{-1}$ this relaxation times dramatically increases, owing to it being increasingly governed by energy relaxation, which is characterized by $\tau_E$.  Both $\tau_E$ and $\tau_{th}$ vary significantly---although more gradually than $\tau_H$---with increasing $\alpha$.  The dependencies of these two relaxation times on $\alpha$ allow us to identify three characteristic regions of carrier relaxation: (i) a low excitation region with very little dependence on $\alpha$, (ii) an intermediate region in which $\tau_{th}$ dramatically decreases and $\tau_E$ increases, and (iii) a high-excitation region where the 2T model is approximately valid.  The two boundaries---designated $\alpha_1$ and $\alpha_2$---that separate these three regions are given to reasonable accuracy in terms of the normalized parameters of our model by Eqs.~(\ref{48}) and (\ref{52}). 


\appendix

\section{Approximate dynamics for $ep$ scattering only}
\label{Appendix D}

\begin{figure}[b]
\centerline{\includegraphics[scale=0.54]{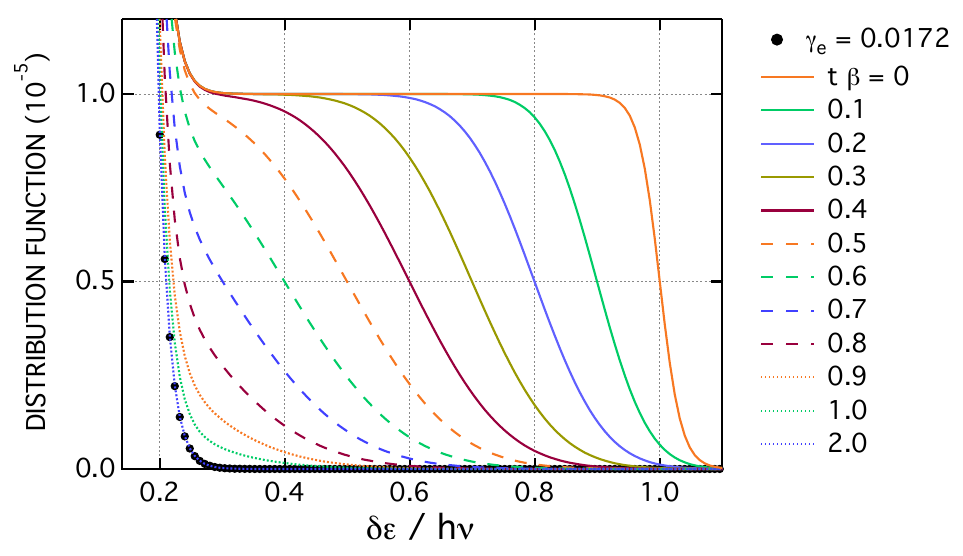}}
\caption{Time series of distribution function $f(\delta \ep,t \beta)$ during relaxation for $ep$ scattering only.  Here the normalized excitation level and phonon temperature are $\alpha = 1 \times 10^{-5}$ and $\gamma = 0.0172$, respectively. }
\label{FigD1}
\end{figure}

Here we compute analytic approximations to $\tau_H^{ep}$, $\tau_E^{ep}$, and $\tau_{th}^{ep}$, respectively the hot-electron, energy, and thermalization relaxation times, in the limit of purely $ep$ scattering ($\delta = 0$), respectively.  Along the way we gain insight into the evolution of $f(\ep,t)$ under this condition.

We begin by deriving an approximate equation that governs an excited distribution in the two regions of energy (sufficiently far from $\ep_F$) where the equilibrium distribution $f_{\scriptscriptstyle \! F\!D}(\ep,T_p)$ can be treated as constant (either 1 or 0).  As we see below, for our standard values of $\gamma = 0.0172$ and $\alpha = 10^{-5}$, these regions are defined by $|\delta \ep / h \nu| \gtrsim 0.3$.  We first write the distribution function as the sum
\begin{equation}
\label{D1}
f(\ep,t) = f_{\scriptscriptstyle \! F\!D}(\ep,T_p) + \phi(\ep,t),
\end{equation}
where $\phi(\ep,t)$ is the deviation of $f(\ep,t)$ from the equilibrium distribution.  If we (i) substitute this form of $f(\ep,t)$ into the $ep$ collision term as given by Eq.~(\ref{4}), (ii) recall $f_{\scriptscriptstyle \! F\!D} (1- f_{\scriptscriptstyle \! F\!D}) = - f'_{\scriptscriptstyle \! F\!D} \, k_B T_p$, and (iii) assume this is the only collision term in the BTE, then we straightforwardly obtain for $\phi(\ep,t)$ the partial differential equation
\begin{align}
\label{D2}
\frac{\partial \phi}{\partial t} =& \, K_{ep} \bigg[  - 2 f'_{\scriptscriptstyle \! F\!D} \, \phi  + (1 - 2 f_{\scriptscriptstyle \! F\!D}) \frac{\partial \phi}{\partial \ep}  \nonumber \\
					 &   + k_B T_p \, \frac{\partial^2 \phi}{\partial \ep^2} +  \phi \, \frac{\partial \phi}{\partial \ep} \bigg].
\end{align}
For energies where $f_{\scriptscriptstyle \! F\!D}$ can be taken as constant this equation readily simplifies to
\begin{equation}
\label{D3}
\frac{\partial \phi}{\partial t} = \mp \, K_{ep} \, \frac{\partial \phi}{\partial \ep} + K_{ep} k_B T_p \, \frac{\partial^2 \phi}{\partial \ep^2}.
\end{equation}
Here we have also ignored the nonlinear term in Eq.~(\ref{D2}), owing to it being quadratic in the (small) deviation $\phi$.  The minus and plus signs on the right side of this equation correspond to the regions where $\delta \ep$ is negative and positive, respectively.  

The interpretation of Eq.~(\ref{D3}) is straightforward.  We first note the right side comprises two terms, one convective ($\propto \partial \phi / \partial \ep$) and one diffusive ($\propto \partial^2 \phi / \partial \ep^2$).  If the diffusive term were absent, then the convective term would simply drive the distribution to smaller values of $|\delta \ep|$ at the rate $K_{ep}$.  Physically, this convective flow arises from spontaneous phonon emission, which causes excited electrons and holes to each move towards the Fermi energy (whereupon they recombine).  Conversely, if the convective term were absent, then the diffusive term would serve to increase the width of the distribution.  Physically, this energy spreading arises from the sum of phonon absorption and stimulated emission.

The combination of convection and diffusion is evident in the time series of distribution functions shown in Fig.~\ref{FigD1}.  We note this sequence of distributions functions are from the same set as those shown in part (b) of Fig.~\ref{Fig3}.   Interestingly, pure convection is evident at the point in the distribution where $f = \alpha /2$, which is initially located at $\delta \ep / h \nu = 1$.  Because $f = \alpha /2$ is an inflection point and $\phi(\ep)$ is antisymmetric about this point, the diffusive term has no effect upon this point of the distribution.  Thus, the point $f = \alpha /2$ simply moves towards lower energies at the normalized rate $\beta = K_{ep} / h \nu$.  As the graph shows, this behavior occurs until $t \beta \approx 0.7$, after which the effect of $f_{\scriptscriptstyle \! F\!D}$ not being constant comes into play.  The consequence of diffusion is also evident as the time-dependent increase in the width of the distribution's high-energy edge. 

We note Gusev and Wright also consider the evolution of the distribution when $ep$ scattering is solely at play \cite{Gusev1998}.  Having derived an equation similar to Eq.~(\ref{D2}) [but for the whole distribution function $f(\ep,t)$], they state the diffusive term is only important for energies such that $\delta \ep \sim k_B T_p$.  They then proceed to work with the equivalent of Eq.~(\ref{D3}), but with only the convective term.  Our results displayed in Fig.~\ref{FigD1} show such an approximation has its limitations.  However, neglect of the diffusive term affords quite simple calculations of $\tau_H^{ep}$, $\tau_E^{ep}$, and $\tau_{th}^{ep}$, as we now show.
  
To estimate these relaxation times we begin by finding an approximate solution for $\phi(\ep,t)$.  First, we note the solution to Eq.~(\ref{D3}) under neglect of the diffusive term is simply
\begin{equation}
\label{D4}
\phi(\delta \ep,0) =  \left\{ \begin{array}{rl}
\phi(\delta \ep - K_{ep} \, t, 0) & \, \delta \ep < 0  \\
\phi(\delta \ep + K_{ep} \, t, 0) & \, \delta \ep > 0 
\end{array} \right. ,
\end{equation}
Second, because $k_B T_p \ll h \nu$, the low-energy and high-energy cutoffs of $\phi(\ep,t)$ are rapid on the scale of $h \nu$. We thus approximate the initial distribution function as 
\begin{equation}
\label{D5}
\phi(\delta\ep,0) = \alpha \left\{ \begin{array}{rl}
-\Theta(h \nu + \delta \ep) & \, \delta \ep < 0  \\
\Theta(h \nu - \delta \ep) & \, \delta \ep > 0 
\end{array} \right. ,
\end{equation}
where $\Theta$ is the Heaviside function.  With this initial condition we immediately have
\begin{equation}
\label{D6}
\phi(\delta\ep,t) = \alpha \left\{ \begin{array}{rl}
-\Theta(h \nu - K_{ep} \, t + \delta \ep) & \, \delta \ep < 0  \\
\Theta(h \nu - K_{ep} \, t - \delta \ep) & \, \delta \ep > 0 
\end{array} \right. .
\end{equation}
As mentioned above, this analytic approximation for $\phi(\delep,t)$ is valid whenever $f_{\scriptscriptstyle \! F\!D}(\ep,T_p)$ can be treated as 0 ($\delep > 0$) or 1 ($\delep < 0$).  However, in our discussion below we shall -- whenever convenient -- assume Eq.~(\ref{D6}) is valid for all values of $\delep$.  This assumption is strictly only valid if $T_p$ for the underlying FD distribution is zero.  However, it leads to reasonably accurate results as long as $\gamma = k_B T_p / h \nu \ll 1$.

As defined in Sec.~\ref{SubSecVC2}, the hot-carrier relaxation time is the time it takes the number of excited electrons with energy above $\delta \ep / h \nu = 0.5$ to decay to $1/e$ of its initial value.  With the approximations at hand this electron number can be written as
\begin{align}
\label{D7}
N_{\scriptscriptstyle >h \nu \!/ 2}(t) &= \alpha g_0 \int_{h \nu / 2}^{h \nu - K_{ep} t}  d(\delta \ep) \nonumber \\
							   &= \alpha g_0 \, (h \nu / 2 - K_{ep} \, t).
\end{align} 
From this equation it follows the hot-electron relaxation time is given by 
\begin{equation}
\label{D8}
\tau_H^{ep}  \beta = \frac{1}{2} \, (1 - 1/e) = 0.3161.
\end{equation}

As also defined in Sec.~\ref{SubSecVC2}, the energy relaxation time is the time it takes the energy in the excited distribution to fall to $1/e$ of its initial value.  Given Eq.~(\ref{D6}), the excess energy in the excited carrier distribution is given by   
\begin{align}
\label{D9}
\langle \ep \rangle_{\scriptscriptstyle \! L}(t) &= 2 \alpha g_0 \int_0^{h \nu - K_{ep} t}  d(\delta \ep) \, \delta \ep \nonumber \\
							   &= \alpha g_0 \, (h \nu - K_{ep} \, t)^2.
\end{align} 
This expression decreases to $1/e$ of its initial value at the time $\tau_E^{ep}$ given by
\begin{equation}
\label{D10}
\tau_E^{ep}  \beta = 1- 1/ \sqrt{e} = 0.3935.
\end{equation}

In Sec.~\ref{SubSecVC2} we see that these analytic estimations for $\tau_H^{ep}$ and $\tau_E^{ep}$ are perhaps more accurate than one might have anticipated.  This high degree of accuracy can be understood by considering the set of distribution functions displayed in Fig.~\ref{FigD1}.  The simplicity of the results resides in the fact that over the energy range of consequence ($\delta \ep / h \nu \ge 0.3$) the effect of diffusion is antisymmetric about the purely convective point $f = \alpha / 2$.  As far as $\tau_H^{ep} \beta \approx 0.32$ is concerned, we see up that to $t \beta = 0.3$ the lower diffusive tail extends only slightly below $\delta \ep / h \nu = 0.5$.  Consequently, diffusion does not dramatically affect the number of hot carriers for $t < \tau_H$.  Concerning $\tau_E^{ep} \beta \approx 0.39$, we see that up to $t \beta = 0.4$ the effects of diffusion only extend down to $\delta \ep / h \nu \approx 0.3$, while below this energy the distribution has yet to evolve.  Therefore, the diffusive term also has very little effect upon the energy remaining in the carriers for $t < \tau_E$.

We now analytically approximate the thermalization time $\tau_{th}^{ep}$.  As also defined in Sec.~\ref{SubSecVC2}, $\tau_{th}$ is the time it take the difference in $d\langle \ep \rangle / dt$ between the laser-excited distribution and an energy-equivalent FD distribution to decrease to $1/e$ of its initial value.  For the laser-excited distribution we calculate $d\langle \ep \rangle / dt$ starting with $\langle \ep \rangle_{\scriptscriptstyle \! L}$ given by Eq.~(\ref{D9}), which yields
\begin{equation}
\label{D11}
\frac{d \langle\ep \rangle_{\scriptscriptstyle \! L}}{dt} = - 2 K_{ep} \alpha g_0 \, (h \nu - K_{ep} t).
\end{equation}
The equivalent expression for a FD distribution is given by Eq.~(\ref{4c}).  Because we want to compare laser-excited and FD distributions with identical energy, we now express $T_e$ in terms of $\langle \ep \rangle_{\scriptscriptstyle \! L}$.  Equating this energy with that of a FD distribution \cite{Ashcroft1976}
\be{\eq}
\label{D12A}
\langle \ep \rangle_{\scriptscriptstyle \! F\!D} = \frac{\pi^2}{6} g_0 \, k_B^2 (T_e^2 - T_p^2). 
\en{\eq}
we find
\begin{equation}
\label{E13}
k_B^2 T_e^2 = \frac{6}{\pi^2} \frac{\langle \ep \rangle_{\scriptscriptstyle \! L}}{g_0} + k_B^2T_p^2, 
\end{equation}
which allows us to rewrite Eq.~(\ref{4c}) as
\begin{equation}
\label{D14}
\frac{d \langle\ep \rangle_{\scriptscriptstyle \! F\!D}}{dt} = - K_{ep} g_0 \Bigg[  \sqrt{\frac{6}{\pi^2} \frac{\langle \ep \rangle_{\scriptscriptstyle \! L}}{g_0} + k_B^2T_p^2} - k_B T_p  \Bigg].
\end{equation}
Our present interest is the limit of low excitation, in which case Eq.~(\ref{D14}) simplifies to
\begin{align}
\label{D15}
\frac{d \langle\ep \rangle_{\scriptscriptstyle \! F\!D}}{dt} &= - \frac{3}{\pi^2} K_{ep} \frac{\langle \ep \rangle_{\scriptscriptstyle \! L}}{k_B T_p}  \nonumber \\
		&= - \frac{3}{\pi^2} K_{ep} \alpha g_0 \, \frac{(h \nu - K_{ep} t)^2}{k_B T_p}.
\end{align}
Taking the difference between $d\langle \ep \rangle_{\scriptscriptstyle \! F\!D} / dt$ and $d \langle \ep \rangle_{\scriptscriptstyle \! L} / dt$ as given by Eqs.~(\ref{D15}) and (\ref{D11}) we find $\tau_{th}^{ep}$ is determined by the quadratic equation
\begin{equation}
\label{D16}
\frac{3}{\pi^2} \frac{1}{\gamma} (1 -  \tau_{th}^{ep}\beta)^2 - 2 (1 - \tau_{th}^{ep}\beta) = \frac{1}{e} \bigg( \frac{3}{\pi^2} \frac{1}{\gamma} - 2 \bigg).
\end{equation}
We note for our conditions of interest the term quadratic in ($1- \tau_{th}^{ep} \beta)$ is significantly larger than the linear term.  Therefore, $\tau_{th}^{ep} \beta$ is close to $\tau_E^{ep} \beta$ [which is solely controlled by this same quadratic term, see Eq.~(\ref{D9})].  Indeed, for $\gamma =$ 0.0086, 0.0172, and 0.0344, the solutions to Eq.~(\ref{D16}) are $\tau_{th}^{ep} \beta =$ 0.3819, 0.3694, and 0.3415, respectively, which are not far from the analytic result 0.3935 for $\tau_E^{ep} \beta$ [see Eq.~(\ref{D10})].  These analytic values for $\tau_{th}^{ep} \beta$ are also reasonably close to the corresponding numerically calculated values 0.3944, 0.3848, and 0.3766 [see Table \ref{table2a}].

\vspace{-0.5cm}

\bibliography{AlkaliMetals}

\begin{thebibliography}{111}
\bibitem [{Sup()}]{SuppMat}%
  \BibitemOpen
  \href@noop {} {}See {S}upplemental {M}aterial at [URL will be inserted by
  publisher] for details concerning (i) the experimental data presented in
  Fig.~1 and 11, (ii) least-squares analysis of our relaxation-time data, and
  (iii) the initial time dependence of $f(\epsilon,t)$\BibitemShut {NoStop}%
\end{thebibliography}

\end{document}


\title{Supplemental Material for ``Ultrafast relaxation dynamics of excited carriers in metals:  Simplifying the intertwined dependencies upon scattering strengths, phonon temperature, photon energy, and excitation level''}

\author{D. M. Riffe}

\affiliation{Physics Department, Utah State University, Logan, UT 84322, USA}

\author{Richard B. Wilson}

\affiliation{Mechanical Engineering Department and Materials Science and Engineering Department, University of California, Riverside, CA 92521, USA}


\maketitle

\def\theequation{S\arabic{equation}}
\renewcommand\thesection{S\Roman{section}}

\section{$\bdivd$ and $\alpha$ Data in Figs.~1 and 11}

In Figs.~1 and 11 of the paper we plot concurrent values of the normalized interaction strength ratio
\be{\eq}
\label{1}
\bdivd = \frac{K_{ep}}{K_{ee} (h \nu)^3}
\en{\eq}
and normalized excitation level
\be{\eq}
\label{2}
\alpha = \frac{u_d}{(h \nu)^2 g_v}
\en{\eq}
relevant to a wide variety experiments on Cu, Ag, Au, and Al.  In each experiment the sample---a thin film or bulk sample at room temperature---is excited by sub-ps laser pulses.  In most experiments several values of the incident fluence $F_{i}$ are utilized, although some experiments are carried out at a single fluence.  We now describe how we infer the values of $\bdivd$ and $\alpha$ relevant to each experiment.  

Values of $\bdivd$ are straightforwardly obtained, as all experiments report the photon energy $h \nu$ utilized.  With this parameter and the $K_{ep}$ and $K_{ee}$ values reported in Table I, $\bdivd$ is obtained from Eq.~(\ref{1}).

More effort is required to deduce values of $\alpha$.  As Eq.~(\ref{2}) indicates, we also need the electronic density of states $g_v$ (at the Fermi level) and the energy density $u_d$ absorbed by the sample when excited by a single laser pulse.  For $g_v$ we use values calculated by Papaconstantopolous \cite{papaconstantopoulos1986}.  The absorbed energy density is related to $F_i$ (which is usually reported) via
\be{\eq}
\label{3}
u_d = \frac{F_i A}{d},
\en{\eq}
where $A$ is the absorptivity (fraction of incident energy absorbed) and $d$ is the depth over which the absorbed energy is initially deposited in the sample.  In most cases we calculate $A$ using thin-film optics relations for normal-incidence radiation \cite{Reitz1993}, although when required we utilize Fresnel relations for non-normal incidence \cite{Jackson1975}.  The length $d$ is possibly one of three quantities.  If the sample is thin enough, then carrier transport that occurs during the laser pulse will ensure that the initial energy is spread fairly uniformly from the front to the back of the sample.  In this case $d = d_s$, where $d_s$ is the sample thickness.  However, it may be that transport is not fast enough to enable such uniform heating.  For each experiment we thus calculate the 1D diffusion length  $d_{\scriptscriptstyle D} = d_m \sqrt{\Delta t \, (3 \tau_s)^{-1}}$.  Here $\Delta t$ is the laser-pulse duration, $\tau_s$ is the average  single-carrier scattering time, and $d_m = v_F \, \tau_s$ is the carrier mean free path. We take $\tau_s$ to be the scattering time (from $ep$ and $ee$ interactions) for an electron with energy $\delta \ep = h \nu / 2$, which is given by $\tau_s^{-1} = (\tau_{ep}(T_p))^{-1} + (\tau_{ee}(h \nu / 2))^{-1}$.  A third length scale is also important, the intensity skin depth $d_{\scriptscriptstyle I}$ of the incident radiation.  If the larger of $d_{\scriptscriptstyle D}$ and $d_{\scriptscriptstyle I}$ is smaller than $d_s$, then then $d$ is taken to be that larger value.  Otherwise, we assume $d = d_s$.  In more mathematical notation, $d$ is determined by
\be{\eq}
\label{4}
d = \min \! \big( d_s, \max(d_{\scriptscriptstyle I}, d_{\scriptscriptstyle D}) \big).
\en{\eq}

Strictly speaking, the thin-film optics relations \cite{} we use to calculate the absorptivity $A$ are only valid in the low-excitation (linear-response) limit.  At high levels of excitation stimulated emission should be included when determining the absorbed energy density.  We include in approximate fashion the effect of stimulated emission by using---instead of Eq.~(\ref{2})---the relation
\be{\eq}
\label{5}
\alpha = \frac{1- e^{-2 \frac{u_d}{(h \nu)^2 g_v}}}{2},
\en{\eq}
where $u_d$ is still calculated via the linear-response relation Eq.~(\ref{3}).  Equation~(\ref{5}) is derived under the assumption that excited electrons are not subsequently excited by another photon. Notice at low excitation $\big($small values of $u_d/ [(h \nu)^2 g_v]$$\big)$ Eq.~(\ref{5}) simplifies to Eq.~(\ref{2}), but at high excitation $\alpha$ is limited to values less than $1/2$, as expected.

In Tables \ref{table1} - \ref{table4} we list values of $\bdivd$ and the range of $\alpha$ values appropriate to the experiments on Cu \cite{Elsayed1987,Hohlfeld1995,Hohlfeld1996,Papadogiannis1997,Bartoli1997,Farztdinov1999,Bonn2000,Kruglyak2005,Ligges2009,Ligges2011,Shen2015,Nakamura2016,Mo2018,Obergfell2020,Suemoto2021}, Ag \cite{Hohlfeld1996,Kruglyak2005,Ligges2011,Suemoto2021,Riffe1993,Groeneveld1995,Fatti1998,Fatti2000,Owens2010,Er2013}, Au \cite{Groeneveld1995,Hohlfeld1996,Bonn2000,Fatti2000,Kruglyak2005,Ligges2009,Ligges2011,Nakamura2016,Suemoto2021,Schoenlein1987,Brorson1987,Elsayed-Ali1991,Fann1992,Juhasz1992,Juhasz1993,Girardeau-Montaut1993,Sun1993,Sun1994,wang1994,Hohlfeld1997,Maznev1997,Moore1999,Hibara1999,Hostetler1999,Smith2000,Smith2001,Guo2001,Ibrahim2004,Hopkins2007,Ma2010,Hopkins2011,Chen2011a,DellaValle2012,Kolomenskii2013,Hopkins2013,Guo2014,White2014,Chase2016,Sokolowski-Tinten2017,Heilpern2018,Yao2018,Sielcken2020,Lu2020,Tomko2021,Grauby2022}, and Al \cite{Hostetler1999,Riffe1993,Tas1994,Stevens2005,Park2005,Nie2006,Ma2013,Waldecker2016} represented in Figs.~1 and 11.

\renewcommand{\thetable}{S\arabic{table}}

\begin{table}[h]
\caption{Values of $\bdivd$ and range of $\alpha$ values (indicated by $\alpha_{\scriptscriptstyle \min}$ and $\alpha_{\scriptscriptstyle \max}$) in experimental investigations of room-temperature Cu excited by sub-ps laser pulses.  Experiments with only $\alpha_{\scriptscriptstyle \min}$ indicated were carried out at a single excitation level.  \label{table1}}
\vspace{0.2cm}
\begin{tabular}{l@{\hspace{0.5cm}}l@{\hspace{0.5cm}}l@{\hspace{0.5cm}}l@{\hspace{0.05cm}}r@{\hspace{0.5cm}}r@{\hspace{0.5cm}}r@{\hspace{0.5cm}}r@{\hspace{0.5cm}}r@{\hspace{0.5cm}}r}

\multicolumn{4}{c} {\textbf{Cu}}		 \\	

\hline
\hline  
    
Study 	&	$\bdivd$  	& $\alpha_{\scriptscriptstyle \min}$ 	&  $\alpha_{\scriptscriptstyle \max}$    \\

\hline

\rule{0pt}{2.5ex}Elsayed-Ali (1987) [\onlinecite{Elsayed1987}]	& $6.55 \times 10^{-4}$		& $4.75 \times 10^{-3}$		& $3.73 \times 10^{-2}$	\\

Hohlfeld (1995) [\onlinecite{Hohlfeld1995}]	& $6.71 \times 10^{-4}$		& $6.52 \times 10^{-2}$		& $1.22 \times 10^{-1}$	\\

Hohlfeld (1996) [\onlinecite{Hohlfeld1996}]	& $6.87 \times 10^{-4}$		& $1.10 \times 10^{-1}$		& $1.97 \times 10^{-1}$	\\

Papadogiannis (1997) [\onlinecite{Papadogiannis1997}]	& $4.19 \times 10^{-5}$		& $3.91 \times 10^{-4}$		& $5.85 \times 10^{-3}$	\\

Bartoli (1997) [\onlinecite{Bartoli1997}]			& $5.94 \times 10^{-4}$		& $1.75 \times 10^{-2}$		& $1.17 \times 10^{-1}$	\\

Farztidinov (1999) [\onlinecite{Farztdinov1999}]		& $3.13 \times 10^{-5}$		& $2.85 \times 10^{-3}$		& $4.11 \times 10^{-2}$	\\

									& $3.52 \times 10^{-4}$		& $6.28 \times 10^{-3}$		& $8.64 \times 10^{-2}$	\\

									& $4.09 \times 10^{-4}$		& $7.43 \times 10^{-2}$		& $1.01 \times 10^{-1}$	\\

Bonn (2000) [\onlinecite{Bonn2000}]			& $1.76 \times 10^{-4}$		& $8.47 \times 10^{-3}$		& $1.68 \times 10^{-2}$	\\

Kruglyak (2005) [\onlinecite{Kruglyak2005}]	& $1.34 \times 10^{-3}$		& $1.57 \times 10^{-3}$		& 	\\

Ligges (2009/11) [\onlinecite{Ligges2009,Ligges2011}]			& $1.76 \times 10^{-4}$		& $5.30 \times 10^{-3}$		& 	\\

Shen (2015) [\onlinecite{Shen2015}]			& $1.41 \times 10^{-3}$		& $1.14 \times 10^{-3}$		& $2.09 \times 10^{-2}$	\\

Nakamura (2016) [\onlinecite{Nakamura2016}]			& $3.00 \times 10^{-3}$		& $2.82 \times 10^{-2}$		& 	\\

Mo (2018) [\onlinecite{Mo2018}]			& $1.76 \times 10^{-4}$		& $9.05 \times 10^{-3}$		& $4.21 \times 10^{-2}$	\\

Obergfell (2020) [\onlinecite{Obergfell2020}]	& $1.41 \times 10^{-3}$		& $1.12 \times 10^{-3}$		& $6.52 \times 10^{-2}$	\\

Suemoto (2021) [\onlinecite{Suemoto2021}]	& $3.06 \times 10^{-3}$		& $3.23 \times 10^{-3}$		& $1.59 \times 10^{-2}$	\\

\hline
\hline

\end{tabular}
\smallskip
\end{table}

\begin{table}[h]
\caption{Values of $\bdivd$ and range of $\alpha$ values (indicated by $\alpha_{\scriptscriptstyle \min}$ and $\alpha_{\scriptscriptstyle \max}$) in experimental investigations of room-temperature Ag excited by sub-ps laser pulses.  Experiments with only $\alpha_{\scriptscriptstyle \min}$ indicated were carried out at a single excitation level.  \label{table2}}
\vspace{0.2cm}
\begin{tabular}{l@{\hspace{0.5cm}}l@{\hspace{0.5cm}}l@{\hspace{0.5cm}}l@{\hspace{0.05cm}}r@{\hspace{0.5cm}}r@{\hspace{0.5cm}}r@{\hspace{0.5cm}}r@{\hspace{0.5cm}}r@{\hspace{0.5cm}}r}

\multicolumn{4}{c} {\textbf{Ag}}		 \\	

\hline
\hline  
   
Study 	&	$\bdivd$  	& $\alpha_{\scriptscriptstyle \min}$ 	&  $\alpha_{\scriptscriptstyle \max}$    \\

\hline

\rule{0pt}{2.5ex}Riffe (1993) [\onlinecite{Riffe1993}]	& $3.96 \times 10^{-4}$		& $2.82 \times 10^{-3}$		& $1.04 \times 10^{-1}$	\\

Groneveld (1995) [\onlinecite{Groeneveld1995}]	& $3.96 \times 10^{-4}$		& $3.05 \times 10^{-5}$		& $1.31 \times 10^{-4}$	\\

Hohlfeld (1996) [\onlinecite{Hohlfeld1996}]	& $4.16 \times 10^{-4}$		& $4.88 \times 10^{-2}$		& $9.29 \times 10^{-2}$	\\

Del Fatti (1998) [\onlinecite{Fatti1998}]	& $1.04 \times 10^{-3}$		& $1.46 \times 10^{-5}$		& $1.82 \times 10^{-3}$	\\

Del Fatti (2000) [\onlinecite{Fatti2000}]	& $1.04 \times 10^{-3}$		& $1.45 \times 10^{-5}$		& $5.78 \times 10^{-4}$	\\

Kruglyak (2005) [\onlinecite{Kruglyak2005}]	& $8.11 \times 10^{-4}$		& $1.64 \times 10^{-3}$		& 	\\

Owens (2010) [\onlinecite{Owens2010}]	& $4.57 \times 10^{-4}$		& $4.61 \times 10^{-3}$		& $1.82 \times 10^{-2}$	\\

	& $2.77 \times 10^{-4}$		& $1.77 \times 10^{-2}$		& 	\\
	
Ligges (2011) [\onlinecite{Ligges2011}]			& $1.06 \times 10^{-4}$		& $2.45 \times 10^{-3}$		& 	\\

Oguz Er (2013) [\onlinecite{Er2013}]	& $3.16 \times 10^{-5}$		& $1.02 \times 10^{-2}$		& $1.60 \times 10^{-2}$	\\

Suemoto (2021) [\onlinecite{Suemoto2021}]	& $1.85 \times 10^{-3}$		& $3.21 \times 10^{-3}$		& $1.59 \times 10^{-2}$	\\

\hline
\hline

\end{tabular}
\smallskip
\end{table}

\begin{table}[h]
\caption{Values of $\bdivd$ and range of $\alpha$ values (indicated by $\alpha_{\scriptscriptstyle \min}$ and $\alpha_{\scriptscriptstyle \max}$) in experimental investigations of room-temperature Au excited by sub-ps laser pulses.  Experiments with only $\alpha_{\scriptscriptstyle \min}$ indicated were carried out at a single excitation level.  \label{table3}}
\vspace{0.2cm}
\begin{tabular}{l@{\hspace{0.5cm}}l@{\hspace{0.5cm}}l@{\hspace{0.5cm}}l@{\hspace{0.05cm}}r@{\hspace{0.5cm}}r@{\hspace{0.5cm}}r@{\hspace{0.5cm}}r@{\hspace{0.5cm}}r@{\hspace{0.5cm}}r}

\multicolumn{4}{c} {\textbf{Au}}		 \\	

\hline
\hline  
   
Study 	&	$\bdivd$  	& $\alpha_{\scriptscriptstyle \min}$ 	&  $\alpha_{\scriptscriptstyle \max}$    \\

\hline

\rule{0pt}{2.5ex}Schoenlein (1987) [\onlinecite{Schoenlein1987}]	& $4.47 \times 10^{-4}$		& $7.34 \times 10^{-4}$		& $7.29 \times 10^{-3}$	\\

Brorson (1987) [\onlinecite{Brorson1987}]	& $4.47 \times 10^{-4}$		& $1.51 \times 10^{-3}$		& $2.94 \times 10^{-3}$	\\

Elsayed-Ali (1991) [\onlinecite{Elsayed-Ali1991}]	& $4.16 \times 10^{-4}$		& $9.72 \times 10^{-4}$		& $1.57 \times 10^{-2}$	\\

Fann (1992) [\onlinecite{Fann1992}]	& $5.47 \times 10^{-4}$		& $4.27 \times 10^{-3}$		& $1.06 \times 10^{-2}$	\\

Juhasz (1992) [\onlinecite{Juhasz1992}]	& $4.16 \times 10^{-4}$		& $1.38 \times 10^{-4}$		& 	\\

Juhasz (1993) [\onlinecite{Juhasz1993}]	& $4.26 \times 10^{-4}$		& $1.24 \times 10^{-4}$		& $2.43 \times 10^{-4}$	\\

Girardeau-Montaut (1993) [\onlinecite{Girardeau-Montaut1993}]	& $2.73 \times 10^{-5}$		& $2.22 \times 10^{-4}$		& $8.85 \times 10^{-4}$	\\

Sun (1993) [\onlinecite{Sun1993}]	& $1.79 \times 10^{-3}$		& $4.32 \times 10^{-5}$		& $9.18 \times 10^{-3}$	\\

Sun (1994) [\onlinecite{Sun1994}]	& $1.22 \times 10^{-3}$		& $7.79 \times 10^{-5}$		& 	\\

							& $2.16 \times 10^{-3}$		& $1.02 \times 10^{-4}$		& 	\\

Wang (1994) [\onlinecite{wang1994}]	& $4.47 \times 10^{-4}$		& $1.85 \times 10^{-2}$		& $2.16 \times 10^{-1}$	\\

Groneveld (1995) [\onlinecite{Groeneveld1995}]	& $4.26 \times 10^{-4}$		& $2.73 \times 10^{-5}$		& $1.18 \times 10^{-4}$	\\

Hohlfeld (1996) [\onlinecite{Hohlfeld1996}]	& $4.47 \times 10^{-4}$		& $7.89 \times 10^{-2}$		& $1.45 \times 10^{-1}$	\\

Hohlfeld (1997) [\onlinecite{Hohlfeld1997}]	& $1.14 \times 10^{-4}$		& $4.75 \times 10^{-3}$		& $1.22 \times 10^{-2}$	\\

Maznev (1997) [\onlinecite{Maznev1997}]	& $9.15 \times 10^{-4}$		& $9.84 \times 10^{-2}$		& 	\\

Moore (1999) [\onlinecite{Moore1999}]	& $9.15 \times 10^{-4}$		& $6.58 \times 10^{-5}$		& $6.28 \times 10^{-3}$	\\

Hibara (1999) [\onlinecite{Hibara1999}]	& $1.14 \times 10^{-4}$		& $2.21 \times 10^{-3}$		& $1.29 \times 10^{-2}$	\\

Hostetler (1999) [\onlinecite{Hostetler1999}]	& $8.81 \times 10^{-4}$		& $5.92 \times 10^{-4}$		& $1.06 \times 10^{-3}$	\\

Bonn (2000) [\onlinecite{Bonn2000}]			& $1.14 \times 10^{-4}$		& $1.16 \times 10^{-2}$		& $2.29 \times 10^{-2}$	\\

Del Fatti (2000) [\onlinecite{Fatti2000}]	& $1.12 \times 10^{-3}$		& $1.44 \times 10^{-5}$		& $5.78 \times 10^{-4}$	\\

Smith (2000) [\onlinecite{Smith2000}]	& $8.81 \times 10^{-4}$		& $2.01 \times 10^{-3}$		& $4.81 \times 10^{-3}$	\\

Smith (2001) [\onlinecite{Smith2001}]	& $9.15 \times 10^{-4}$		& $2.84 \times 10^{-4}$		& $1.42 \times 10^{-3}$	\\

Guo (2001) [\onlinecite{Guo2001}]			& $9.15 \times 10^{-4}$		& $1.02 \times 10^{-1}$		& 	\\

Ibrahim (2004) [\onlinecite{Ibrahim2004}]		& $9.15 \times 10^{-4}$		& $1.95 \times 10^{-4}$		& $2.39 \times 10^{-3}$	\\

Kruglyak (2005) [\onlinecite{Kruglyak2005}]	& $8.72 \times 10^{-4}$		& $1.20 \times 10^{-3}$		& 	\\

Hopkins (2007) [\onlinecite{Hopkins2007}]		& $9.15 \times 10^{-4}$		& $1.47 \times 10^{-4}$		& $3.88 \times 10^{-3}$	\\

Ma (2010) [\onlinecite{Ma2010}]			& $9.15 \times 10^{-4}$		& $6.67 \times 10^{-5}$		& $2.00 \times 10^{-4}$	\\

Hopkins (2011) [\onlinecite{Hopkins2011}]		& $8.65 \times 10^{-4}$		& $1.00 \times 10^{-3}$		& $7.83 \times 10^{-3}$	\\

Chen (2011) [\onlinecite{Chen2011a}]			& $1.14 \times 10^{-4}$		& $2.66 \times 10^{-2}$		& $1.11 \times 10^{-1}$	\\


Ligges (2009/11) [\onlinecite{Ligges2009,Ligges2011}]			& $1.14 \times 10^{-4}$		& $8.56 \times 10^{-3}$		& 	\\

Della Valle (2012) [\onlinecite{DellaValle2012}]	& $1.24 \times 10^{-3}$		& $1.53 \times 10^{-3}$		& 	\\

Kolomenskii (2013) [\onlinecite{Kolomenskii2013}]	& $9.15 \times 10^{-4}$		& $1.51 \times 10^{-2}$		& $4.25 \times 10^{-2}$	\\

										& $1.14 \times 10^{-4}$		& $2.06 \times 10^{-2}$		& 	\\
										
Hopkins (2013) [\onlinecite{Hopkins2013}]	& 	$1.14 \times 10^{-4}$		& $1.47 \times 10^{-3}$		& $1.85 \times 10^{-2}$	\\

Guo (2014) [\onlinecite{Guo2014}]	& 		$9.15 \times 10^{-4}$		& $1.77 \times 10^{-3}$		& $7.06 \times 10^{-3}$	\\
										
White (2014) [\onlinecite{White2014}]		& $9.15 \times 10^{-4}$		& $3.32 \times 10^{-5}$		& 	\\

Chase (2016) [\onlinecite{Chase2016}]		& $9.15 \times 10^{-4}$		& $9.15 \times 10^{-3}$		& 	\\

Nakamura (2016) [\onlinecite{Nakamura2016}]	& $1.95 \times 10^{-3}$		& $7.41 \times 10^{-3}$		& $2.00 \times 10^{-2}$	\\

Sokolowski-Tinten (2017) [\onlinecite{Sokolowski-Tinten2017}]		& $1.14 \times 10^{-4}$		& $2.53 \times 10^{-3}$		& $2.64 \times 10^{-2}$	\\

Heilpern (2018) [\onlinecite{Heilpern2018}]	& $3.09 \times 10^{-3}$		& $1.37 \times 10^{-3}$		& 	\\

Yao (2018) [\onlinecite{Yao2018}]			& $9.15 \times 10^{-4}$		& $2.76 \times 10^{-3}$		& $1.44 \times 10^{-2}$	\\

Sielcken (2020) [\onlinecite{Sielcken2020}]	& $4.59 \times 10^{-2}$		& $5.46 \times 10^{-2}$		& $1.12 \times 10^{-1}$	\\

Lu (2020) [\onlinecite{Lu2020}]				& $9.15 \times 10^{-4}$		& $1.09 \times 10^{-1}$		& 	\\

Tomko (2021) [\onlinecite{Tomko2021}]		& $2.51 \times 10^{-4}$		& $2.86 \times 10^{-3}$		& 	\\

Suemoto (2021) [\onlinecite{Suemoto2021}]	& $1.99 \times 10^{-3}$		& $3.86 \times 10^{-3}$		& $1.89 \times 10^{-2}$	\\

Grauby (2022) [\onlinecite{Grauby2022}]		& $9.15 \times 10^{-4}$		& $1.12 \times 10^{-3}$		& $8.79 \times 10^{-2}$	\\


\hline
\hline

\end{tabular}
\smallskip
\end{table}

\begin{table}[h]
\caption{Values of $\bdivd$ and range of $\alpha$ values (indicated by $\alpha_{\scriptscriptstyle \min}$ and $\alpha_{\scriptscriptstyle \max}$) in experimental investigations of room-temperature Al excited by sub-ps laser pulses.  Experiments with only $\alpha_{\scriptscriptstyle \min}$ indicated were carried out at a single excitation level.  \label{table4}}
\vspace{0.2cm}
\begin{tabular}{l@{\hspace{0.5cm}}l@{\hspace{0.5cm}}l@{\hspace{0.5cm}}l@{\hspace{0.05cm}}r@{\hspace{0.5cm}}r@{\hspace{0.5cm}}r@{\hspace{0.5cm}}r@{\hspace{0.5cm}}r@{\hspace{0.5cm}}r}

\multicolumn{4}{c} {\textbf{Al}}		 \\	

\hline
\hline  
   
Study 	&	$\bdivd$  	& $\alpha_{\scriptscriptstyle \min}$ 	&  $\alpha_{\scriptscriptstyle \max}$    \\

\hline

\rule{0pt}{2.5ex}Riffe (1993) [\onlinecite{Riffe1993}]	& $1.01 \times 10^{-3}$		& $7.54 \times 10^{-3}$		& $1.71 \times 10^{-1}$	\\

Tas (1994) [\onlinecite{Tas1994}]	& $1.07 \times 10^{-3}$		& $3.02 \times 10^{-4}$		& 	\\

Hostetler (1999) [\onlinecite{Hostetler1999}]	& $2.08 \times 10^{-3}$		& $8.93 \times 10^{-3}$		& $1.22 \times 10^{-2}$	\\

Stevens (2005) [\onlinecite{Stevens2005}]	& $2.08 \times 10^{-3}$		& $9.43 \times 10^{-4}$		& 	\\

Park (2005) [\onlinecite{Park2005}]	& $2.08 \times 10^{-3}$		& $2.39 \times 10^{-2}$		& 	\\

Nie (2006) [\onlinecite{Nie2006}]	& $2.08 \times 10^{-3}$		& $1.36 \times 10^{-2}$		& 	\\

Ma (2013) [\onlinecite{Ma2013}]	& $2.16 \times 10^{-3}$		& $5.75 \times 10^{4}$		& 	\\

Waldecker (2016) [\onlinecite{Waldecker2016}]		& $2.16 \times 10^{-3}$		& $1.35 \times 10^{-2}$		& $8.27 \times 10^{-2}$	\\

\hline
\hline

\end{tabular}
\smallskip
\end{table}

\section{Analysis of Numerical Relaxation-Time Data}

In Sec. IV.A.2 of the paper we discuss the fitting of our numerical data for $\tau_H \beta$, $\tau_E \beta$, and $\tau_{th} \beta$ vs $\bdivd$ to Eqs.~(30), (36) and (40)/(41), respectively.  Here we provide details of the fitting procedure.  Because normalized relaxation times $\tau \beta$ can vary by several orders of magnitude over the relevant range of $\bdivd$, we utilize $\log(\tau \beta)$ in our analysis.  Specifically, our goodness-of-fit function is defined to be
\be{\eq}
\label{10}
\chi^2 = \sum_i \big[ \log \! \big(\mathbb{F}(\XX,(\bdivd)_i\big) - \log(\tau_i \beta_i) \big]^2,
\en{\eq}
where  $\mathbb{F}$ is the fitting function and $\XX$ represents the set of fitting parameters.  In the analysis of $\tau_H \beta$ and $\tau_{th} \beta$ separate fittings are done for each value of $\gamma$ (= 0.0086, 0.0172, and 0.0344).  In analysis of $\tau_E \beta$ the data obtained at all three values of $\gamma$ are simultaneously fit, and so in this case $\mathbb{F} = \mathbb{F}\big(\XX,\gamma_i,(\bdivd)_i\big)$.

The numerical data and fits for all three relaxation times are shown in Fig.~\ref{Fig1}.  For $\tau_H \beta$ the value of the fitted exponent $a$ in Eq.~(30) is indicated in each panel of Fig.~\ref{Fig1}.  Notice it is quite close to 0.95 for all three cases.  For $\tau_{th} \beta$ the exponent $b$ in Eq.~(40) depends upon $\bdivd$, as expressed by Eq.~(41),
$$
b = \frac{b_0 + b_1 \, \beta/\delta}{1 + b_1 \, \beta / \delta}.
$$
The exponent $b$ for each value of $\gamma$ is plotted vs $\bdivd$ in Fig.~8.  Fitted values of $b_0$ and $b_1$ are reported in Table~\ref{table5}. For $\tau_E \beta$ the best-fit parameters of Eq.~(36) are presented in Table IV.

\begin{table}[h]
\caption{Fitted values of $b_0$ and $b_1$ in Eq.~(41) for three values of $\gamma$.  \label{table5}}

\vspace{0.2cm}

\begin{tabular}{l@{\hspace{0.6cm}}l@{\hspace{0.6cm}}l@{\hspace{0.0cm}}l@{\hspace{0.05cm}}r@{\hspace{0.5cm}}r@{\hspace{0.5cm}}r@{\hspace{0.5cm}}r@{\hspace{0.5cm}}r@{\hspace{0.5cm}}r}

\hline
\hline  
   
$\gamma$ 	&	$b_0$  	& $b_1$ 	   \\

\hline

\rule{0pt}{2.5ex}0.0086	& $0.4173$		& $0.4811$		\\

0.0172	& $0.5145$		& $0.5168$			\\

0.0344	& $0.6475$		& $0.7379$			\\

\hline
\hline

\end{tabular}
\smallskip
\end{table}

We have also fit our data to determine the dependence of $\tau_{th} \delta$ on $\gamma$ in the $\beta \rightarrow 0$ limit ($ee$ scattering only).  For this analysis we have also calculated $\tau_{th} \delta$ at three more values of $\gamma$:  0.01, 0.02, and 0.0274.  These data and a power-law fit are shown in Fig.~\ref{Fig2}.  As the figure shows, $\tau_{th} \delta = 3.33 +0.059 \, \gamma^{-1.83}$ provides an excellent description of the calculated relaxation times.  For comparison Fig.~\ref{Fig2} also displays the Gusev and Wright \cite{Gusev1998} result, which is given by Eq.~(42) as
$$
\tau_{th}^{\scriptscriptstyle G}  \delta = \bigg( \! \frac{2}{\pi} \! \bigg)^{\!\! 5} \bigg[  \frac{\ln(2)}{\gamma}  \bigg]^{\! 2}.
$$
This equation yields values that are within a factor of two of our numerical results.

\renewcommand{\thefigure}{S\arabic{figure}}

\begin{figure*}[h]
\centerline{\includegraphics[scale=0.8]{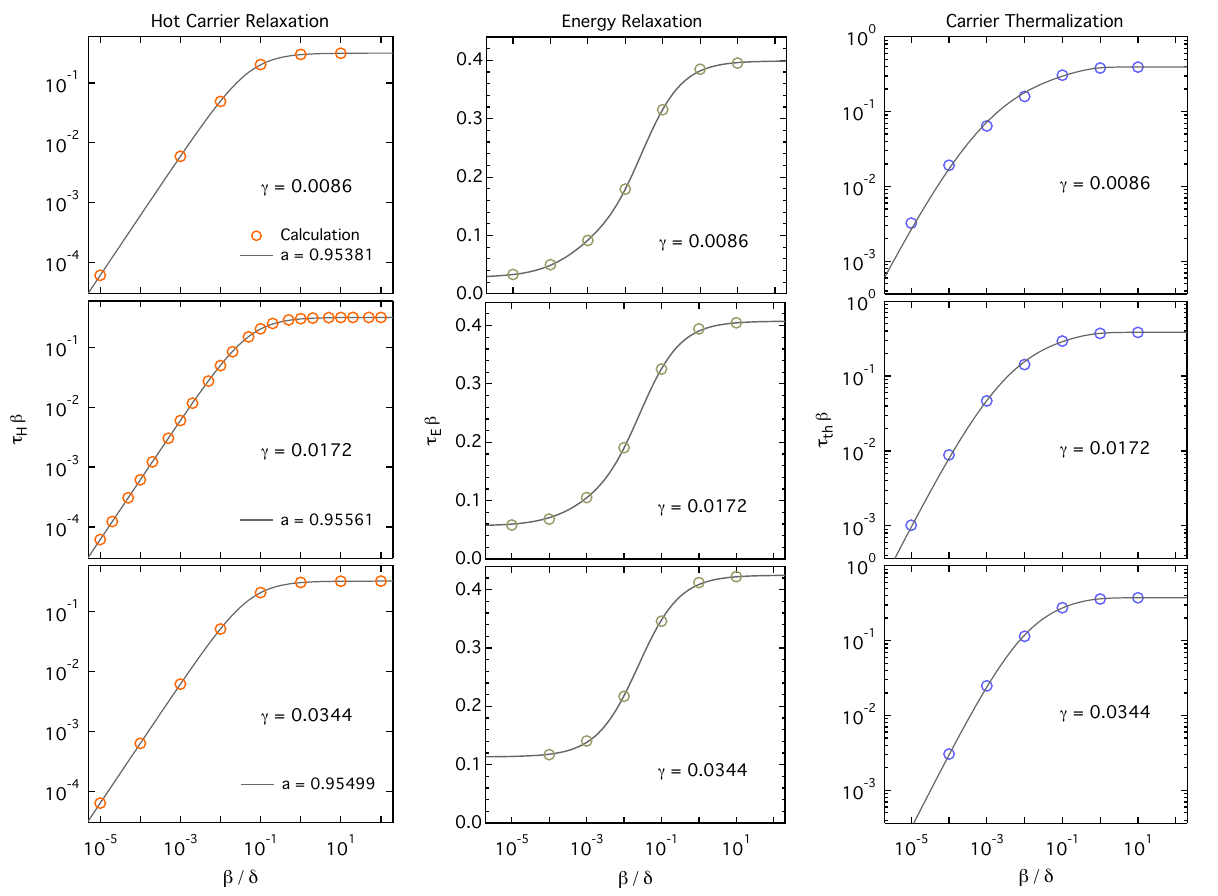}}
\caption{Numerical relaxation-time data (circles) and least-squares fits (smooth curves) of those data.  All relaxations time are normalized by $\beta = K_{ep}/(h \nu)$, the normalized $ep$ interaction strength.}
\label{Fig1}
\end{figure*}

\begin{figure*}[h]
\centerline{\includegraphics[scale=0.5]{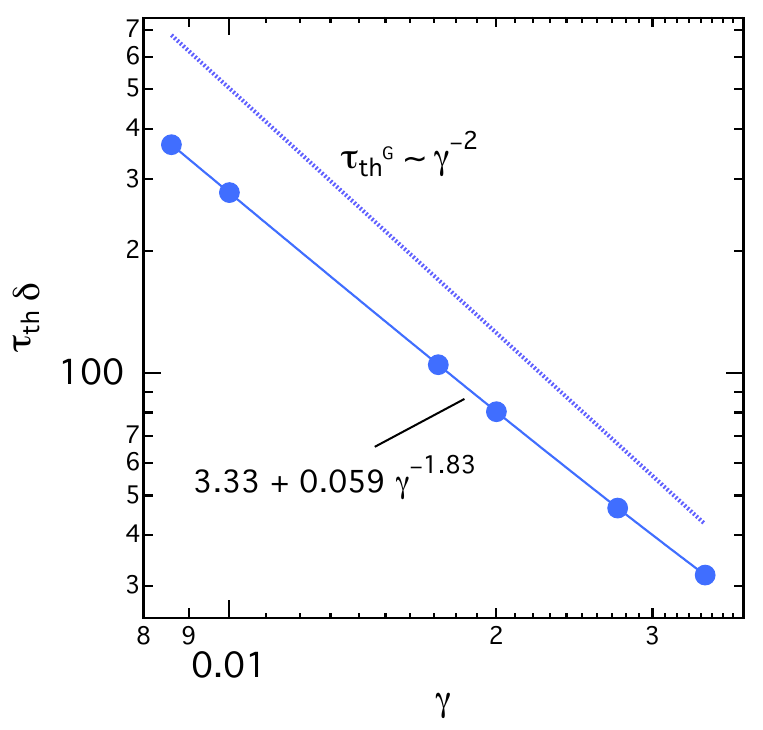}}
\caption{Numerical values of $\tau_{th} \delta$ vs $\gamma$ (circles) in the $\beta \rightarrow 0$ limit and least squares fit (solid line) to a power law.  The $\tau_{th}^{\scriptscriptstyle G}$ result of Gusev and Wright \cite{Gusev1998} is also shown (dotted line).}
\label{Fig2}
\end{figure*}

\section{Initial time dependence of $f(\epsilon,t)$}

Here we derive an approximate expression for the initial time dependence of the distribution function $f(\epsilon,t)$.  As can be ascertained from Figs.~3 and 4 in the paper, the early-time ($t \delta \lesssim 2$) dynamics of the distribution are dominated by electron-electron scattering when $\beta / \delta \le 10^{-2}$, a range that includes all metals in Table I when excited by photon energies in the near-IR, visible, or uv.  For low excitation levels we are thus justified in describing the initial time dependence of the distribution function with an approximate, linearized $ee$ scattering integral derived by Tas and Maris \cite{Tas1994}. We note their scattering integral was also obtained assuming a constant density of states and $k_B T_p \ll h \nu$.  Using that scattering integral we obtain the integro-differential equation
\begin{equation}
\label{E8}
\frac{d\phi(\delta\ep,t)}{dt} = - K_{ee} \bigg[ \frac{(\delta \ep)^2}{2} \phi(\delta \ep,t)  - 3 \! \int_{\delta\ep}^\infty  d(\delta\ep') \, (\delta\ep' - \delta\ep) \phi(\delta\ep',t)  \bigg].
\end{equation}
Recall from the Appendix $\phi(\delta \ep,t) = f(\delta \ep,t) - f_{\scriptscriptstyle \! F\!D}(\delta \ep,T_p)$.  The first term on the right side of Eq.~(\ref{E8}) describes scattering out of states with energy $\delta \ep$ [and is consistent with the single-carrier relaxation time given by Eq.~(9)], while the second describes electron excitation into these same states.  We now set $t = 0$ in Eq.~(\ref{E8}), use the approximate expression for $\phi(\delta \ep,0)$ as given by Eq.~(A5), and thence evaluate the integral to find the initial rate of change in $\phi(\delta \ep, t)$ is given by 
\begin{equation}
\label{E9}
\frac{d\phi(\delta\ep,t)}{dt}\bigg|_{t=0} = K_{ee} \alpha \bigg[ (\delta \ep)^2 - 3 h\nu \, \delta\ep + \frac{3}{2} (h \nu)^2  \bigg].
\end{equation}
To be clear, this expression is valid for $0 \le \delta \ep \le h \nu$ [and for $\delep > h \nu$ the function $\phi(\delep,t)$ is zero].  For small values of $\delta \ep$ the quantity in brackets in Eq.~(\ref{E9}) is positive, indicating that initially $\phi$ initially increases.  Conversely, for large values of $\delta \ep$ the quantity in brackets is negative, indicating an initial decrease in $\phi$.  These two behaviors are separated by the energy
\begin{equation}
\label{E10}
\delep_0 = \big( 3 - \sqrt{3} \big)/2  \, \, h \nu = 0.634 \, h \nu.
\end{equation} 
As mentioned in Sec.~IV.A.1, in part (a) of Fig.~3 and in five panels of Fig.~4 we have marked $\delep_0 / h \nu$ with a vertical arrow.  As can be seen in these figures, for $\delta \ep < \delta \ep_0$ ($\delta \ep > \delta \ep_0$) the distribution function indeed initially increases (decreases).


\newpage

\section*{References}

\bibliography{AlkaliMetals}